\documentclass[fleqn,usenatbib]{mnras}

\usepackage{newtxtext,newtxmath}
\usepackage{lineno}

\usepackage[T1]{fontenc}

\DeclareRobustCommand{\VAN}[3]{#2}
\let\VANthebibliography\thebibliography
\def\thebibliography{\DeclareRobustCommand{\VAN}[3]{##3}\VANthebibliography}
\usepackage{bm}

\usepackage{graphicx}	% Including figure files
\usepackage{amsmath}	% Advanced maths commands
\title[Revealing asymmetry in disc continuum emission]{Revealing asymmetry on midplane of proto-planetary disc through modelling of axisymmetric emission: methodology}

\author[M. Aizawa et al.]{
Masataka Aizawa,$^{1,2,3}$\thanks{E-mail: masataka.aizawa.sw35@vc.ibaraki.ac.jp}
Takayuki Muto,$^{4,5,6}$
Munetake Momose$^{3}$
\\
$^{1}$Tsung-Dao Lee Institute, Shanghai Jiao Tong University, Shengrong Road 520, 201210 Shanghai, P. R. China\\
$^{2}$ RIKEN Cluster for Pioneering Research, 2-1 Hirosawa, Wako, Saitama 351-0198, Japan \\
$^{3}$College of Science, Ibaraki University, 2-1-1 Bunkyo, Mito, Ibaraki 310-8512, Japan \\
$^{4}$Division of Liberal Arts, Kogakuin University, 1-24-2 Nishi-Shinjyuku, Shinjyuku-ku, Tokyo 163-8677, Japan\\
$^{5}$Leiden Observatory, Leiden University, P.O. Box 9513, NL-2300 RA Leiden, The Netherlands\\
$^{6}$Department of Earth and Planetary Sciences, Tokyo Institute of Technology, 2-12-1 Oh-okayama, Meguro-ku, Tokyo 152-8551, Japan\\
}

\date{Accepted XXX. Received YYY; in original form ZZZ}

\pubyear{2024}
  
\begin{document}
\label{firstpage}
\pagerange{\pageref{firstpage}--\pageref{lastpage}}
\maketitle

%\linenumbers

\begin{abstract}
This study proposes an analytical framework for deriving the surface brightness profile and geometry of a geometrically-thin axisymmetric disc from interferometric observation of continuum emission. Such precise modelling facilitates the exploration of faint non-axisymmetric structures, such as spirals and circumplanetary discs. As a demonstration, we simulate interferometric observations of geometrically-thin axisymmetric discs. The proposed method can reasonably recover the injected axisymmetric structures, whereas Gaussian fitting of the same data yielded larger errors in disc orientation estimation. To further test the applicability of the method, it was applied to the mock data for $m=1,2$ spirals and a point source, which are embedded in a bright axisymmetric structure. The injected non-axisymmetric structures were reasonably recovered except for the innermost parts, and the disc geometric parameter estimations were better than Gasussian fitting. The method was then applied to the real data of Elias 20 and AS 209, and it adequately subtracted the axisymmetric component, notably in Elias 20, where substantial residuals remained without our method. We also applied our method to continuum data of PDS 70 to demonstrate the effectiveness of the method. We successfully recovered emission from PDS 70 c consistently with previous studies, and also tentatively discovered new substructures. The current formulation can be applied to any data for disc continuum emission, and aids in the search of spirals and circumplanetary discs, whose detection is still limited. 
\end{abstract}
\begin{keywords}
radio continuum: planetary systems --methods: data analysis --protoplanetary discs 
\end{keywords}

%%%%%%%%%%%%%%%%%%%%%%%%%%%%%%%%%%%%%%%%%%%%%%%%%%%%%%%%%%%%
\section{Introduction}
%%%%%%%%%%%%%%%%%%%%%%%%%%%%%%%%%%%%%%%%%%%%%%%%%%%%%%%%%%%%

High-resolution imaging of proto-planetary discs by the Atacama Large Millimeter/submillimeter Array (ALMA) has revolutionized our view of planet formation. The spatial resolutions, down to $\sim 0.01\arcsec$, have enabled the identification of rings and gap structures in the disc continuum emission  \citep{2015ApJ...808L...3A,andrews2018, huang_ring_2018}. The mechanisms for invoking such annular structures have been under debate, and multiple explanations have been provided, such as a disc-planet interaction \citep{lin1979,goldreich1980,dong2018}, snowline \citep{zhang2015}, sintering \citep{okuzumi2016}, secular gravitational instability \citep{takahashi2014}, and non-ideal MHD effects \citep{flock2015}. Nevertheless, the disc-planet interaction is 
independently supported by recent discoveries of kinematic signals of embedded planets \citep{pinte2018,pinte_2020,pinte_2023}. Here, the signals, usually referred to as  velocity ``kinks'', appear as localized deviations from the Keplerian motion of discs in velocity maps, and there have been dozens of such detections. 

In contrast to multiple detections of rings, gaps, and kinks, the direct detection of embedded planets and the circumplanetary discs within gaps remains limited. Currently, there are a few cases presenting the robust detection of embedded planets in gaps, such as, PDS 70 \citep{keppler2018, wagner2018,benisty2021}, AB Aur \citep{currie2022}, HD 169142 \citep{reggiani2014,hammond2023}, and MWC 758 \citep{wagner2023}. These embedded planets have been somewhat selectively identified around (pre-)transitional discs with the ages greater than 4 Myr. In addition, \cite{andrews2021} searched for circumplanetary discs in gaps of DSHARP discs \citep{andrews2018}, whose typical age is 1 Myr; however, no compelling case remains. 

Another supportive evidence for the planetary hypothesis can be obtained from a planetary spiral, which is excited by planetary gravity. However, despite the discovery of many spirals in discs \citep{muto2012, grady2013,huang_spiral_2018}, only the spirals in systems with detected embedded planets are highly likely to be produced by the planetary gravity. The other spirals can also originate from non-planetary mechanisms, for example, gravity from stellar companions, gravitational instability or stellar flyby. Indeed, the spiral in HD 100453 A is a notable example, where its M-dwarf companion is thought to be the driver for the structure \citep{wagner2015,ruobing2016}. 

Recently, \cite{speedie_obs_2022} searched for planetary spirals in 10 discs with kinematic signatures of embedded planets; however, they did not find any conclusive case. The limited detection of circumplanetary discs and spirals might indicate that the current sensitivities are still inadequate for most of the cases, or certain annular structures may have been caused via non-planetary mechanisms. 

Nevertheless, an observational effort to search for circumplanetary discs and spirals is still essential for testing the planetary hypothesis. However, one obstacle to their detection is that these signals appear as small perturbations to the bright disc emission. This is similar to the situation of searching for a needle in a haystack. Thus, the precise modelling of the background disc emission is necessary to subtract its contribution. 

One reasonable approximation for the background emission is an axisymmetric disc. However, the modelling of a vertically extended axisymmetric disc is difficult because of the geometric effect (e.g., shadows) and additional parameters to be solved (e.g., radial disc heights). Such geometric effects are significant at optical to infrared wavelengths, because of their sensitivity to vertically extended small dust grains. On the other hand, the disc emission at the radio band is well approximated by a geometrically thin disc model with no vertical structure owing to its sensitivity to large grains, which are more settled to the disc midplane. This approximation is to a certain extent justified by the observations of many clear concentric structures of discs in radio continuum emissions  \citep[e.g.,][]{2015ApJ...808L...3A, andrews2018}.

Recently, assuming a geometrically thin and axisymmetric disc, \cite{jennings2020} proposed an algorithm for deriving a radial surface brightness profile of continuum emission in the radio band. Their method can resolve annular structures finer than the standard imaging technique, CLEAN, using simulated and real data \citep{jennings2020,jennings2022,jennings_taurus_2022}. Using their method, \cite{andrews2021} also searched for circumplanetary disc emission from DSHARP data by subtracting the axisymmetric structures of discs. 

Although the method proposed in \cite{jennings2020} can reasonably recover the brightness profile of the disc, there is still room for improvement. Specifically, in their modelling, the geometric parameters of the axisymmetric model, such as an orientation and a central position of a disc, and hyperparameters for the Gaussian Process model that is used to prevent overfitting must be fixed. This limitation can be problematic, because a slight change in geometric parameters of the disc model, for example, several mas in a central position, can introduce the apparent non-axisymmetric structures in the image with their axisymmetric structure being subtracted \citep{andrews2021}. Such false structures can be degenerate with the real structures, and may disrupt the detection of spirals and circumplanetary discs. Till date, in the frameworks proposed for axisymmetric disc models, these geometric parameters or hyperparameters for regularization have been manually tuned \citep{jennings2020,andrews2021}, or estimated through parameterized models \citep{zhang2015,tazzari2017,jennings2020, kanagawa2021} and image-based analysis \citep{huang_ring_2018}.  

Therefore, this study proposed an analytical framework to derive all of the parameters for a geometrically-thin axisymmetric disc and hyperparameters for the Gaussian Process kernel from observations assuming a geometrically thin disc. This study employed the formulation for the inverse modelling presented in \cite{kawahara2020}, which recovered a planetary surface map and its spin and orbital geometry from planetary reflection light. Interestingly, the problem considered in that study is mathematically similar to the current problem. Thus, we can apply their methodology for the current problem. We demonstrate the feasibility of the current method via its application to mock and real data. In addition, we perform injection and recovery experiments for circumplanetary discs and spirals, and discuss the applicability and limitations of the method.

The remainder of this paper is organized as follows. In Sec \ref{sec:form}, the method for estimating an axisymmetric structure from visibilities is formulated.  In Sec \ref{sec:assym}, the methodology for studying non-axisymmetric features in a residual image is discussed. In Sec \ref{sec:sim}, the feasibility of the proposed method for mock observations is investigated. In Sec \ref{sec:sim_non}, circumplanetary discs and spirals are injected to mock data, and the ability to recover the structures is examined. In Sec \ref{sec:apptoreal1}, our method was also applied to the real data for Elias 20 and AS 209. Sec \ref{sec:pds70} presents an analysis of the PDS 70 data, demonstrating the recovery of circumplanetary emission. Finally, the conclusions and future improvements are presented in Sec \ref{sec:summ}. 
 %%%%%%%%%%%%%%%%%%%%%%%%%%%%%%%%%%%%%%%%%%%%%%%%%%%%%%%%%%%%
 
\section{Analytical formulation for parameters for a geometrically thin axisymmetric disc} \label{sec:form}

\subsection{Model}
We model a geometrically thin axisymmetric disc, whose parameters are composed of the brightness profile 
$\bm{a}$ and its geometry $\bm{g}$. Here, $\bm{a} = \{I(r_{k})\}$ for $k=1, 2, ..., N$ is a vector obtained by discretizing the radial surface brightness profile $I(r)$, and $r_{k}$ is the $k$-th collocation point of the Fourier-Bessel series:
 \begin{eqnarray}
     r_{k} = R_{\rm out}\frac{j_{0k}}{j_{0N}}, 
 \end{eqnarray}
where $R_{\rm out}$ is the outer boundary of $I(r)$ with $I(r>R_{\rm out})=0$, and $j_{0k}$ is the $k$-th root of the zero-order Bessel function of the first kind $J_{0}(r)$;  $J_{0}(j_{0k}) = 0$. 

The disc geometry is specified by the disc orientation and the central position: $\bm{g} = (\Delta x_{\rm cen}, \Delta y_{\rm cen},\cos i, {\rm PA})$. Here, the disc centre $(\Delta x_{\rm cen}, \Delta y_{\rm cen})$ is assumed to be shifted from a phase centre of the observation, and the disc orientation is specified by the position angle PA and  inclination $i$. PA is defined as the angle of the major axis measured counter-clockwise from the north direction, and $i$ is defined as the angle between the line of sight and the  axis normal to the disc plane. Further, the formulation adopts $\cos i$, which represents the aspect ratio for the disc ellipse.

\subsection{Visibilities for axisymmetric disc} \label{sec:forward}
\subsubsection{Face-on disc} \label{sec:face-on}
As the simplest problem, we consider the face-on disc with no positional offset from the phase centre; $\bm{g} = (\Delta x_{\rm cen} = 0\arcsec, \Delta y_{\rm cen}= 0\arcsec, \cos i=1, {\rm PA}=0)$. Visibilities for an azimuthally symmetric geometrically thin disc with a brightness profile $I(r)$ are expressed as the Hankel transformation \citep{2017isra.book.....T,jennings2020}:  
\begin{eqnarray}
V(u,v) &=& \int_{-\infty}^{\infty} \int_{-\infty}^{\infty} I(x, y) \exp(-2\pi j (ux + vy)) \bm{dx}  \label{eq:von_zer}\\
&=&2 \pi \int_{0}^{\infty} J_{0}(2 \pi r q_{\cos i =1}) I(r) r dr,  \label{eq:v_rad} 
\end{eqnarray}
where $V(u,v)$ is the complex visibility at a spatial frequency $(u,v)$, 
$q_{\cos i =1} = \sqrt{u^{2} + v^{2}}$ is the deprojected spatial frequency for the face-on disc, and $J_{0}$ is the zero-order Bessel function of the first kind.

We assume $M$ observational spatial frequencies $\{\bm{u}, \bm{v}\}_{j} = (u_{j}, v_{j})$, where $j=1, 2, ..., M$. At these spatial frequencies, we compute model visibilities, denoted by  $(\bm{V}=\{V_{j}(u_{j}, v_{j})\}$. 
Using equation (\ref{eq:v_rad}) and assuming $\bm{a}$, the model visibilities $\bm{V}$ are expressed as follows \citep[e.g.,][]{jennings2020}: 
\begin{eqnarray}
     \bm{V} = \bm{H}\bm{a},  \label{eq:V_Ha}
\end{eqnarray}
where the matrix for the Hankel transformation $\bm{H}$ is expressed as follows: 
\begin{eqnarray}
   \{\bm{H}\}_{j,k} &=&  \frac{4 \pi R_{\rm out}^{2}}{j_{0 (N+1)}^{2} J_{1}^{2}(j_{0k})} J_{0} \left(2 \pi q_{j, \cos i =1} r_{k} \right), \label{eq:v_ha} \\
   q_{j, \cos i =1} &=& \sqrt{u_{j}^{2} + v_{j}^{2}}, \label{eq:q}
\end{eqnarray}
where $J_{1}$ is the first-order Bessel function of the first kind. 

\subsubsection{Inclined disc}
%-----------------------------Figure Start---------------------------
\begin{figure*}
\begin{center}
\includegraphics[width=0.85\linewidth]{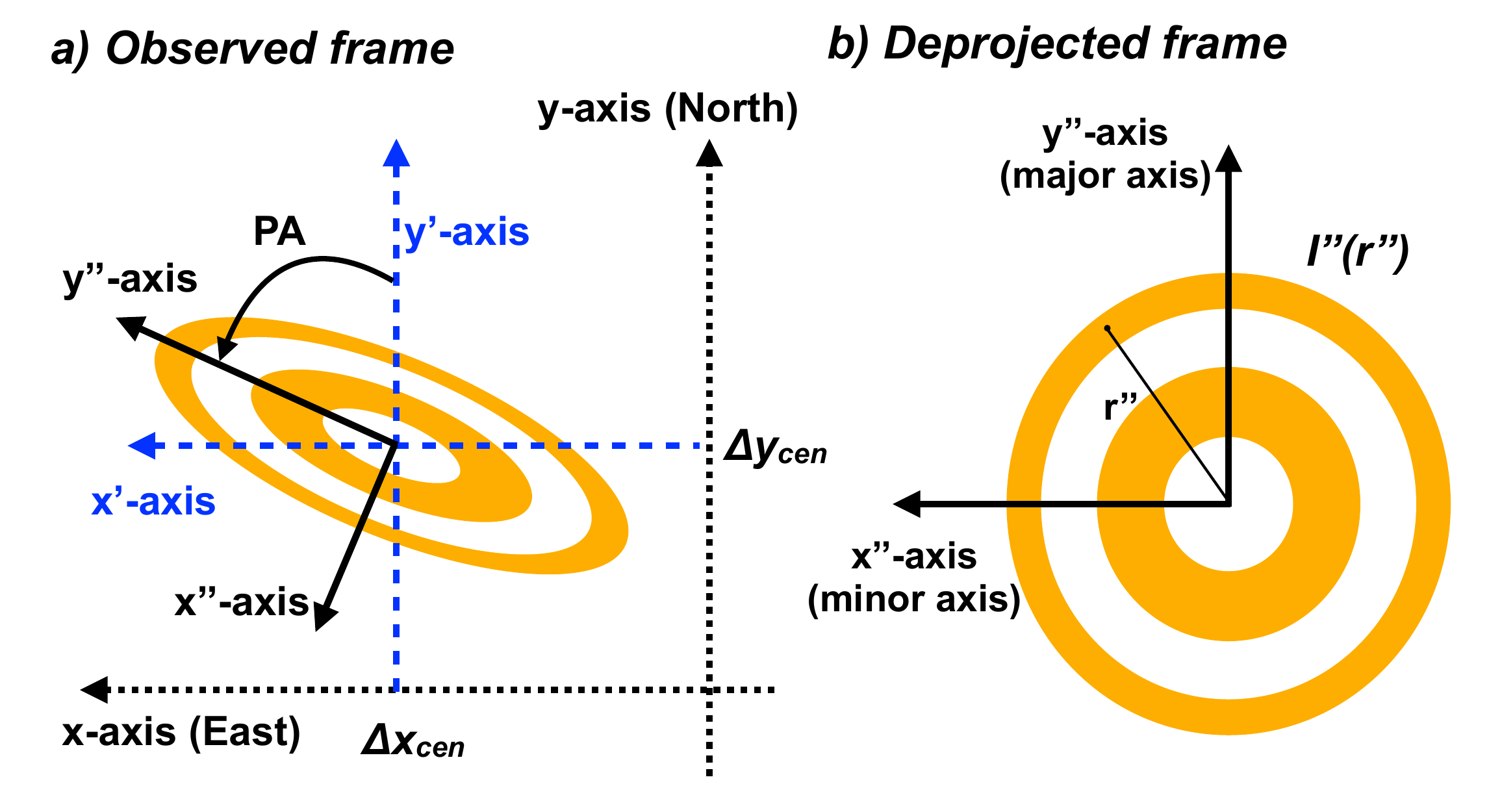}
\end{center}
\caption{Schematic of a disc in observed (a) and deprojected frames (b). The disc is inclined with $\cos i$ and oriented by PA with respect to the $x$-axis in the projected frame.  The origin of $(x,y)$ is set to be the phase centre, and $(x', y')$ is shifted from  $(x,y)$  by $(\Delta x_{\rm cen}, \Delta y_{\rm cen})$. $(x'',y'')$ is the coordinate for the deprojected frame, and the brightness profile is assumed to be only radially dependent on $I''(r'')$. }
\label{fig:disc_def}
\end{figure*}
%-----------------------------Figure End------------------------------
The computations of model visibilities can be simply extended to the case of the inclined disc. For the calculation, we prepare three different coordinates $(x, y)$, $(x', y')$, and $(x'', y'')$; Fig. \ref{fig:disc_def} shows a schematic of the coordinates and disc. Here, $(x, y)$ is the observational coordinate system with the phase centre being the origin, $(x', y')$ is the shifted coordinate system with the disc centre being the origin, and $(x'', y'')$ is the deprojected coordinate system according to (PA, $i$).  $(x, y)$ and $(x', y')$ are related as $(x, y) = ( x'+ \Delta x_{\rm cen} , y'+ \Delta y_{\rm cen} )$, and $(x', y')$ and $(x'', y'')$ are converted as follows:
\begin{eqnarray}
\left(
\begin{array}{c}
   x''\\
   y''
\end{array}
\right)
=
\left(
\begin{array}{cc}
    1/\cos i& 0 \\
    0 & 1
\end{array}\right)
\left(
\begin{array}{cc}
\cos ({\rm PA})& -\sin ({\rm PA})\\
    \sin ({\rm PA})& \cos({\rm PA})
\end{array}\right)
\left(
\begin{array}{c}  
x'\\
 y'
\end{array}\right)
\end{eqnarray}

The brightness profiles for the coordinates $I(x, y)$, $I'(x', y')$, and $I''(x'', y'')$ are assumed to be 
invariant to coordinate transformations:
\begin{eqnarray}
 I(x, y) =    I'(x', y')  = I''(x'', y''). 
\end{eqnarray}
This assumption is a mathematical hypothesis for modeling purposes, and it is not necessarily consistent with the physical picture of what would actually occur if the disc were observed from the face-on view\footnote{Nevertheless, the assumption is physically consistent when the disc is a geometrically-thin, optically thick disc.}.

Assuming the brightness profile in a deprojected frame $I''(r'')$ with $r''\equiv \sqrt{x''^{2} +y''^{2}}$, we aimed to compute the model visibilities. First, we transform equation (\ref{eq:von_zer}) as follows:
\begin{eqnarray}
    V(u,v) &=& \int_{-\infty}^{\infty}\int_{-\infty}^{\infty} I(x, y) \exp(-2\pi j (ux + vy)) \bm{dx}  \nonumber \\ 
    &=& \exp(-2\pi j (u\Delta x_{\rm cen} + v\Delta y_{\rm cen} )) \nonumber \\
    &&\int_{-\infty}^{\infty}\int_{-\infty}^{\infty} I'(x', y') \exp(-2\pi j (u x' +  v y' )) \bm{dx'},  \nonumber  \\
    &=& \exp(-2\pi j (u\Delta x_{\rm cen} + v\Delta y_{\rm cen} )) \nonumber \\
    &&\int_{-\infty}^{\infty}\int_{-\infty}^{\infty} I''(x'', y'')\exp(-2\pi j (u' x'' +  v' y'' )) \nonumber\\
    &&|\cos i| \bm{dx''},  \label{eq:vis_eq_2}
\end{eqnarray} 
where we define $(u', v')$ as follows: 
\begin{eqnarray}
\left(
\begin{array}{c}
   u'\\
   v'
\end{array}\right)
=\left(
\begin{array}{cc}
    \cos i& 0 \\
    0 & 1
\end{array}\right)
\left(
\begin{array}{cc}
    \cos ({\rm PA})& -\sin ({\rm PA})\\
    \sin ({\rm PA})& \cos({\rm PA})
\end{array}\right)
\left(
\begin{array}{cc}
   u\\
   v
\end{array}\right). \label{eq:uv_convert}
\end{eqnarray}

Using equation (\ref{eq:v_rad}), equation (\ref{eq:vis_eq_2}) is reduced as follows:
\begin{eqnarray}
    V(u, v) &=2 \pi|\cos i| \exp(-2\pi j (u\Delta x_{\rm cen} + v\Delta y_{\rm cen} )) \nonumber \\
    &\times \displaystyle \int_{0}^{\infty} J_{0}(2 \pi r'' q) I''(r'') r'' dr'',  \label{eq:calc}
\end{eqnarray}
where we define a deprojected radial spatial frequency $q$ as follows: 
\begin{eqnarray}
q =\sqrt{u'^{2} +v'^{2}}. 
\end{eqnarray}
Note that $q_{\cos i =1}$ is the special case of $q$ with $(\cos i, {\rm PA}) = (1, 0)$. In the calculation, we assume an axisymmetric disc, whose brightness profile with $I''(x'', y'') = I''(r'')$, and we discretize $I''(r'')$ using $\bm{a}$ in the same manner as that presented in Sec \ref{sec:face-on}. 

 For $\bm{V}_{\rm real}$ and $ \bm{V}_{\rm imag}$, which are defined as real and imaginary parts of $\bm{V}$, respectively, we derive a linear equation as follows: 
\begin{eqnarray}
\left(\begin{array}{c}
    \bm{V}_{\rm real} \\
    \bm{V}_{\rm imag} 
\end{array}\right)
= 
\left(\begin{array}{cc}
 \bm{C}_{\rm real}\bm{H'} \\
 \bm{C}_{\rm imag}\bm{H'} \\
\end{array}\right)
 \bm{a}, \label{eq:linear_inclined}
\end{eqnarray} 
where $\bm{H'}$, $\bm{C}_{\rm real}$, and $\bm{C}_{\rm imag}$ are defined as follows: 
\begin{eqnarray}
    \{\bm{H'}\}_{j,k} &=&  \displaystyle \frac{4 \pi R_{\rm out}^{2}}{j_{0( N+1)}^{2} J_{1}^{2}(j_{0k})} J_{0} \left(2 \pi q_{j} r_{k} \right),  \\
  \{\bm{C}_{\rm real}\}_{j, k} &=& |\cos i| \cos(-2\pi (\Delta x_{\rm cen} u_{j} + \Delta y_{\rm cen} v_{j})) \delta_{j,k}, \nonumber \\ \\ 
 \{ \bm{C}_{\rm imag}\}_{j, k} &=&  |\cos i| \sin(-2\pi (\Delta x_{\rm cen} u_{j} + \Delta y_{\rm cen} v_{j})) \delta_{j,k}. \nonumber \\ 
\end{eqnarray}

\subsection{Inverse modelling} \label{sec:inverse}
On the basis of the forward modelling in Sec \ref{sec:forward}, we consider inverse modelling for brightness profile $\bm{a}$ and geometric parameters
$\bm{g} =(\Delta x_{\rm cen}, \Delta y_{\rm cen}, \cos i$, {\rm PA}) from visibilities.  

\subsubsection{Data with Gaussian noise}
Let us assume that there are $M$ visibilities $\bm{d}_{\rm obs} = \{d_{{\rm obs}, i} (u_{i}, v_{i})\}$ ($i=1, 2, ..., M$), with the noise of each visibility data obtained via the standard deviation of $\bm{\sigma}_{\rm obs} = \{\sigma_{{\rm obs}, i} \}$.  We separate the real and imaginary parts of the visibilities as $ \bm{d}_{\rm real}\equiv \mathfrak{Re}(\bm{d}_{\rm obs})$ and $\bm{d}_{\rm imag} \equiv  \mathfrak{Im}( \bm{d}_{\rm obs})$, respectively. We assume that observational noises for $\bm{d}_{\rm real}$ and  $\bm{d}_{\rm imag}$  obey the multivariate normal distribution. 
Here, the multivariate normal distribution for a $N$-dimensional random vector $\bm{x}$ is expressed as follows: 
 \begin{eqnarray}
    \mathcal{N}(\bm{x}|\bm{\mu}, \bm{\Sigma}) = \frac{1}{(2\pi)^{N/2}  |{\rm det} \bm{\Sigma}|^{1/2}} \exp \left( -\frac{1}{2} (\bm{x} - \bm{\mu})^{T} \Sigma^{-1} (\bm{x} - \bm{\mu}) \right), 
\end{eqnarray}
where $\bm{\mu}$ is the mean vector and $\bm{\Sigma}$ is the covariance matrix. For $\bm{d}_{\rm real}$ and  $\bm{d}_{\rm imag}$, we assume the covariance matrix to be 
$\bm{\Sigma} = {\bm{\Sigma}_{\rm obs}} \equiv \{\delta_{i, j}/\sigma_{i}^{2}\}$. 

\subsubsection{Gaussian prior on brightness profile}
In interferometric observations, a brightness profile $\bm{a}$ is generally under-constrained. If it is directly obtained by solving a linear equation, for example, equation (\ref{eq:V_Ha}), the solution can diverge, leading to overfitting. This can be avoided by introducing regularization in optimization. Specifically, we assume the Gaussian Process kernel with a set of hyperparameters $\bm{\theta}=(\alpha, \gamma)$ on $\bm{a}$ as follows: 
\begin{eqnarray}
    p(\bm{a} | \bm{\theta}) &=& \mathcal{N}(\bm{a} | \bm{0},
    \bm{\Sigma}_{a}(\bm{\theta})), \label{eq:a_prior} 
\end{eqnarray}
where we adopt the radial basis function (RBF) kernel $ \bm{\Sigma}_{a}(\bm{\theta})$, which is expressed as;
\begin{eqnarray}
   (\bm{\Sigma}_{a} (\bm{\theta}))_{j,k} &=&  \alpha k_{RBF}(|r_{j} - r_{k}|; \gamma) +   \alpha \epsilon \delta_{ jk},   \label{eq:Sigma_a}\\
k_{RBF}(|r_{j} - r_{k}|; \gamma) &=& \displaystyle \exp \left ( -\frac{||r_{j} - r_{k}||^{2}}{2 \gamma^{2}}\right), 
\end{eqnarray}
where $\alpha$ determines the relative weight of the prior information, and $\gamma$ determines the spatial scale for regularization. The second term in Eq (\ref{eq:Sigma_a}) is the small identity matrix, which stabilizes the computation of the inverse matrix $\Sigma_{a}(\theta)$ \citep{kawahara2022}. We adopted  $\epsilon =10^{-8}$ in this paper.

This prior tends to promote smooth solutions, and can aid in preventing overfitting. A previous study by \cite{jennings2020} adopted parameters for controlling the Gaussian Process prior on powers of the brightness profiles in the frequency domain. However, in this study, we adopted the parameters for regulating the profile in the spatial domain to render the discussion simpler. 

\subsubsection{Likelihood function for geometry and hyperparameters}\label{sec:like_geo}
Following \cite{kawahara2020}, we derive the likelihood function $p( \bm{d} | \bm{g}, \bm{\theta} )$. We briefly overview their formulation and apply it to the current problem. For further details, see equations (14)-(21) and Appendix A.1-A.3 in the previous study. 

We introduce the data vector $\bm{d}$ as follows: 
\begin{eqnarray}\bm{d}=
\left(\begin{array}{c}
 \bm{d}_{\rm real} \\
 \bm{d}_{\rm imag}
\end{array}\right)
\end{eqnarray}

Recalling Bayes' theorem, we obtain
\begin{eqnarray}
  p(\bm{a}| \bm{d},  \bm{g} , \bm{\theta})  p(\bm{d}|  \bm{g} , \bm{\theta} ) = p(\bm{d}|\bm{a},  \bm{g} , \bm{\theta}) p(\bm{a}| \bm{g} , \bm{\theta} ).
\end{eqnarray}
This can be rewritten as follows:
\begin{eqnarray}
    p(\bm{d}|  \bm{g} , \bm{\theta} ) = \frac{p(\bm{d}|\bm{a},  \bm{g} , \bm{\theta}) p(\bm{a}| \bm{g} , \bm{\theta} )}{p(\bm{a}| \bm{d},  \bm{g} , \bm{\theta})} 
     = \frac{p(\bm{d}|\bm{a},  \bm{g} ) p(\bm{a}| \bm{\theta} )}{p(\bm{a}| \bm{d},  \bm{g} , \bm{\theta})} , \label{eq:d_g_theta}
\end{eqnarray}
where we use $p(\bm{d}|\bm{a},  \bm{g} , \bm{\theta})  = p(\bm{d}|\bm{a},  \bm{g} )$ and $p(\bm{a}| \bm{g} , \bm{\theta} ) = p(\bm{a}| \bm{\theta} )$.  

In the equation, $p(\bm{d} | \bm{a}, \bm{g} )$ is calculated using equation (\ref{eq:linear_inclined}): 
 \begin{eqnarray}
       p(\bm{d} | \bm{a}, \bm{g} ) &=& p(\bm{d}_{\rm real} | \bm{a},\bm{g}  ) p(\bm{d}_{\rm imag}| \bm{a},\bm{g}  )  \nonumber \\
       &=& \mathcal{N}(\bm{d}_{\rm real} |\bm{C}_{\rm real} \bm{H'} \bm{a}, {\bm{\Sigma}_{\rm obs}} )    \mathcal{N}(\bm{d}_{\rm imag} |\bm{C}_{\rm imag} \bm{H'} \bm{a}, {\bm{\Sigma}_{\rm obs}} )   \nonumber  \\
       &=& \mathcal{N}(\bm{d} | \bar{\bm{H}} \bm{a},\bar{\bm{\Sigma}}_{d} ), \label{eq:likelihood}
 \end{eqnarray}
  where $\bar{\bm{\Sigma}}$ and $ \bar{\bm{H}}$ are defined as follows:
\begin{eqnarray}
\bar{\bm{H}} &= \left(
\begin{array}{r}
 \bm{C}_{\rm real}\bm{H'} \\
 \bm{C}_{\rm imag}\bm{H'} \\
\end{array}
\right),  \\
 \bar{\bm{\Sigma}}_{d}&=  \left(
\begin{array}{rr}
{\bm{\Sigma}_{\rm obs}}  & \bm{O} \\
\bm{O} & {\bm{\Sigma}_{\rm obs}} \\
\end{array}
\right).
\end{eqnarray}

Further, using  equations (\ref{eq:a_prior}) and  (\ref{eq:likelihood}), 
we obtain 
\begin{eqnarray}
p(\bm{a}| \bm{d},  \bm{g} , \bm{\theta}) \propto p(\bm{d}| \bm{a},  \bm{g}) p(\bm{a}| \bm{\theta}) = \mathcal{N}(\bm{d} | \bar{\bm{H}} \bm{a},\bar{\bm{\Sigma}}_{d} ) \mathcal{N}(\bm{a} | \bm{0}, \bm{\Sigma}_{a}(\bm{\theta})).
\end{eqnarray}
As $\bm{a}$ appears as a quadratic form in $p(\bm{a}| \bm{d},  \bm{g} , \bm{\theta}) )$ in the exponent, 
we can obtain $p(\bm{a}| \bm{d},  \bm{g} , \bm{\theta})$ as follows: 
\begin{eqnarray}
p(\bm{a}| \bm{d}, \bm{g} , \bm{\theta}) &= \mathcal{N}(\bm{a} | \bar{\bm{\mu}}_{a|d} ,  \bar{\bm{\Sigma}}_{a|d} ), \label{eq:a_post}
\end{eqnarray}
where we define 
\begin{eqnarray}
\bar{\bm{\mu}}_{a|d} &=&  (\bar{\bm{H}}^{T}  \bar{\bm{\Sigma}}_{d}^{-1} \bar{\bm{H}}  + \bm{\Sigma}_{a}(\bm{\theta})^{-1})^{-1} \bar{\bm{H}}^{T} \bar{\bm{\Sigma}}_{d}^{-1} \bar{\bm{d}}^{T},\\
\bar{\bm{\Sigma}}_{a|d} &=&  (\bar{\bm{H}}^{T}  \bar{\bm{\Sigma}}_{d}^{-1} \bar{\bm{H}}  + \bm{\Sigma}_{a}(\bm{\theta})^{-1})^{-1}.
\end{eqnarray}

Using equations (\ref{eq:d_g_theta}), (\ref{eq:likelihood}), and (\ref{eq:a_post}), we obtain
\begin{eqnarray}
 p(\bm{d}|  \bm{g} , \bm{\theta} ) \propto  \mathcal{N}(\bm{d} | \bar{\bm{H}} \bm{a},\bar{\bm{\Sigma}}_{d} ) /\mathcal{N}(\bm{a} | \bar{\bm{\mu}}_{a|d} ,  \bar{\bm{\Sigma}}_{a|d} ).
\end{eqnarray}
Again,  $\bm{d}$ appears as a quadratic form in $ p(\bm{d}|  \bm{g} , \bm{\theta} )$ in its exponent. Thus, we can obtain the likelihood function $p(\bm{d}|\bm{g} , \bm{\theta} )$ as follows: 
\begin{eqnarray}
p(\bm{d}|  \bm{g} , \bm{\theta} ) =\mathcal{N}(\bm{d}|\bm{0},\bar{\bm{\Sigma}}_{d} +   \bar{\bm{H}} \bm{\Sigma}_{a} \bar{\bm{H}}^{T}). \label{eq:d_b_theta_post}
\end{eqnarray}

\subsubsection{Posterior distribution for geometry and hyperparameters} \label{sec:post_sample}

The posterior distribution for $p(\bm{g} , \bm{\theta} | \bm{d})$ is given by the Bayes' theorem as follows: 
\begin{eqnarray}
    p(\bm{g} , \bm{\theta} | \bm{d}) \propto  p(\bm{d}|\bm{g} , \bm{\theta}) p(\bm{g} , \bm{\theta}) = \mathcal{N}(\bm{d}|\bm{0},\bar{\bm{\Sigma}}_{d} +   \bar{\bm{H}} \bm{\Sigma}_{a} \bar{\bm{H}}^{T}) p(\bm{g} , \bm{\theta}), \label{eq:g_theta_d_final}
\end{eqnarray}

Based on $p(\bm{g} , \bm{\theta} | \bm{d})$, we draw samples for $(\bm{g} , \bm{\theta})$ using an MCMC sampler. The prior distribution $p(\bm{g}, \bm{\theta})$ is assumed as follows. We assume uniform distributions $\mathcal{U}(0, 1)$ and $\mathcal{U}(0, \pi)$ for $\cos i$  and PA, respectively. Further, we consider $\mathcal{U}(-1\arcsec, 1\arcsec)$ for $\Delta x_{\rm cen}$ and $\Delta y_{\rm cen}$. 
In addition, we assume the log-uniform prior from $10^{-4}$ to $10^{4}$ for $\alpha$ and a uniform prior $\mathcal{U}(0.01\arcsec, 0.15\arcsec)$ for $\gamma$. We checked that the likelihood is exceedingly small when $\alpha$ and $\gamma$ fall outside their respective prior ranges, thus confirming the sufficient broadness of the prior. As the computation for $p(\bm{d}|  \bm{g}, \bm{\theta})$ can be time-consuming, we transform the equation to a more efficient form (further details in Appendix \ref{sec:efficient_comp}). 

Moreover, the computation also relies on the number of data points. In Appendix \ref{sec:data_bin}, we introduce and discuss the use of data binning to reduce the computation. Specifically, we prepare two-dimensional linear and log grids, apply them to the simulated visibilities, and compare the binning errors. The binning with the log grid was found to yield overall low errors. A more detailed discussion is presented in Appendix \ref{sec:data_bin}. In the following analysis, we simply adopt the log grid for the analysis.

\subsubsection{Posterior distribution for all of parameters} \label{sec:post_sample_rad} 
We can draw samples for $\{\bm{a}, \bm{g}, \bm{\theta}\}$ by using the conditional formula for the joint probability: $p(\bm{a}, \bm{g} , \bm{\theta} |\bm{d})=p(\bm{a}|\bm{d}, \bm{g} , \bm{\theta}) p(\bm{g} , \bm{\theta} |d)$. Specifically, we can draw a sample $(\bm{g}\dagger , \bm{\theta}\dagger )$ from $p(\bm{g} , \bm{\theta} |d)$ as done in Sec \ref{sec:post_sample}, and subsequently take a sample for $\bm{a}$ from $p(\bm{a}|\bm{d}, \bm{g}\dagger  , \bm{\theta}\dagger )$ using equation (\ref{eq:a_post}). We can iterate this procedure to make samples, and this sampling is indeed equivalent to drawing sample from the joint distribution $p(\bm{a}, \bm{g} , \bm{\theta} |\bm{d})$. 
Using the samples for $\{\bm{a}, \bm{g}, \bm{\theta}\}$, we can also compute statistics for parameters (e.g., mean and standard deviation for $\bm{a}$). 

\subsubsection{Difference in formulation between current study and {\tt frank}} \label{sec:difference} 
We here highlight the key differences between our approach and that of {\tt frank}. Note that the direct comparison of the recovered profiles is also given in Section \ref{sec:apptoreal1}. While both methods vary in their regularization strategies, the distinct difference is that our approach solve all parameters, including brightness profiles, geometric parameters, and hyperparameters, whereas {\tt frank} is limited to solving brightness profiles. In this sense, our formulation can be seen as a natural extension of that of {\tt frank}. 

One advantage of our method is that it can optimize geometric parmaeters and hyperparameters directly from the data, and thus it is less susceptible to human biases caused by manual tuning. Moreover, while the current paper is limited to the simplest model with single frequency and single source, the methodology itself can be easily extended to more complex problems with multi-frequency data or multiple sources. In such cases, manual tuning of optimal parameters can be both challenging and less reliable, making our approach more effective. 

On the other hand, the disadvantage of our approach is that it is more computationally demanding than {\tt frank}. This is because we attempted to solve all parameters, including non-linear parameters, rather than deriving only brightness profiles like {\tt frank}. Furthermore, our current formulation cannot support imposing a non-negative condition on brightness profiles, a feature available in {\tt frank}. This is because the analytical expression for the marginalization over $\bm{a}$ becomes inapplicable under the non-negative condition. Although we can implement the condition by considering the same problem as {\tt frank}, the current study prioritizes simultaneous fitting of all parameters, so we do not implement the function. 

\section{Extract asymmetric features} \label{sec:assym}
This section summarizes the process of extracting and quantifying the non-axisymmetric components from observations using the derived parameters in Sec \ref{sec:inverse}. The method is applied to data analyses in Sec \ref{sec:sim} and \ref{sec:sim_non}. 

\subsection{Making residual images in observed and deprojected frames} \label{sec:residual}
\subsubsection{Making residual image} \label{sec:make_image}
After subtracting the axisymmetric model from visibilities, we expect the residual visibilities to contain only non-axisymmetric components. Therefore, imaging using these residual visibilities can reveal the non-axisymmetric component of the disc \citep{jennings2020,andrews2021}. Based on this principle, we created images to study the non-axisymmetric components. 

Drawing samples from a posterior distribution as described in Sec \ref{sec:form}, we selected the maximum a posteriori probability (MAP) estimate for $\bm{g}$ and $\bm{\theta}$. Subsequently, we randomly drew a brightness profile using equation (\ref{eq:a_post}). For the chosen parameters and the brightness profile, we computed the model visibilities of an axisymmetric disc, and subtracted them from observed visibilities. Then, the phase centres of measurement sets were aligned with the disc centres using {\tt fixvis} in {\tt CASA}. 

For the shifted and subtracted measurement set, we created images using the CLEAN algorithm, which is implemented as ${\tt tclean}$ in {\tt CASA}. In this study, the pixel scale for the image was set as 6 mas, and the image size was $1000 \times 1000$ pixels. In the imaging process, Although it is typical to adopt 0 iteration for CLEAN in making the residual map \citep[e.g.,][]{jennings2020}, we found that employing non-zero iterations for CLEAN result in slight or moderate improvement in the quality of the residual image, especially when there remain substantial residuals. Thus, in this paper, we implemented thresholds with nsigma$=3.5$, which stops the iterations if the maximum residual on the image is below $3.5$ times the root-mean-square error. Further, we adopted the Briggs weighting with a robust parameter of 0.5 and without UV-taper. Appendix \ref{sec:robust_vary} presents a discussion on the effects of changing robust parameters on the recovered residual images, and 0.5 was determined to be the balanced choice. The brightness of the output image from {\tt CASA} was measured in Jy beam$^{-1}$, where 'beam' is the nominal area where the brightness is defined\footnote{In {\tt CASA}, the synthesized beam is approximated using the Gaussian function, and its area is expressed as $\frac{\pi}{4 \rm{ln} 2} \, b_{\rm maj} \,b_{\rm min}$, where $(b_{\rm maj} \,b_{\rm min})$ are the full widths at half maximum of the Gaussian in major and minor axes. }. The standard deviation of the residual image was also measured outside the disc emission.

\subsubsection{Deprojection of residual image}
In addition to the residual image in an observational frame, we created the deprojected residual image using the disc geometry.  Spatial frequencies of the shifted measurement set were converted using the geometric parameter, $(u, v) \rightarrow (u', v')$ in equation (\ref{eq:uv_convert}). Consequently, the image was created in the same manner as Sec \ref{sec:make_image}. This operation aligns the major axis of the image with the $y$-axis (north direction), and stretches the image along the $x$ axis (east direction). This corresponds to the conversion in Figure \ref{fig:disc_def} from a deprojected to a projected frame. 

In the imaging process, we should be cautious that the transformation $(u, v) \rightarrow (u', v')$ apparently decreases the brightness in Jy sr$^{-1}$ by a factor of $\cos i$, while brightness in Jy beam$^{-1}$ remains  unchanged before and after deprojection. This is because the transformation increases the disc area (and beam) while the total flux is unchanged. Throughout the paper, we consistently presented the brightness of images in Jy beam$^{-1}$, which remains constant, and we did not apply any correction by a factor of $\cos i$. 

Although widely used, the current simple deprojection methods provide the true image of the disc viewed face-on only in the case of an infinitesimally thin disc. The other geometric effects such as the vertical structures or shadows cannot be incorporated with our simple operations. 

\subsection{Mode decomposition of residual image in polar direction} \label{sec:polar_fourier}
Spirals are often characterized by dominant modes in a polar direction. We can decompose $I(r,\phi)$ in a polar coordinate $(r, \phi)$ as follows \citep[e.g.,][]{binney2008}:
\begin{eqnarray}
I(r, \phi)  =  \sum_{m=0}  I_{m}(r) \cos (m(\phi  - \phi_{m}(r))). \label{eq:decomp}
\end{eqnarray}
where $I_{m}(r)$ is the amplitude at the $m$-th mode, and $\phi_{m}(r)$ is a phase shift for the $m$-th component. Note that $I_{0}(r)$ corresponds to the brightness profile. The amplitudes and phases can be derived from the residual images $I(x, y)$ as follows: 
\begin{eqnarray}
I_{m}(r) &=& 2 \sqrt{C_{m}^{2} + S_{m}^{2}}, \\
\phi_{m}(r) &=& {\rm atan2}(S_{m}, C_{m}), 
\end{eqnarray}
where ${\rm atan2}$ is the 2-argument arctangent, and $(C_{m}, S_{m})$ is defined as follows: 
\begin{eqnarray}
C_{m} \equiv \frac{1}{2\pi} \int_{0} ^{2 \pi} I(r, \phi)  \cos (m \phi) d \phi, \\
S_{m} \equiv \ \frac{1}{2\pi} \int_{0} ^{2 \pi} I(r, \phi)  \sin (m \phi) d \phi.  \label{eq:cs_calc}
\end{eqnarray}
The values of $(C_{m}, S_{m})$ can be computed for the real data. In this study, we computed them by interpolating a residual image.

\subsection{Extraction of odd and even symmetric components of images} \label{sec:real_imag}
Odd-symmetric and even-symmetric components of an image can be decomposed by using either of each imaginary or real part of the data. We start with decomposing an image into even- and odd-symmetric components. 
\begin{eqnarray}
I(x, y) = I_{\rm even}(x, y) + I_{\rm odd}(x,y), 
\end{eqnarray}
where we define 
\begin{eqnarray}
\displaystyle I_{\rm even}(x, y) &=&\frac{1}{2}(I(x, y) +  I(-x, -y)) \\
&=& \sum_{m= {\rm even}} I_{m}(r) \cos (m(\phi  - \phi_{m}(r))), \\
I_{\rm odd}(x,y) &=&\frac{1}{2 }(I(x, y) -  I(-x, -y)) \\ 
&=&\sum_{m= {\rm odd}} I_{m}(r) \cos (m(\phi  - \phi_{m}(r))), 
\end{eqnarray}
which satisfy $ I_{\rm even}(x,y)  =  I_{\rm even}(-x,-y)$, and $ I_{\rm odd}(x,y)  = -  I_{\rm odd}(-x,-y)$. However, using equation (\ref{eq:von_zer}), the real and imaginary parts are connected to $I_{\rm even}(x, y)$ and $I_{\rm odd}(x, y)$ as follows: 
\begin{eqnarray}
\displaystyle \mathfrak{Re}(V(u,v) )&=&  \int_{-\infty}^{\infty} \int_{-\infty}^{\infty}  I_{\rm even}(x, y)  \exp(-2\pi j (ux + vy)) \bm{dx}, \nonumber\\ \\
 \mathfrak{Im}(V(u,v)) &=& \int_{-\infty}^{\infty} \int_{-\infty}^{\infty} I_{\rm odd}(x,y) \exp(-2\pi j (ux + vy)) \bm{dx}. \nonumber \\
\end{eqnarray}
Thus, we can create $I_{\rm even}(x, y)$ and $I_{\rm odd}(x, y)$ using either the real or imaginary part of the data. Notably, previous studies have already used the imaginary part of the data to extract an odd-symmetric component \citep{hashimoto2021,kanagawa2021}. 

\section{Application to simulated data for geometrically thin axisymmetric disc} \label{sec:sim}
\subsection{Injection and recovery test } \label{sec:in_re_disc}
As a test case, we applied our method to simulated data for an axisymmetric disc to examine the ability to recover input parameters. We used two radial profiles for our test.  The continuum brightness profiles for AS~209 and WaOph~6 of DSHARP observations\footnote{https://bulk.cv.nrao.edu/almadata/lp/DSHARP}\citep{andrews2018} were used.  AS 209 is chosen as the most structured disc in a radial direction, and WaOph 6 is among $m=2$ spiral discs in the DSHARP sample.  Subsequently, we assume the model discs have the geometric parameters of $(\Delta x_{\rm cen}, \Delta y_{\rm cen},\cos i, {\rm PA}, )=( 0\arcsec, 0\arcsec, 0.75, 45^{\circ})$.  We normalize the brightness profiles to render the total fluxes of the AS~209 and WaOph~6 models as 0.288 and 0.161 Jy, respectively.  

To produce mock data with realistic UV-coverage and noise, we incorporated them using the actual data of DSHARP observations.  We downloaded the self-calibrated continuum measurement set for AS~209 and WaOph~6.  The data were averaged at each spectral window to produce a single channel data. Thereafter, time averaging was applied for 30 s. A new measurement set was created using the averaged data to obtain the UV-coverage and the weight at each spatial frequency.  The signals of the model disc at each spatial frequency were calculated via Fourier transforms of the model surface brightness distribution. Consequently, the Gaussian noises were added to the model visibility according to the the weights.  During the analysis, we used $\tt{tb.getcol}$ in {\tt CASA} to export the measurement sets to handle them numerically in {\tt python}\footnote{We used $\it{tb.getcol}$ in {\tt CASA} to export $(u, v)$ from measurement sets, but we found that signs of the outputted arrays for $(u,v)$ were reversed. Specifically, they have opposite signs with respect to those obtained with $\it{plotms}$. In the analysis, we simply flipped signs of $(u, v)$ output from $\tt{tb.getcol}$. }.

The data weights recorded in the measurement sets were overestimated for the entire DSHARP data.  Following Appendix E in \cite{hashimoto2021}, we deprojected the real and imaginary parts of visibilities, and compared the recorded standard deviations and the deviations of observed visibilities from the binned visibilities in the deprojected spatial frequencies. Consequently, we confirmed that the overestimation existed in both the real and imaginary parts of the visibilities, and the degree of overestimation was within $\sim 10$\% for both. Therefore, we manually reduced the recorded weights by 3.44 and 3.66 for AS~209 and WaOph~6, respectively.

To reduce the computational time, we binned the data using a logarithmically spaced grid with $N_{\rm bin}=500$, where the number of grid points was $(2N_{\rm bin}+3)^{2}$. The boundaries of the grids are defined by  the parameters $(x_{\rm min}, x_{\rm max})$, which are introduced in Appendix \ref{sec:data_bin}. We set $(x_{\rm min}, x_{\rm max})$ to $ (10^{2} \lambda, 10^{7} \lambda)$ in our analyses. This binning decreased the number of data points by an approximate factor of 10. Appendix \ref{sec:data_bin} presents the details of data gridding using a log grid, and we show that the choice of $N_{\rm bin}=500$ is acceptable. 

Using equation (\ref{eq:g_theta_d_final}), we drew samples from the posterior distribution for $(\gamma, \log_{10} \alpha, \Delta x_{\rm cen}, \Delta y_{\rm cen}, \cos i, {\rm PA})$. In sampling, we used {\tt emcee}, which implemented the affine-invariant ensemble sampler for Markov Chain Monte Carlo \citep{foreman2013}. The initial parameters for  {\tt emcee} were set to be close to the input parameters. Here, 16 walkers were prepared and then evolved for at least 8,000 steps. For the drawn samples, we discarded the initial 1,250 steps during the burn-in phase. The convergence is predominantly achieved within 2,000-4,000 steps according to the auto-correlation time analysis. Specifically, we have verified that all samples meet the condition $N>50\tau$, where $N$ represents the step size and $\tau$ represents the integrated autocorrelation time for the chain\footnote{ https://emcee.readthedocs.io/en/stable/tutorials/autocorr/.}  

The total computational time for the single disc was approximately 12 hours using 8 cores. This represents a significantly higher computational demand in contrast to frank's approach. This stems from our strategy to retrieve all parameters in a simultaneous manner, as opposed to frank's approach of fixing parameters except for the brightness profiles. 

\begin{figure*}
\begin{center}
\includegraphics[width=0.48\linewidth]{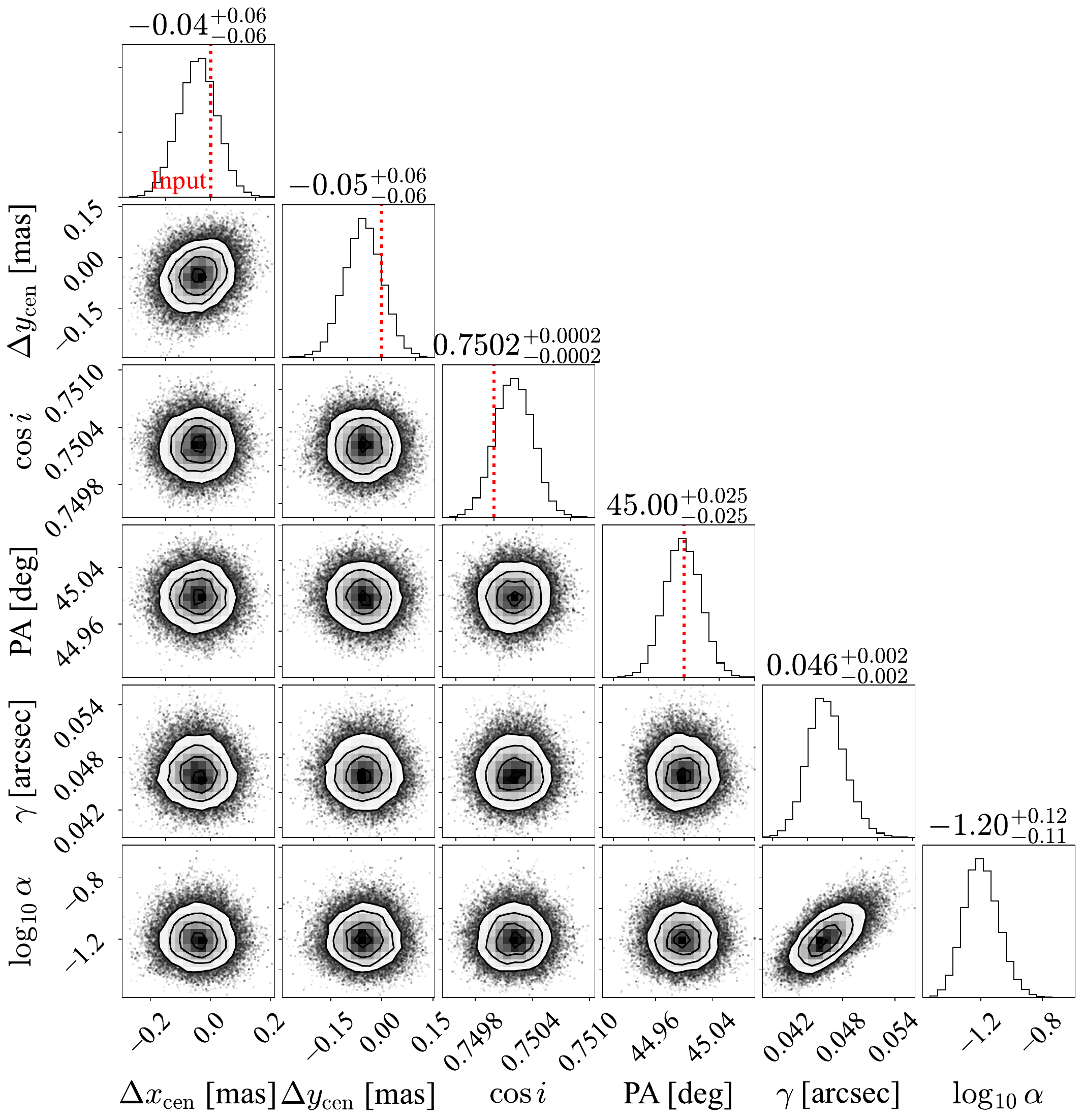}
\includegraphics[width=0.48\linewidth]{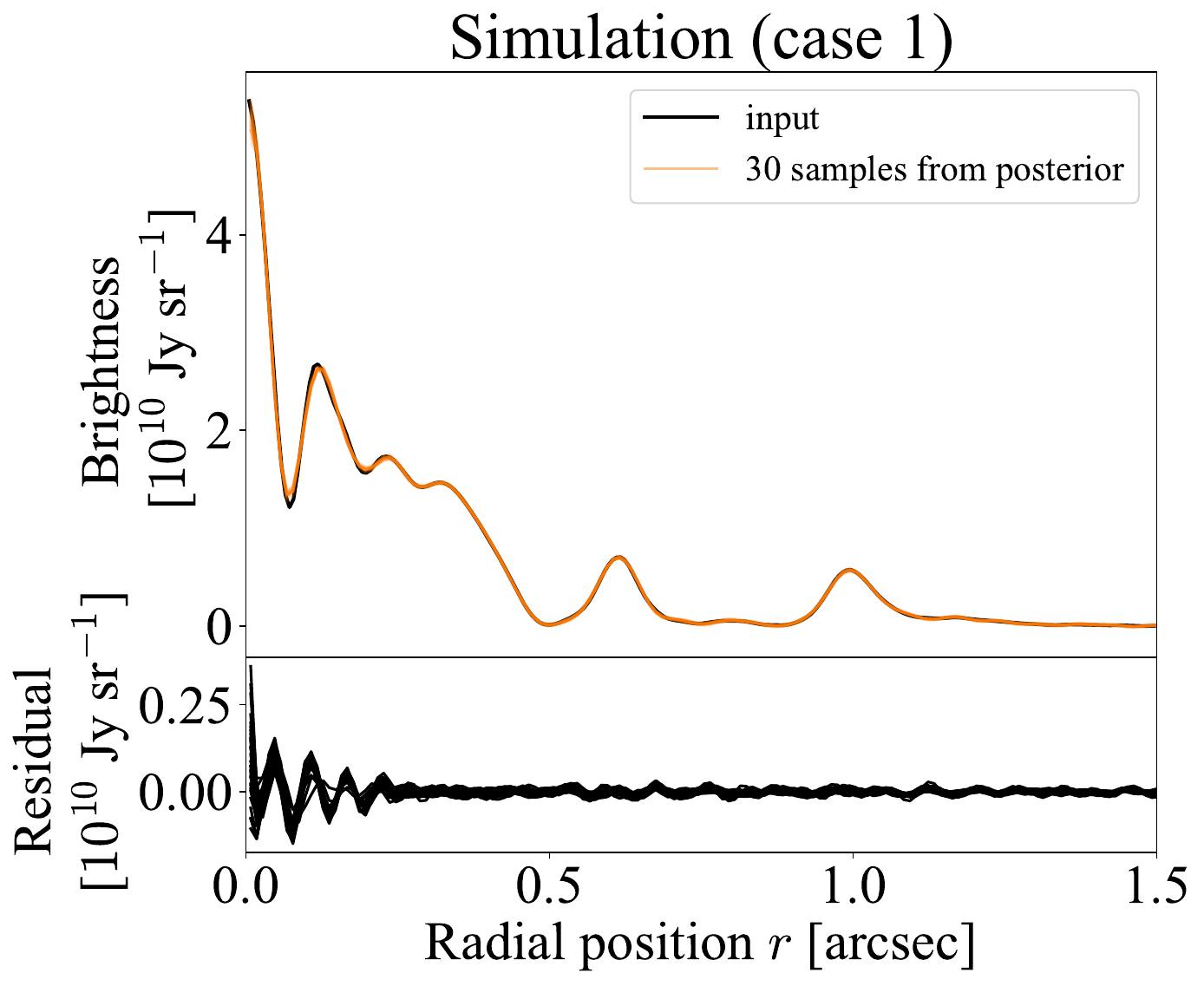}
\includegraphics[width=0.48\linewidth]{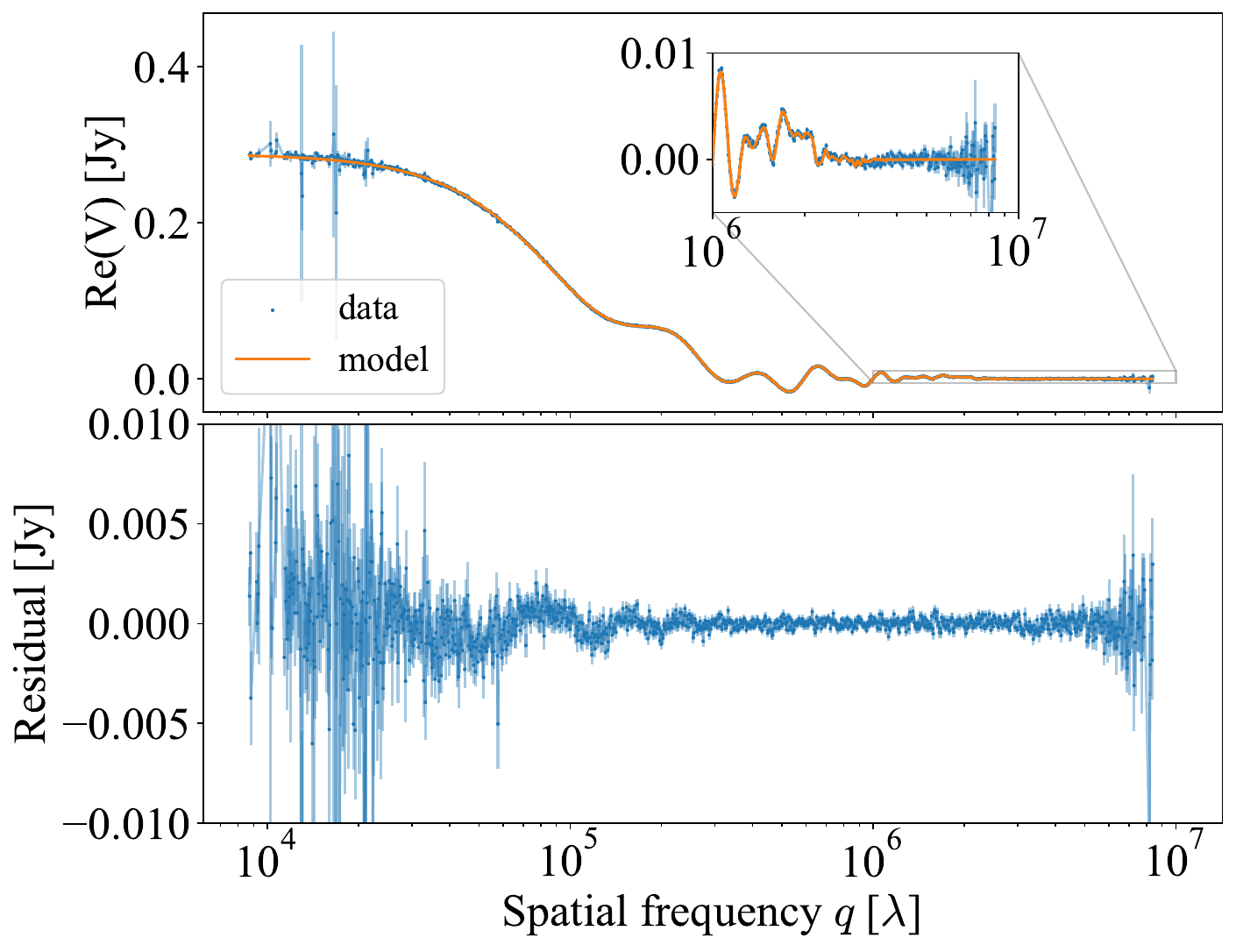}
\includegraphics[width=0.48\linewidth]{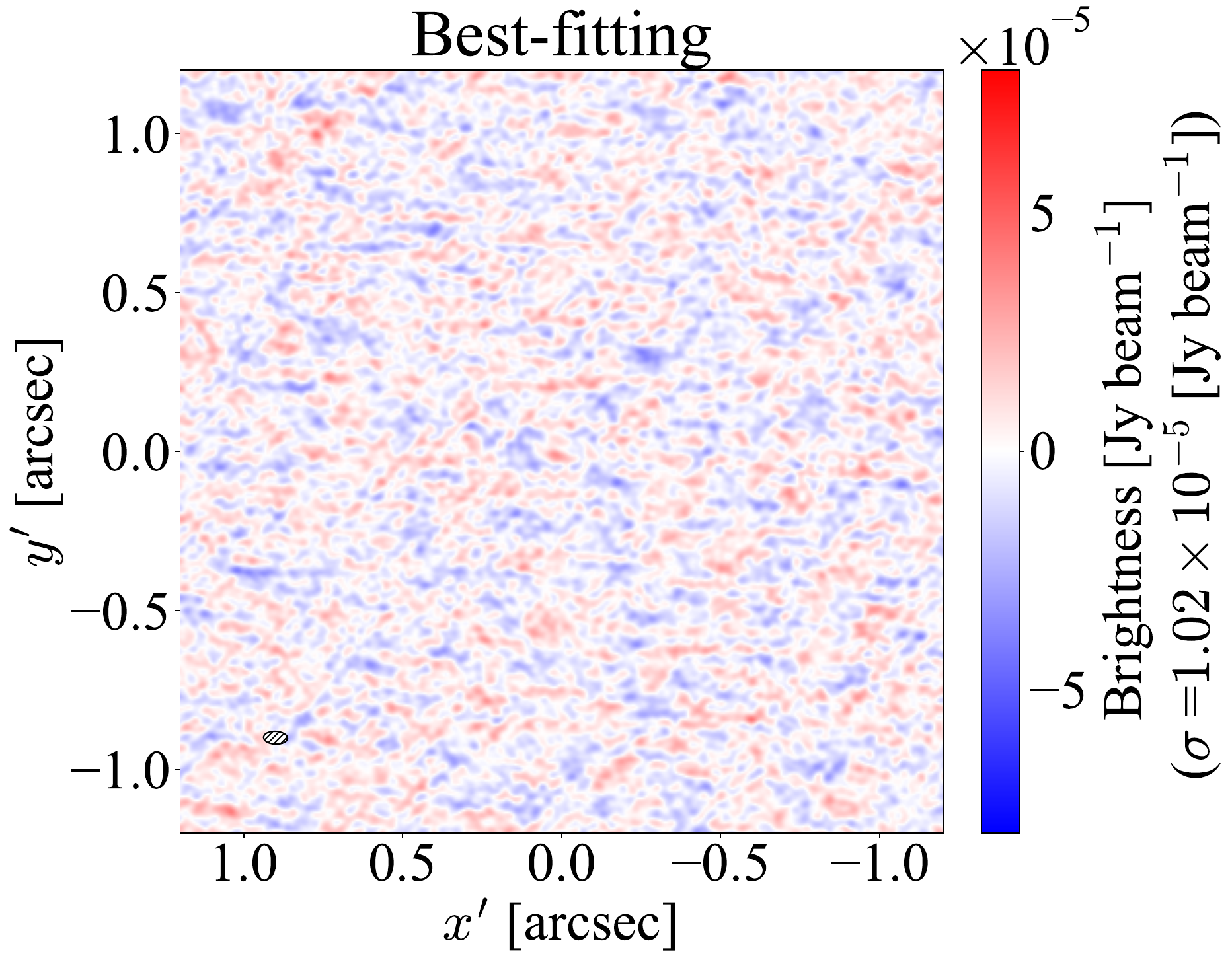}
\end{center}
\caption{Recovery and injection test for mock data of AS 209. (upper-left) Posterior distribution for $(\Delta x_{\rm cen}, \Delta y_{\rm cen}, \cos i, {\rm PA}, \alpha, \gamma)$. Dotted lines indicate the input values. (upper-right) Injected brightness profile, denoted by the black line, and 10 samples drawn from the posterior distribution of brightness profiles, denoted by red lines. At the bottom, the residuals between injected and recovered models are shown.  (lower-left) Simulated visibilities denoted by blue points and model visibilities denoted by an orange line. The visibilities are binned with a logarithmic grid with $N=2000$. In the lower panels, the residuals indicating the difference between the simulated and model visibilities are shown. (lower-right) Residual image produced with the MAP estimate in the observed frame. The synthesized beam size $(0.076\arcsec, 0.040\arcsec)$ is shown at the bottom left.  }
\label{fig:posterior_sim}
\end{figure*}

\begin{figure*}
\begin{center}
\includegraphics[width=0.48\linewidth]{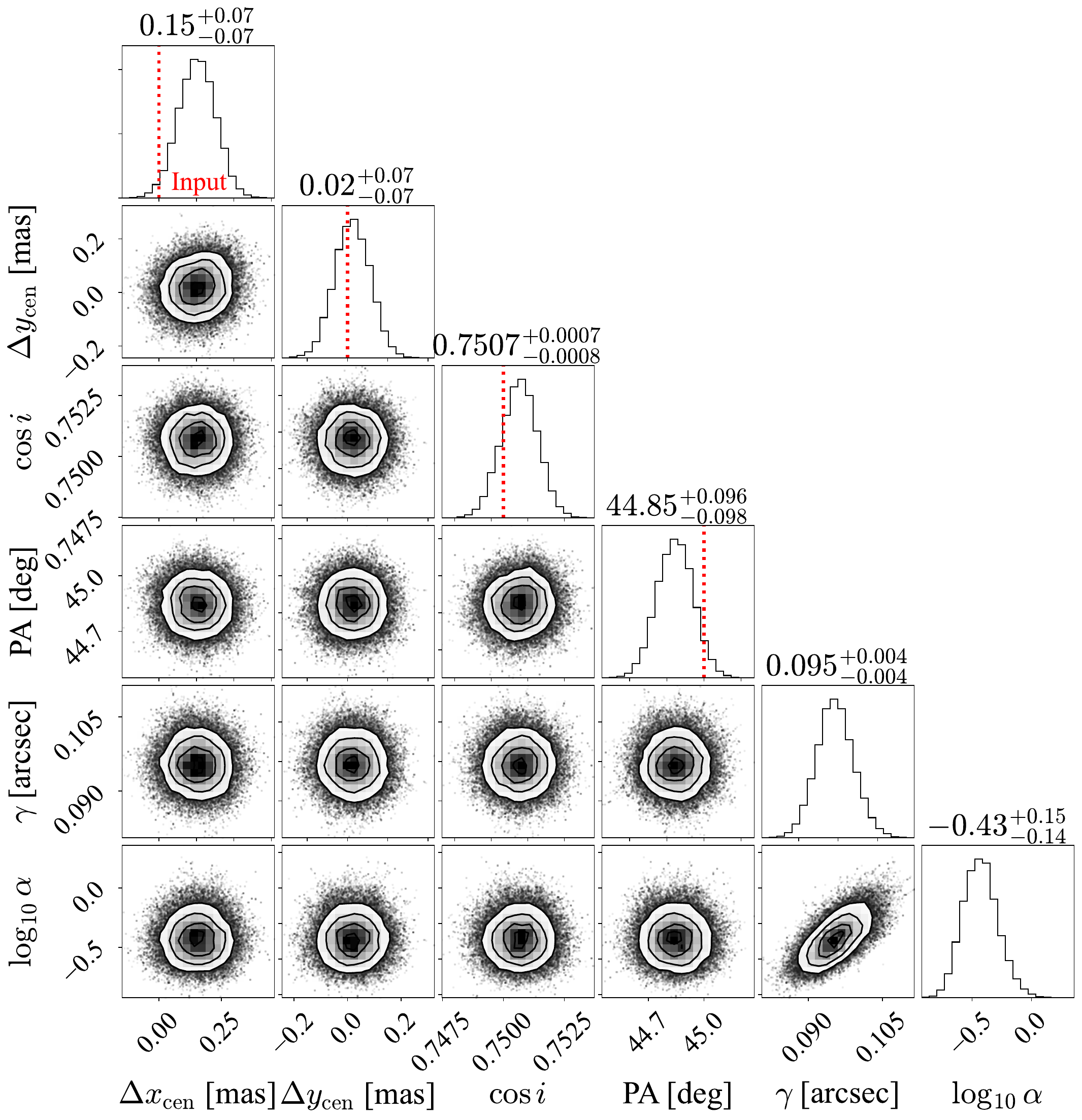}
\includegraphics[width=0.48\linewidth]{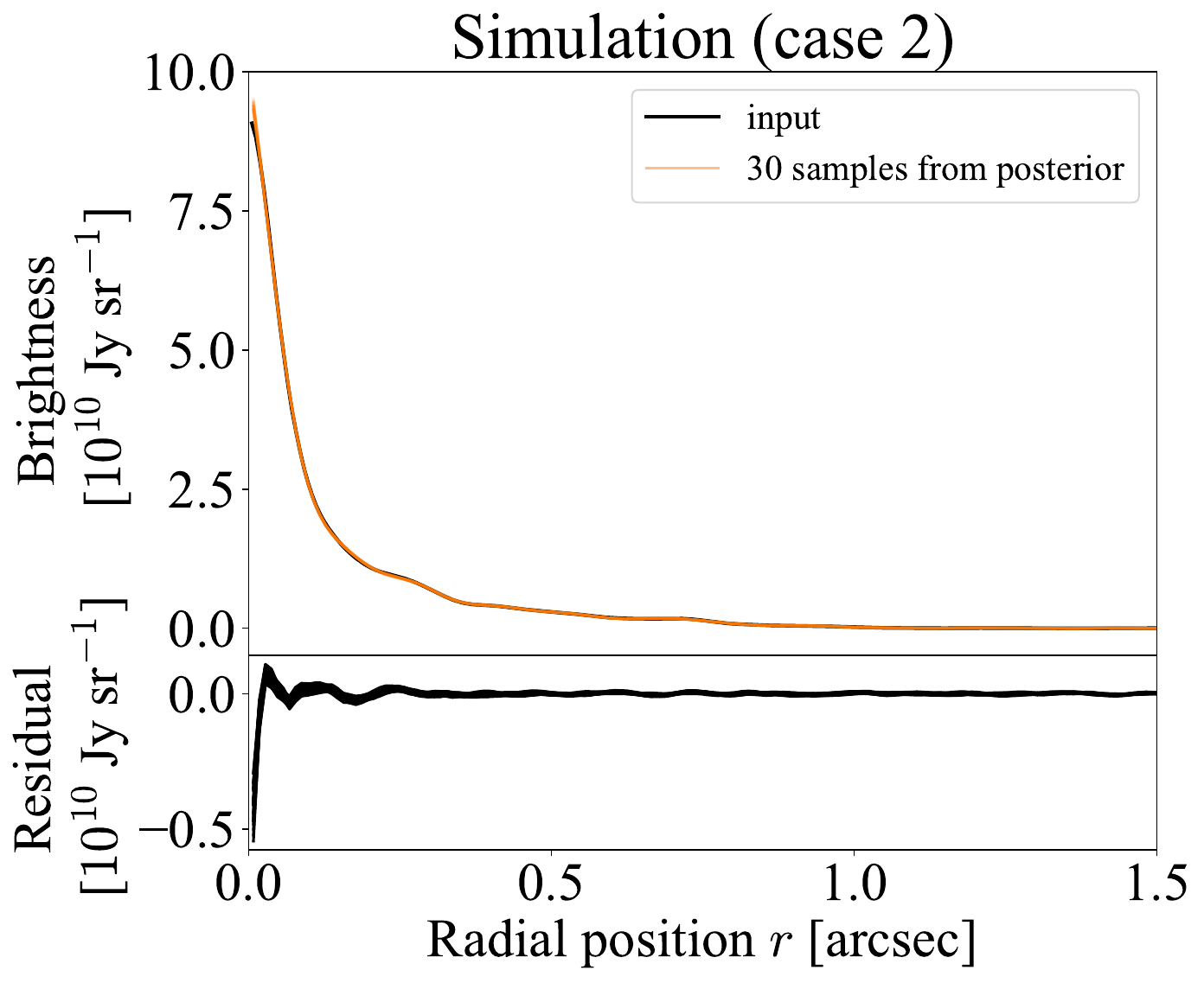}
\includegraphics[width=0.48\linewidth]{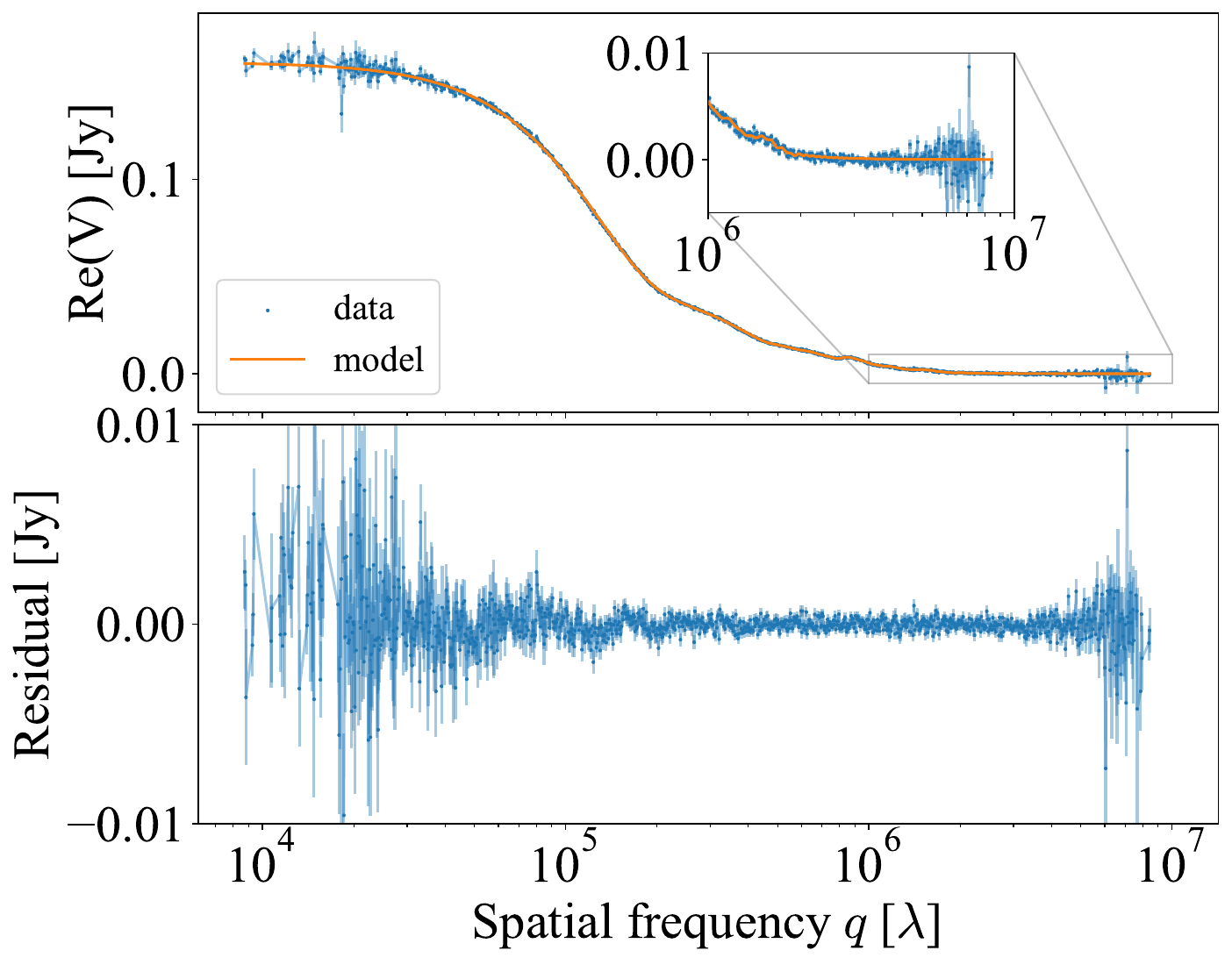}
\includegraphics[width=0.48\linewidth]{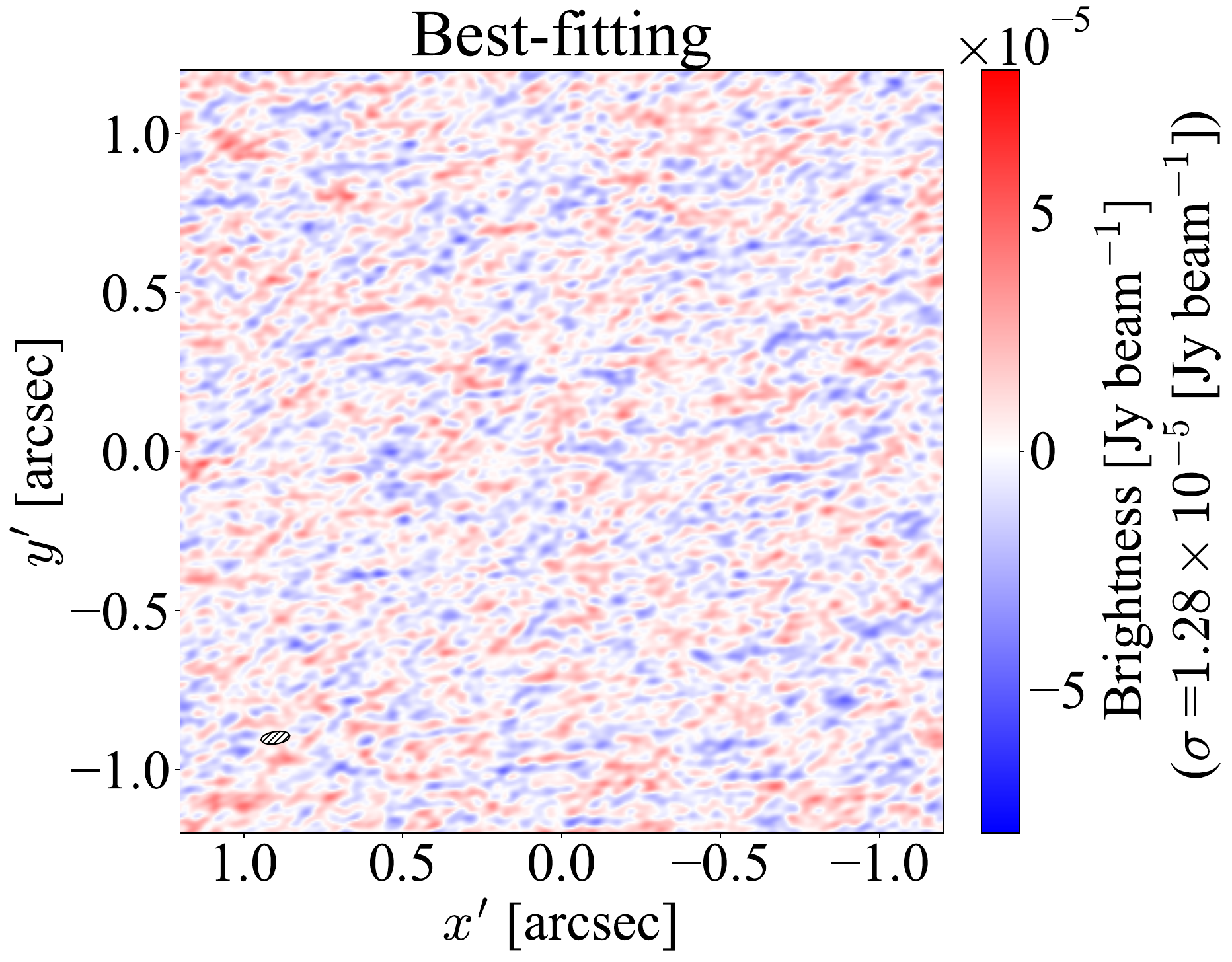}
\end{center}
\caption{Recovery and injection test for mock data of WaOph 6 in Sec \ref{sec:sim}. The format of the figure is same as that of Figure \ref{fig:posterior_sim}. The synthesized beam size $(0.091\arcsec, 0.037\arcsec)$ for the residual image is shown at the bottom left.  }
\label{fig:posterior_sim_waoph6}
\end{figure*}

Figs. \ref{fig:posterior_sim} and \ref{fig:posterior_sim_waoph6} show the recovery and injection test results for the simulated data. The upper-left panels shows the posterior distribution of the parameters for the two simulated cases. All the geometric parameters were reasonably recovered by the current method. 

The analyses resulted in two different length parameters; $\gamma\simeq 0.04 \arcsec$ for the simulated case of AS 209, and $\gamma\simeq 0.1 \arcsec$ for the case of WaOph 6. The differences in $\gamma$ were largely due to the differences in the injected brightness profiles; the injected profile for WaOph 6 was smoother than that of AS 209. The UV-coverage would also affect the length scale, but it remains unclear in the current two cases. The length scale parameter for WaOph 6 was larger than the beam size $(0.091\arcsec, 0.037\arcsec)$, whereas it was smaller than the beam size $(0.076\arcsec, 0.040\arcsec)$ for AS 209. Note that the beam sizes reflect the UV-coverage, but they vary with assumed parameters in {\tt CLEAN}; for example, robust parameter and UV-taper. In Appendix \ref{sec:scaling}, we further study the variations in $\gamma$ by stretching the injected brightness profiles and UV-coverage. In conclusion, the optimal length scales $\gamma$ appear to be determined by multiple factors, including the brightness profiles and UV-coverage (beam size). 

The upper-right panels in Figs. \ref{fig:posterior_sim} and \ref{fig:posterior_sim_waoph6} compare the injected brightness profiles with 10 samples for $\bm{a}$ using the method presented in Sec \ref{sec:post_sample_rad}. The radial profiles were also adequately recovered by the current method. However, the residuals for the brightness profiles were not perfectly consistent with the zero line for both cases, although an oscillating feature was observed in the radial direction. The oscillating length scales for the observed residuals indeed corresponded to the derived length scales $\gamma$, which determined the characteristic scales that could be resolved. Moreover, we also identified the large residuals at the very innermost parts owing to the steep changes in the fluxes in the radial direction. The oscillating features have been also reported in \cite{jennings2020}, which adopted regularization in the frequency domain, and such residuals might be inevitable in the Gaussian Process. 

The lower-left panels in Figs. \ref{fig:posterior_sim} and \ref{fig:posterior_sim_waoph6} show the simulated and model visibilities. The model visibilities were computed by choosing the MAP solution for the geometry and hyperparameters, and the brightness profile was randomly sampled. The residual visibilities are shown in the lower panels. The models were well consistent with the simulated visibilities from the low to high spatial frequencies. However, we observed that the powers of the model visibilities at high frequencies were forcibly suppressed. This cut-off scale was determined by the optimized length scale $\gamma$, which determined the scales below which the oscillating features appeared in the brightness profiles. At the low spatial frequencies, particularly at $q<0.3$ M$\lambda$, we also found oscillating residuals with a length scale of $\sim 0.05$ M$\lambda$. This length scale corresponded to the radius for the outer boundary $R_{\rm out}=2\arcsec$, which limited the model's applicability to the large-scale structure. 

In the lower-right panels in Figs. \ref{fig:posterior_sim} and \ref{fig:posterior_sim_waoph6}, we show the residual images, which are produced by the method presented in Sec \ref{sec:residual}. The residual images did not exhibit noticeable features, demonstrating that we reasonably estimated the injected parameters. 

%-----------------------------Figure Start---------------------------
%-----------------------------Figure End------------------------------

%-----------------------------Figure Start---------------------------
%-----------------------------Figure End------------------------------

\subsection{Residual images constructed from shifted geometric parameters} 
To demonstrate how incorrect determination of geometric parameters affects the resultant residual images, we shifted each geometric parameter before subtracting axisymmetric models from the visibilities. For the simulated case of AS 209, we shifted each geometric parameter from the MAP estimate as follows: $\Delta x_{\rm cen}=5\;{\rm mas}$, $\Delta y_{\rm cen}=5 \; {\rm mas}$, $\Delta  \cos i=0.01$, $\Delta{\rm PA}= 2\; {\rm deg}$. For the simulated case of WaOph 6, we shifted the parameters as follows: $\Delta x_{\rm cen}=2\;{\rm mas}$, $\Delta y_{\rm cen}=2\;{\rm mas}$, $\Delta  \cos i=0.03$, $\Delta {\rm PA}= 3 \;{\rm deg}$. To simulate the realistic situation as much as possible, we assume these incorrect geometric parameters with the hyperparamters from the MAP solution, and draw a brightness profile from the posterior distribution, which nevertheless gives the profile similar to our best-fit model. With the subtracted visibilities made with the new models, we created the residual images using the CLEAN following the method presented in Sec \ref{sec:residual}. We also created images from unsubtracted visibilities for comparison. For that, we adopted 0.04 mJy as the CLEAN threshold and used the multiscale CLEAN algorithm with scale parameters of [0, 30, 120, 360, 720, 1440] mas \citep{rau2011}.

\begin{figure*}
\begin{center}
\includegraphics[width=0.325\linewidth]{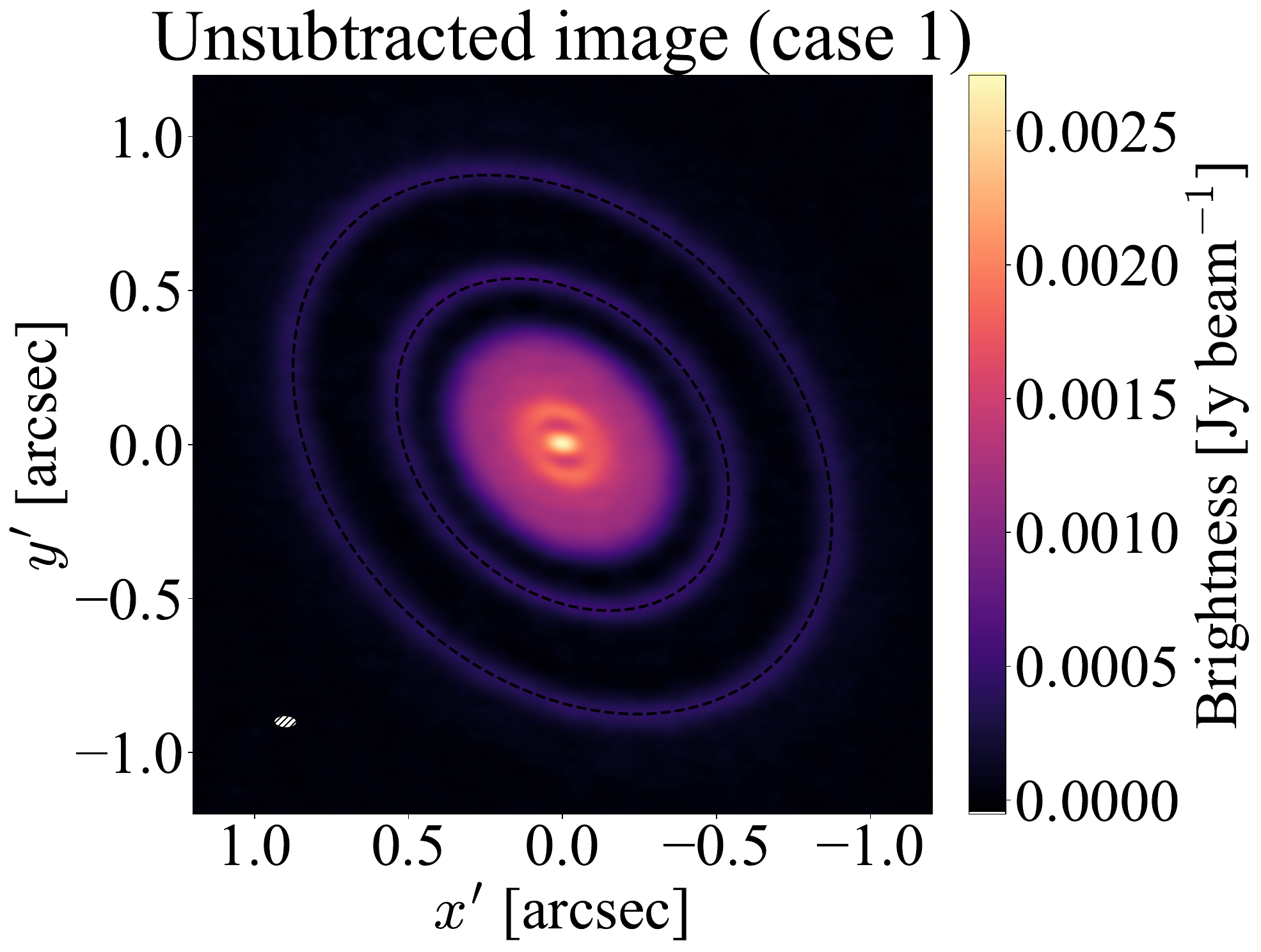}
\includegraphics[width=0.325\linewidth]{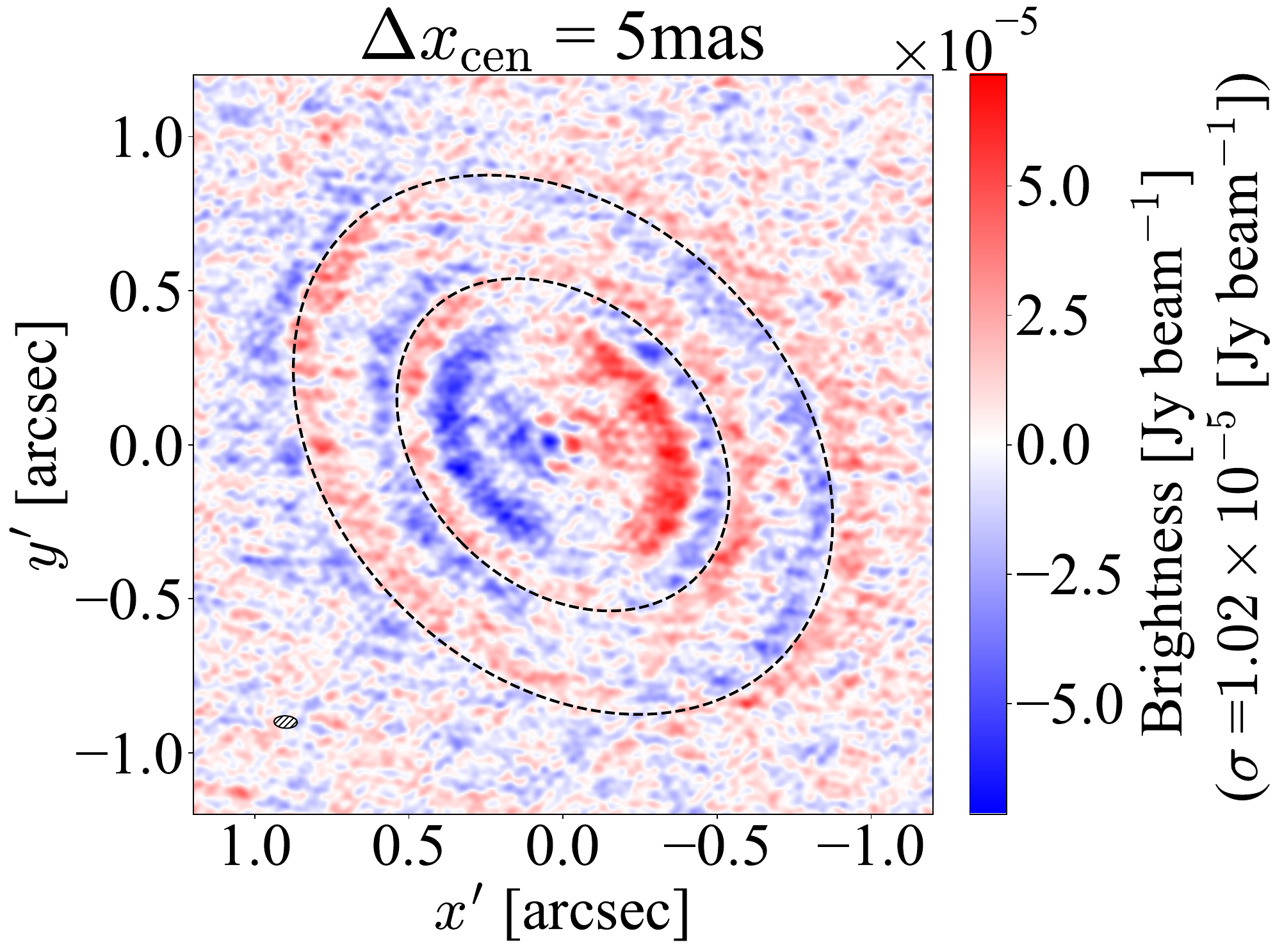}
\includegraphics[width=0.325\linewidth]{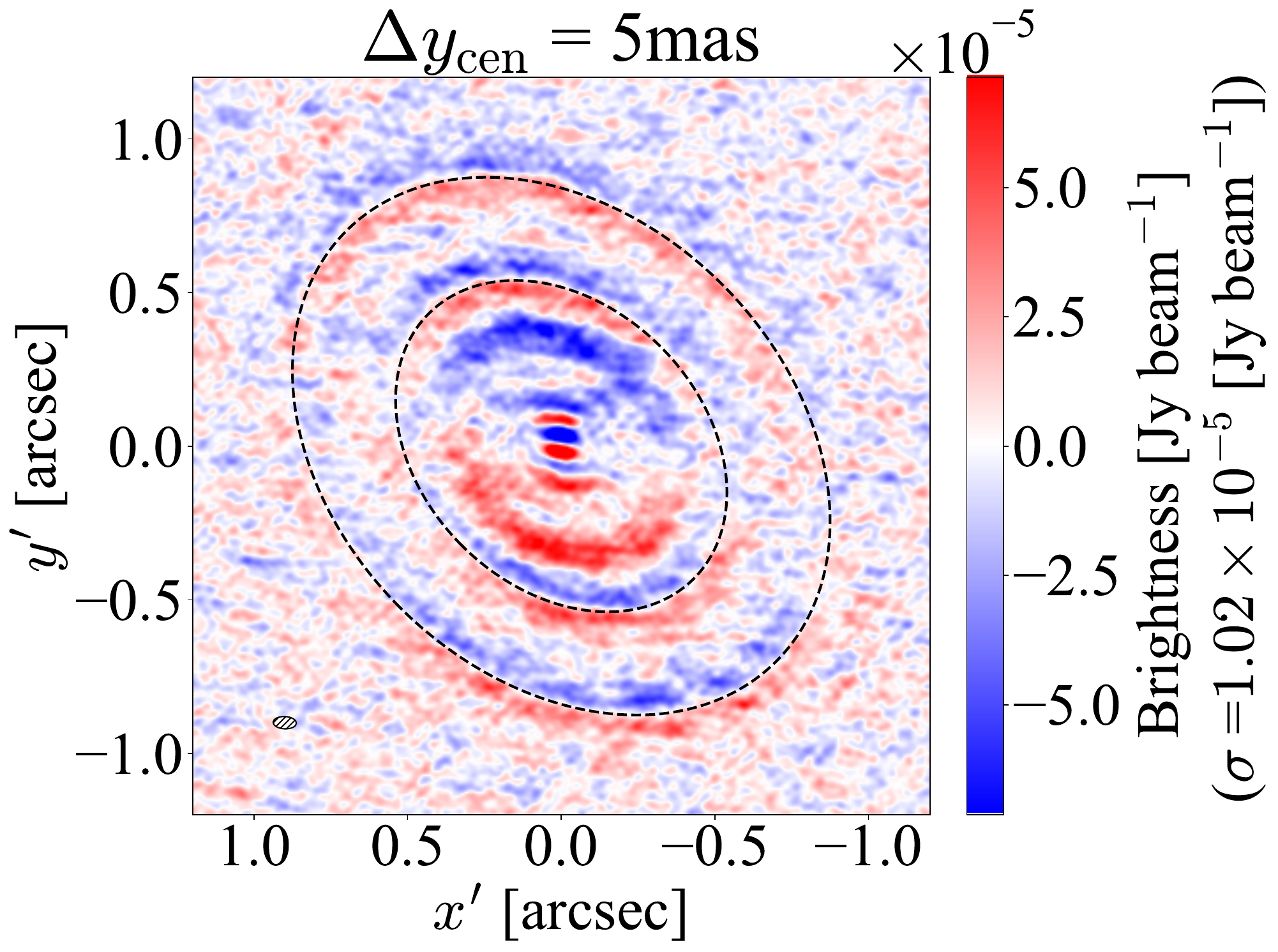}
\includegraphics[width=0.325\linewidth]{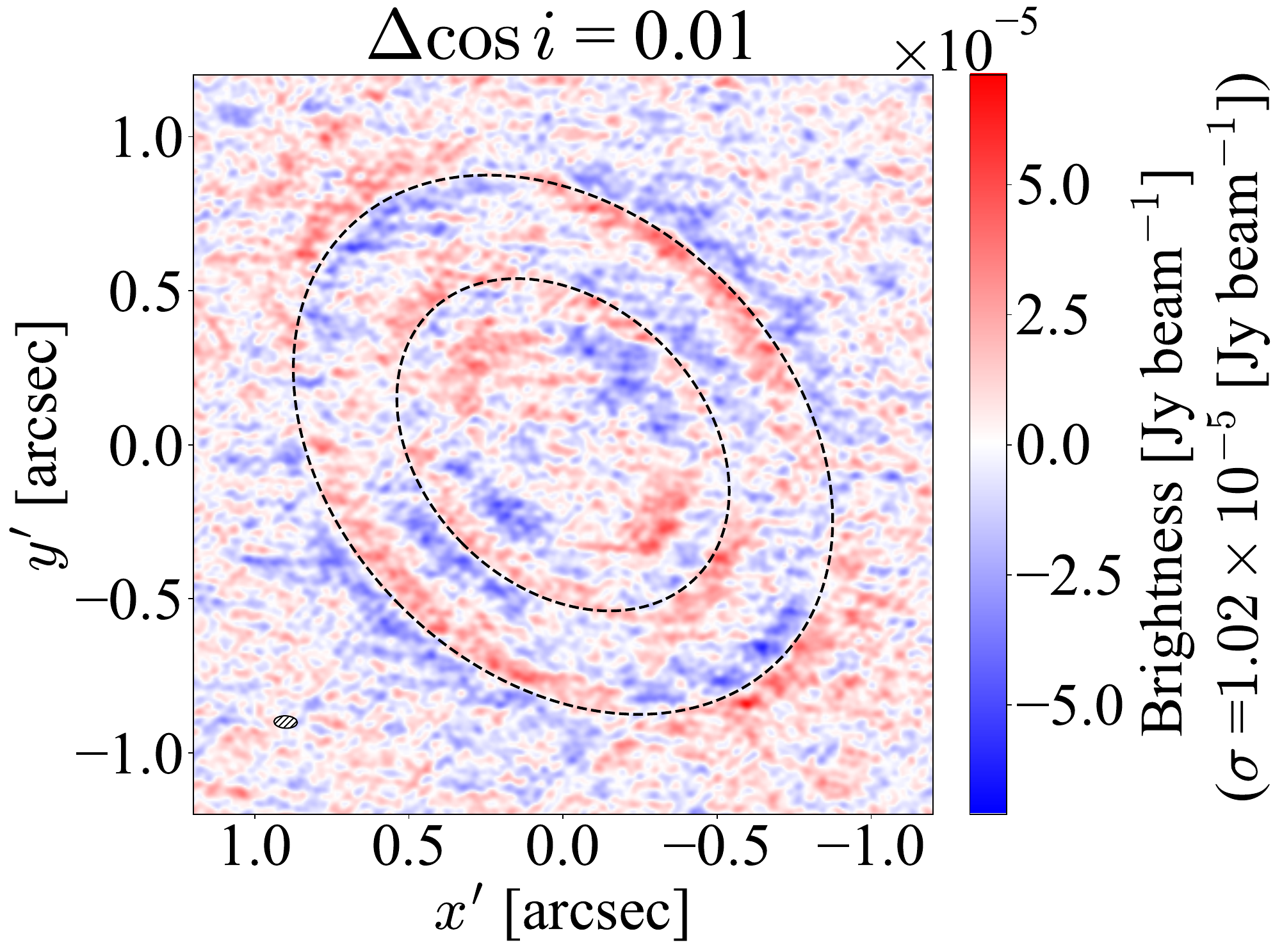}
\includegraphics[width=0.325\linewidth]{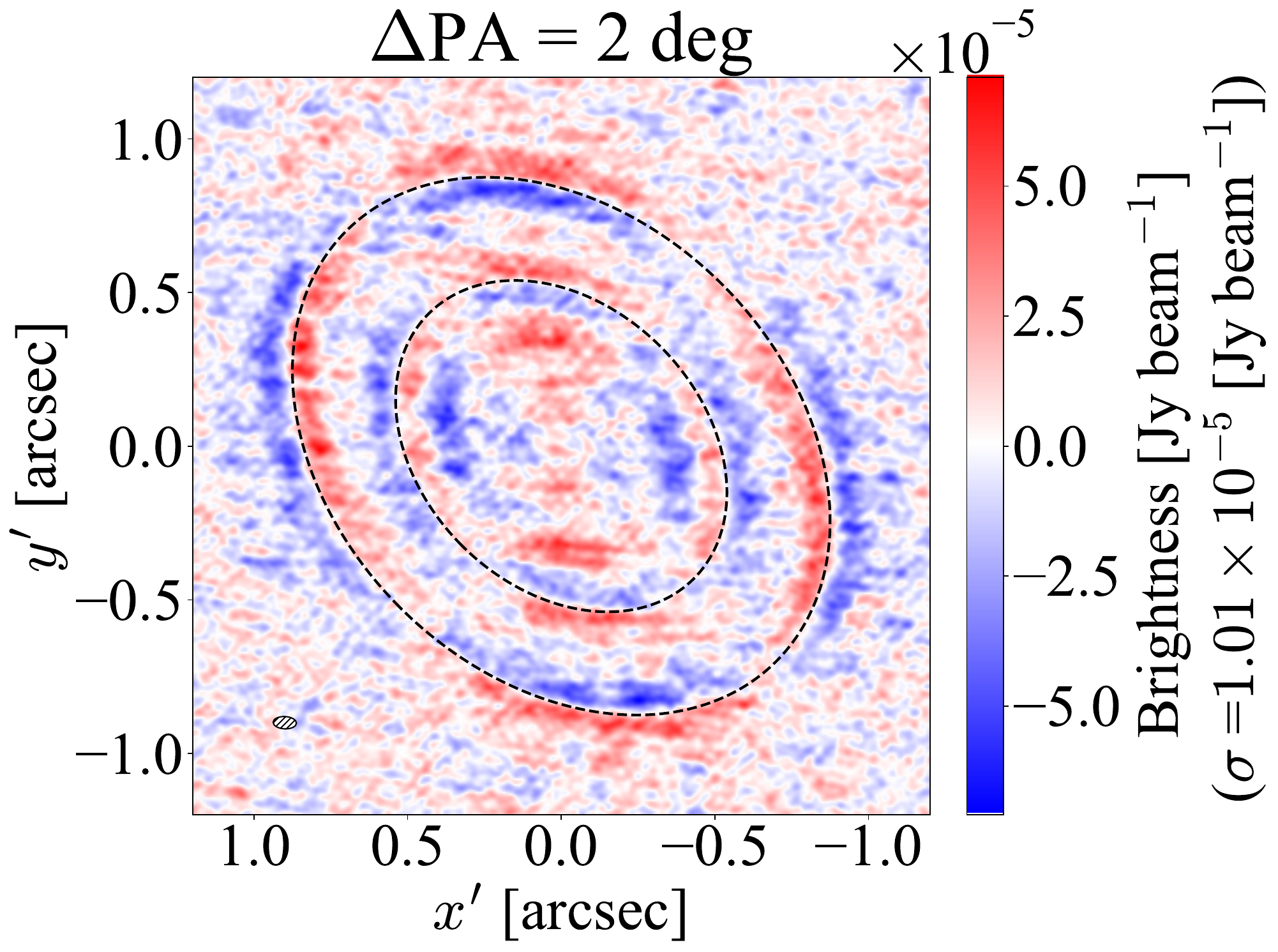}
\end{center}
\caption{Unsubtracted image and residual images produced with incorrect geometric parameters in simulated case for AS 209. The standard deviation for the brightness $\sigma$ is computed in the region outside the disc.  (upper-left panel) The image produced with CLEAN for unsubtracted data. (upper-middle panel) The residual image produced with residual visibilities produced from the MAP estimate. (other panels) The residual images with each of the geometric parameters being shifted. The synthesized beam size $(0.076\arcsec, 0.040\arcsec)$ is shown at the bottom left. }
\label{fig:wrong_geo}
\end{figure*}

\begin{figure*}
\begin{center}
\includegraphics[width=0.32\linewidth]{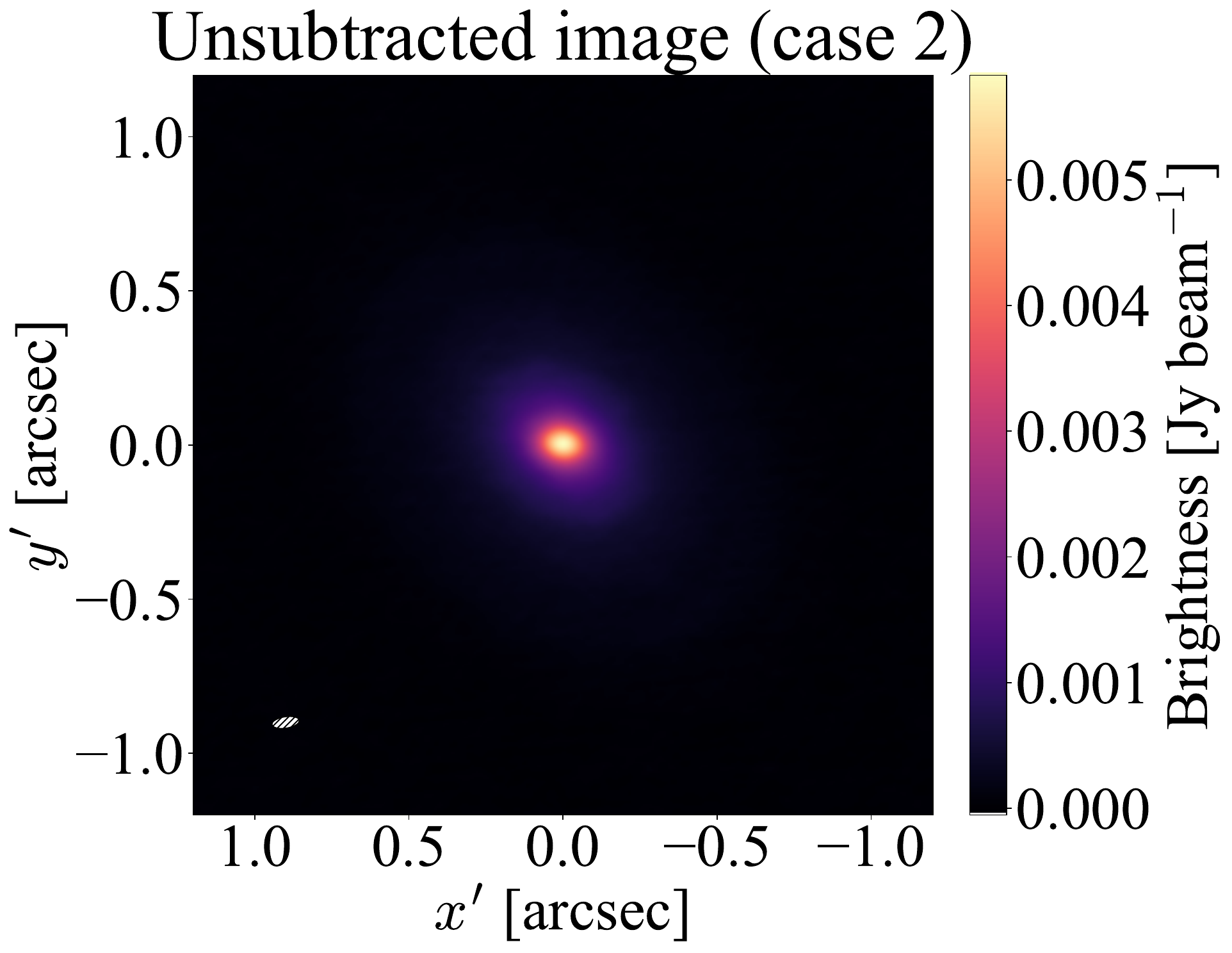}
\includegraphics[width=0.32\linewidth]{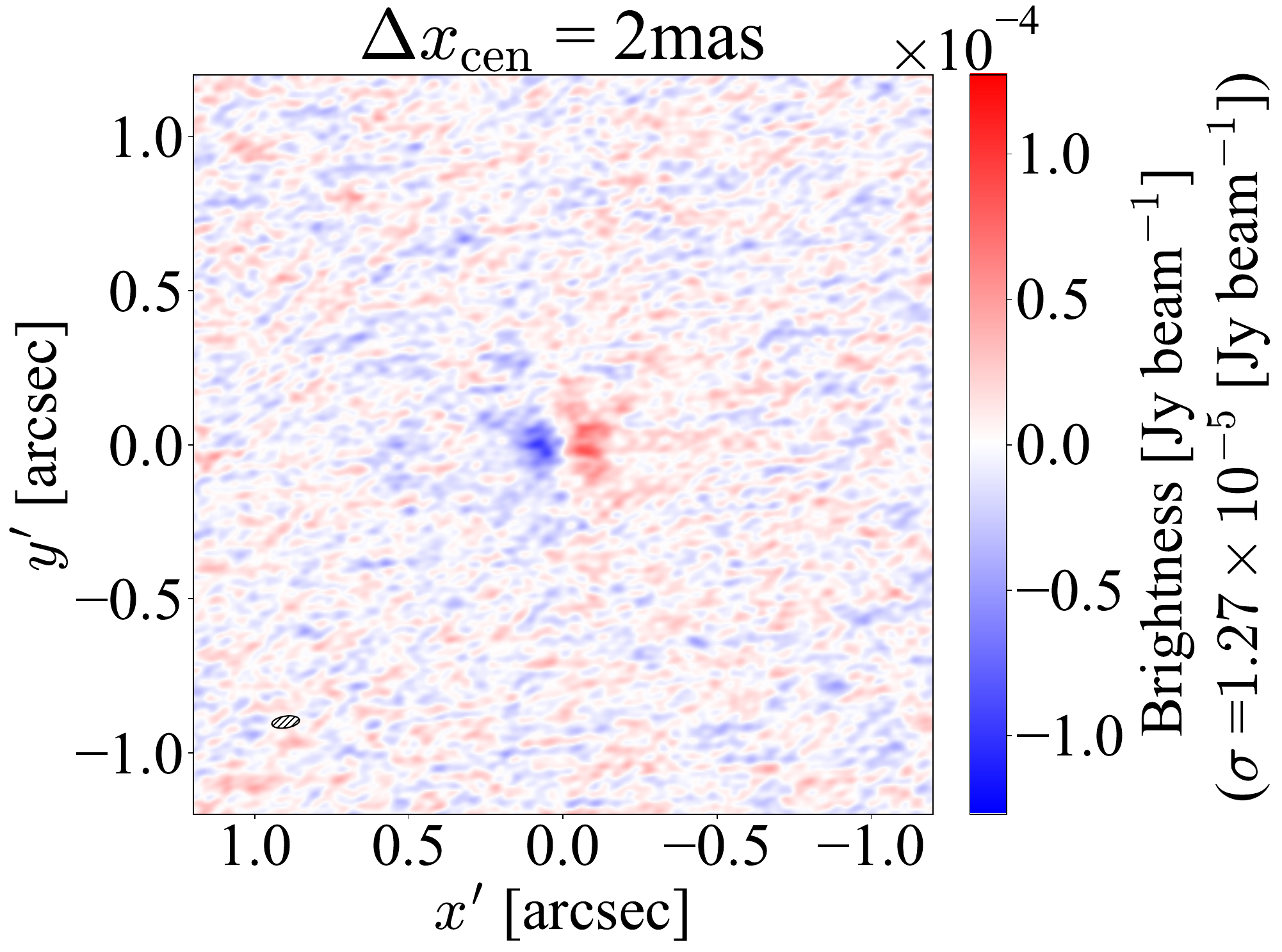}
\includegraphics[width=0.32\linewidth]{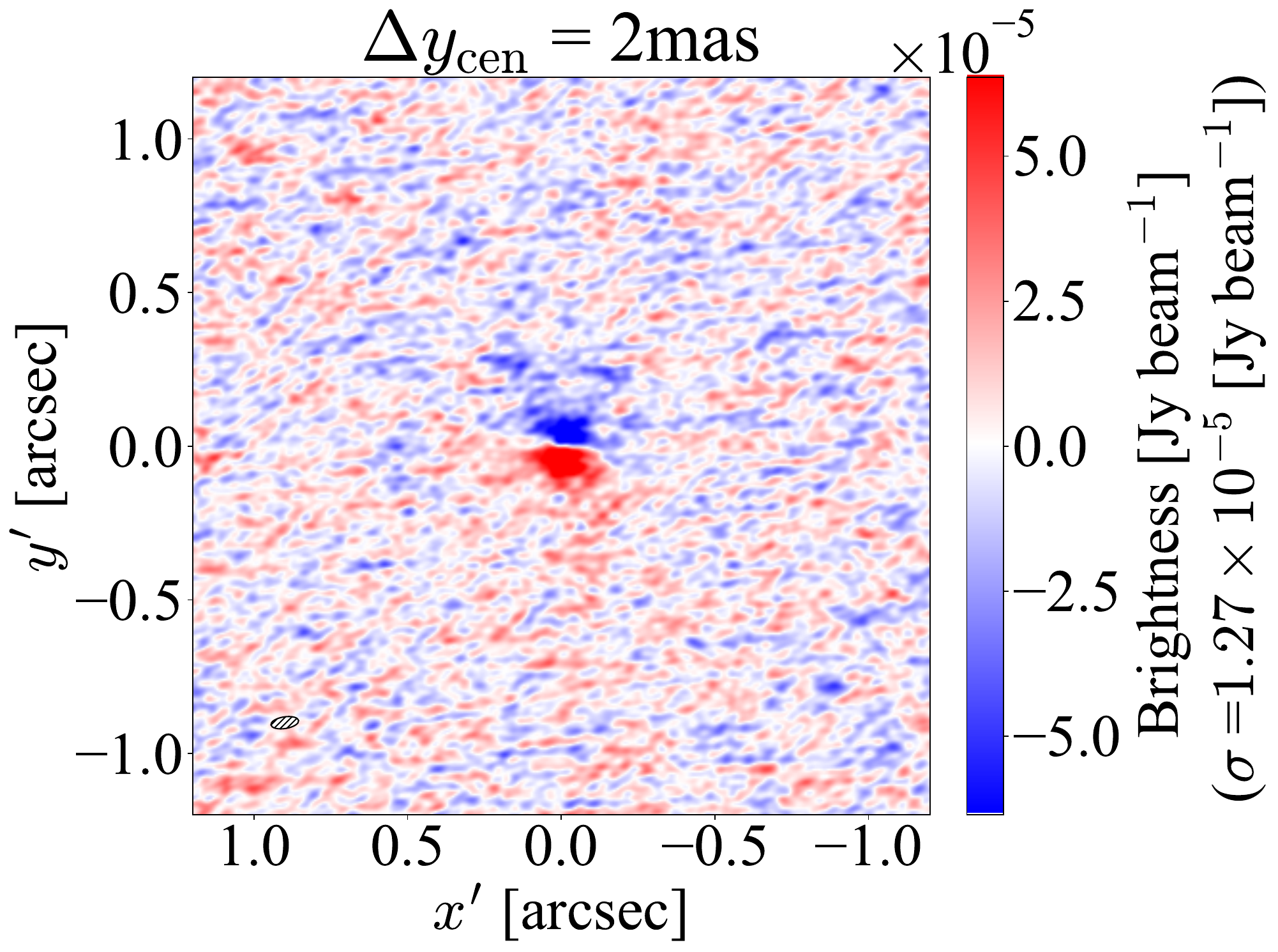}
\includegraphics[width=0.32\linewidth]{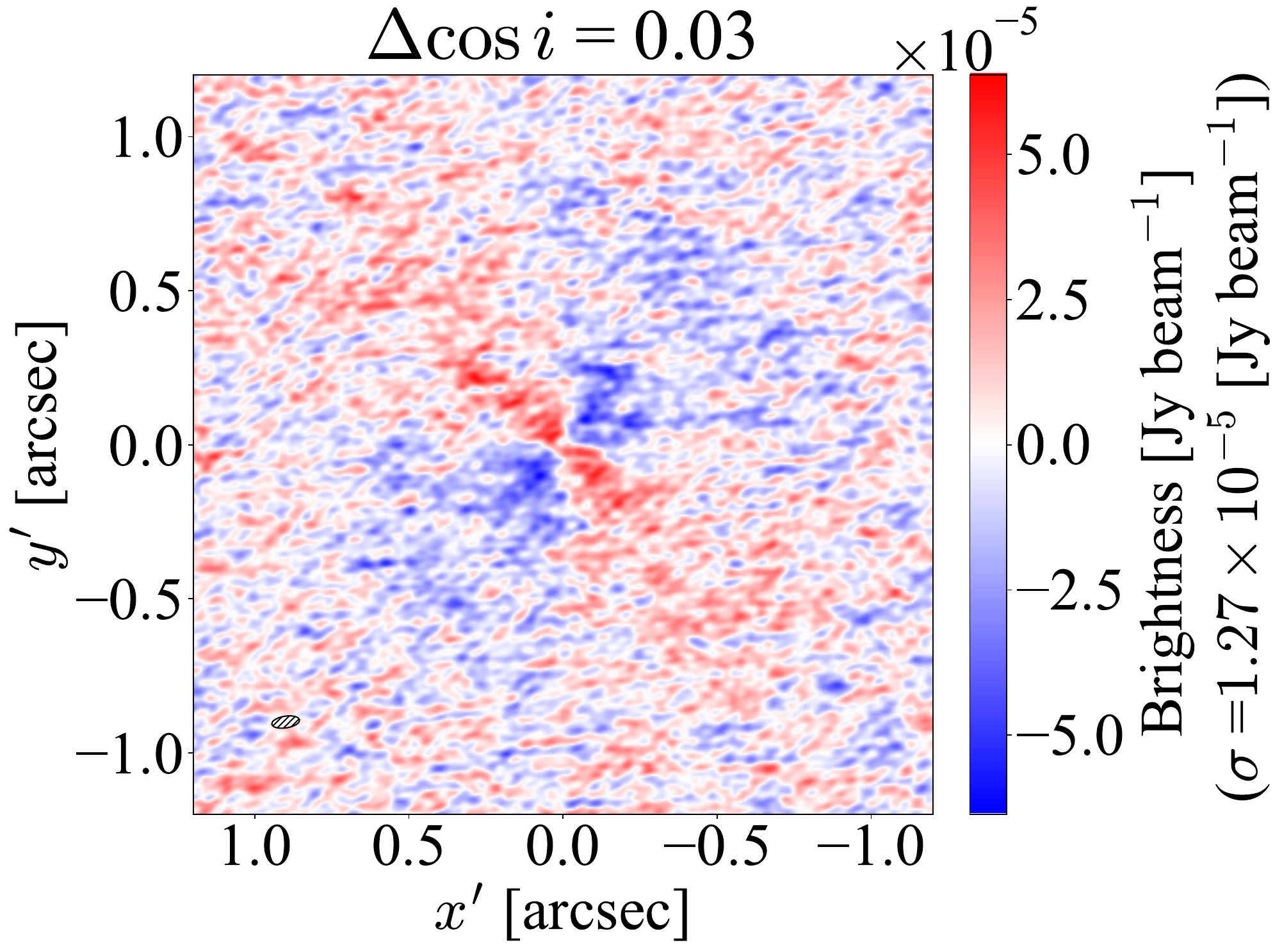}
\includegraphics[width=0.32\linewidth]{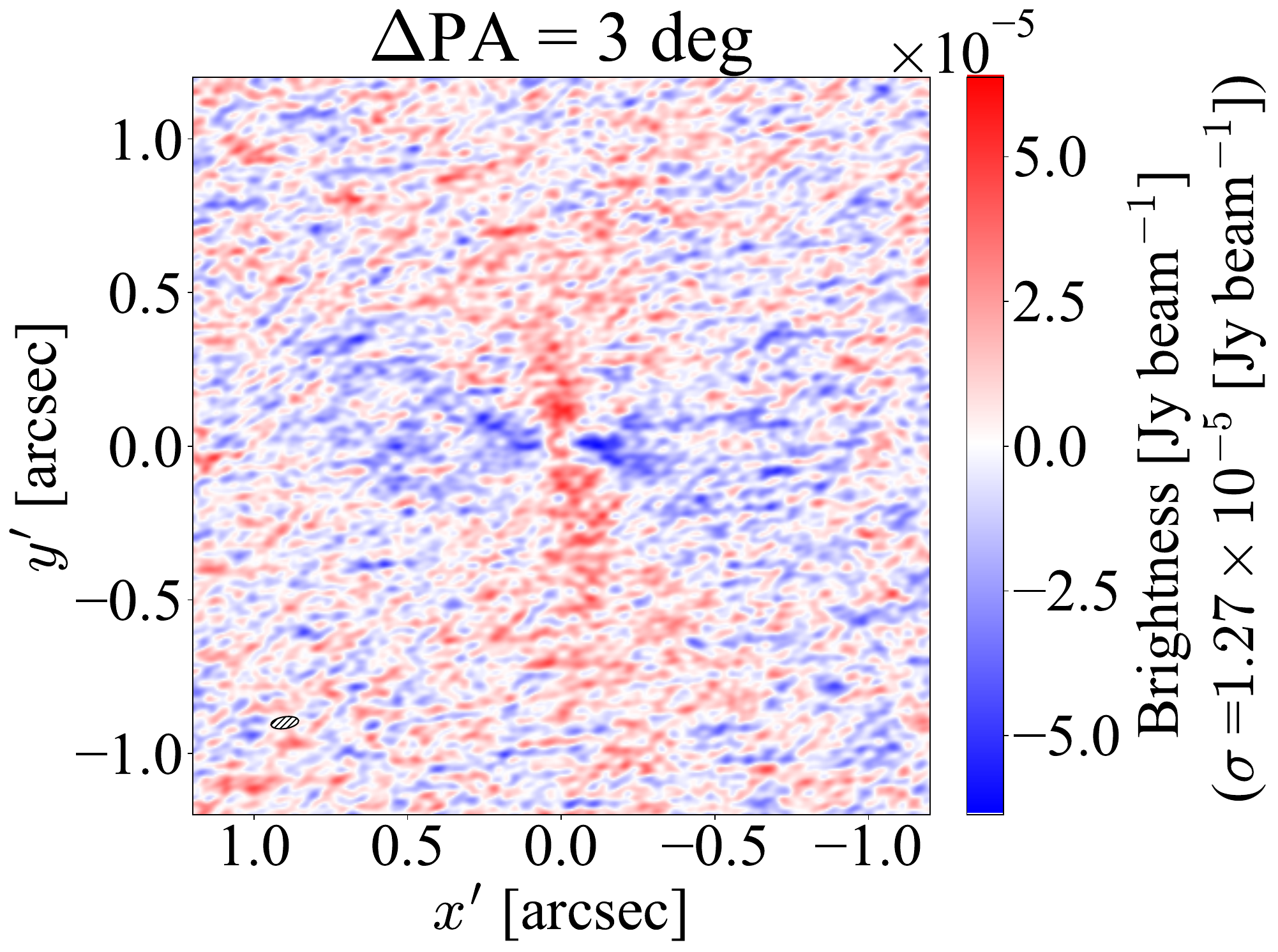}
\end{center}
\caption{Same as Fig. \ref{fig:wrong_geo} albeit for simulated data of WaOph 6.  The synthesized beam size $(0.091\arcsec, 0.037\arcsec)$ is shown at the bottom left. }
\label{fig:wrong_geo_waoph6}
\end{figure*}

Fig. \ref{fig:wrong_geo} and \ref{fig:wrong_geo_waoph6} show the unsubtracted and residual images with their geometric parameters being shifted. They are all shown in observed frames. The residual images based on the best-fitting models did not exhibit any noticeable structure. This demonstrates the feasibility of the current method. However, the residual images with shifted geometric parameters exhibited coherent patterns. These residual images obtained from the current study were overall consistent with those presented in \cite{andrews2021}, wherein the dependence of residual images was studied by shifting the geometric parameters. As shown in the figures, the difference in the injected brightness profiles changed the residual images. We highlight two outer rings for AS 209 by black dashed ellipses in Fig. \ref{fig:wrong_geo}, and we find that the ring locations indeed corresponded to the boundaries, where the signs of the residuals were flipped. However, the brightness profile for WaOph 6 was more continuous than AS 209, and the residual images were less structured. 

\begin{figure*}
\begin{center}
\includegraphics[width=0.45\linewidth]{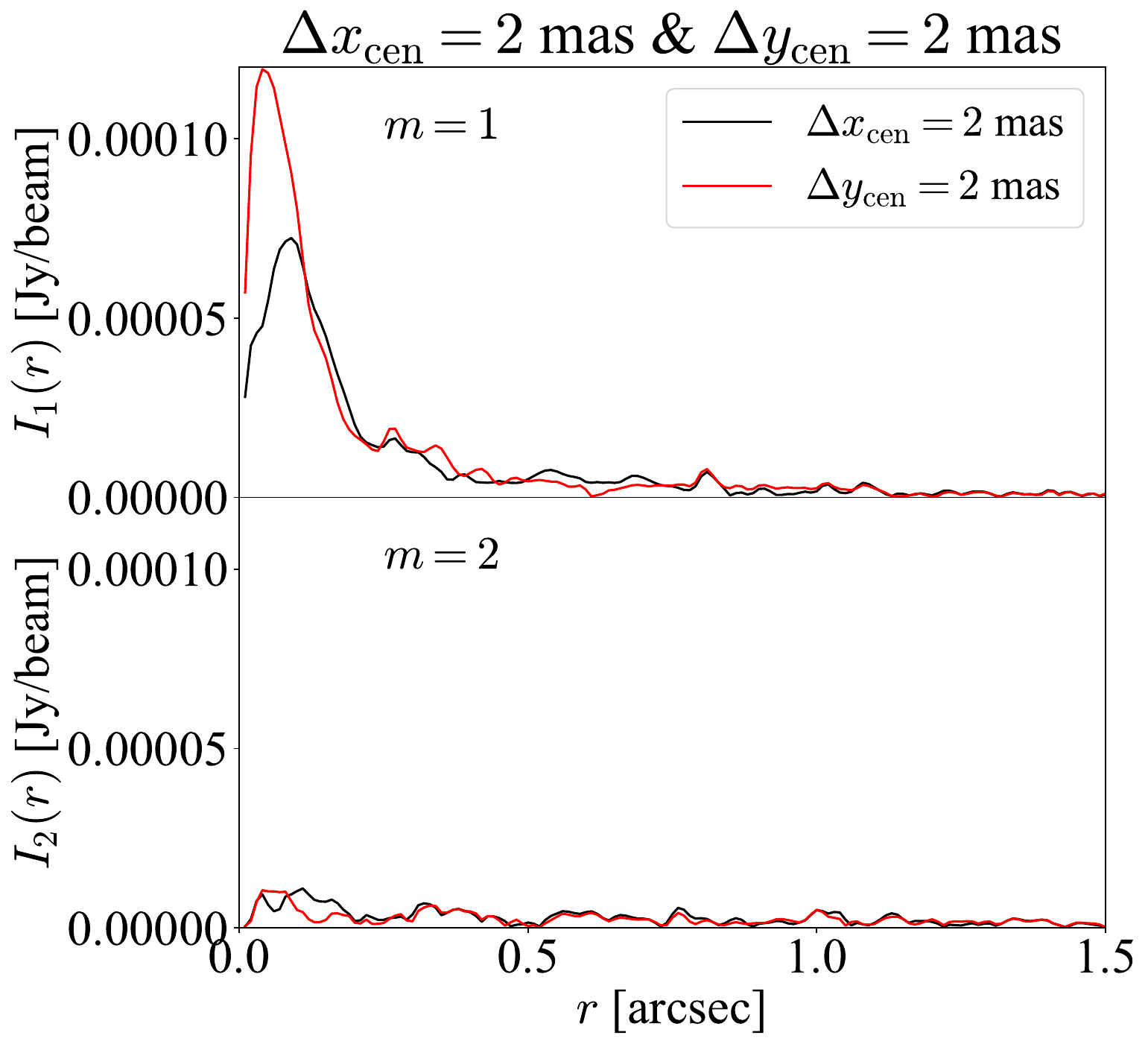}
\includegraphics[width=0.45\linewidth]{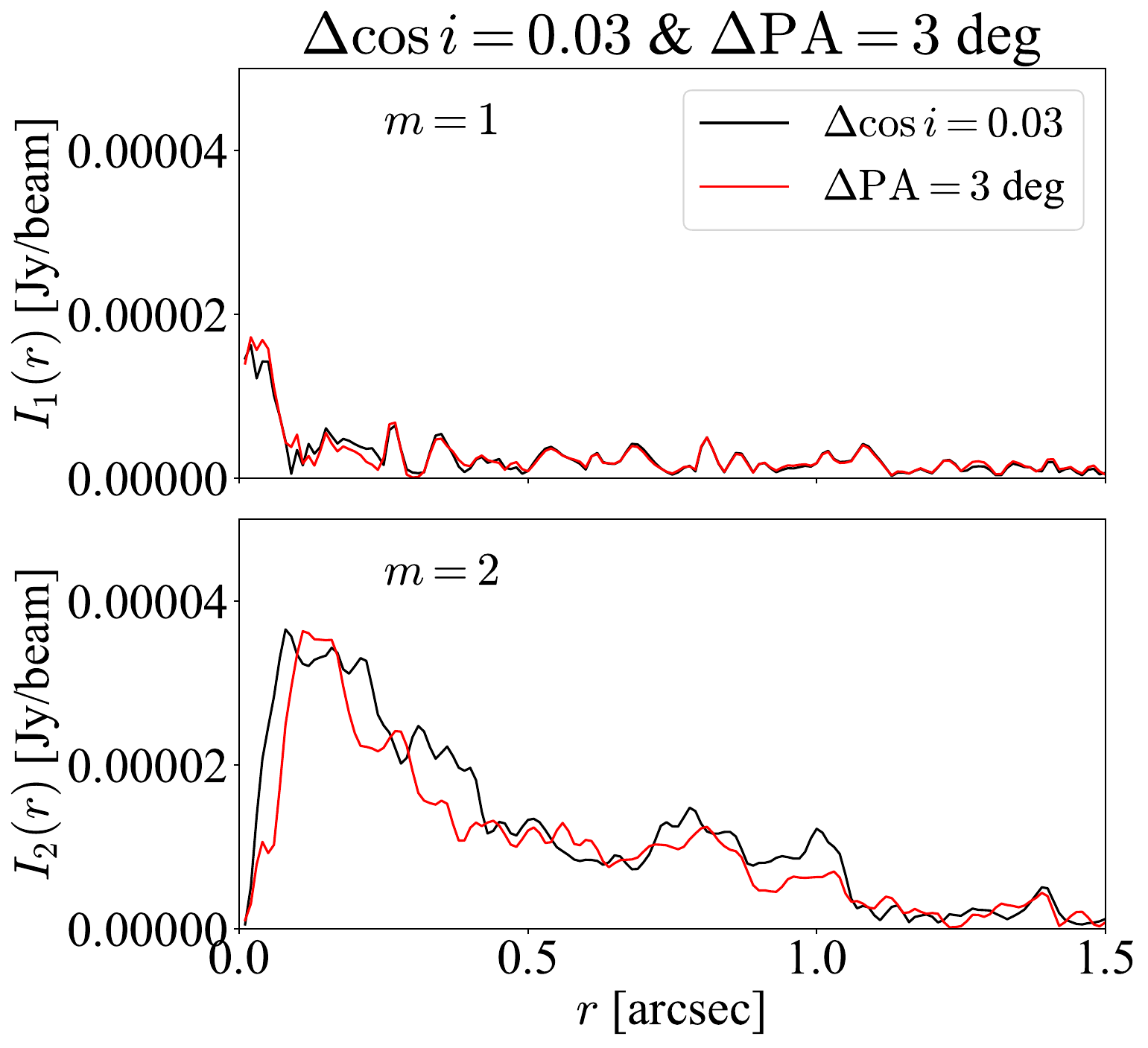}
\end{center}
\caption{$I_{m=1,2}(r)$ computed for residual images in Fig. \ref{fig:wrong_geo_waoph6}. The left and right panels show the results for the shifted central position and the shifted disc orientation, respectively. }
\label{fig:m12_geometry_waoph6}
\end{figure*}

Fig. \ref{fig:m12_geometry_waoph6} shows $I_{m=1,2}(r)$ for the residual images in the simulated case for WaOph6. The shifts in the central positions introduced residuals with the $m=1$ component, whereas those in $\cos i$ and PA introduced the residuals with $m=2$ component. As discussed in the later sections, these residuals can be degenerate with real signals (e.g., spirals); thus they can introduce biases for estimating such non-axisymmetric structures in discs. 
  
As a comparison with the proposed method, we also applied the Gaussian fitting to the visibilities to estimate the geometric parameters. Here, the Gaussian fitting of visibilities corresponds to the fitting of the Gaussian function on the image plane, and it is frequently used for estimating a geometry of a disc. The model was specified by six parameters, where four parameters, $(\Delta x_{\rm cen},\Delta y_{\rm cen}, \cos i, {\rm PA})$ are common to our model. We implemented a posterior sampling for the Gaussian model using {\tt emcee} \citep{foreman2013}. 

\begin{figure*}
\begin{center}
\includegraphics[width=0.42\linewidth]{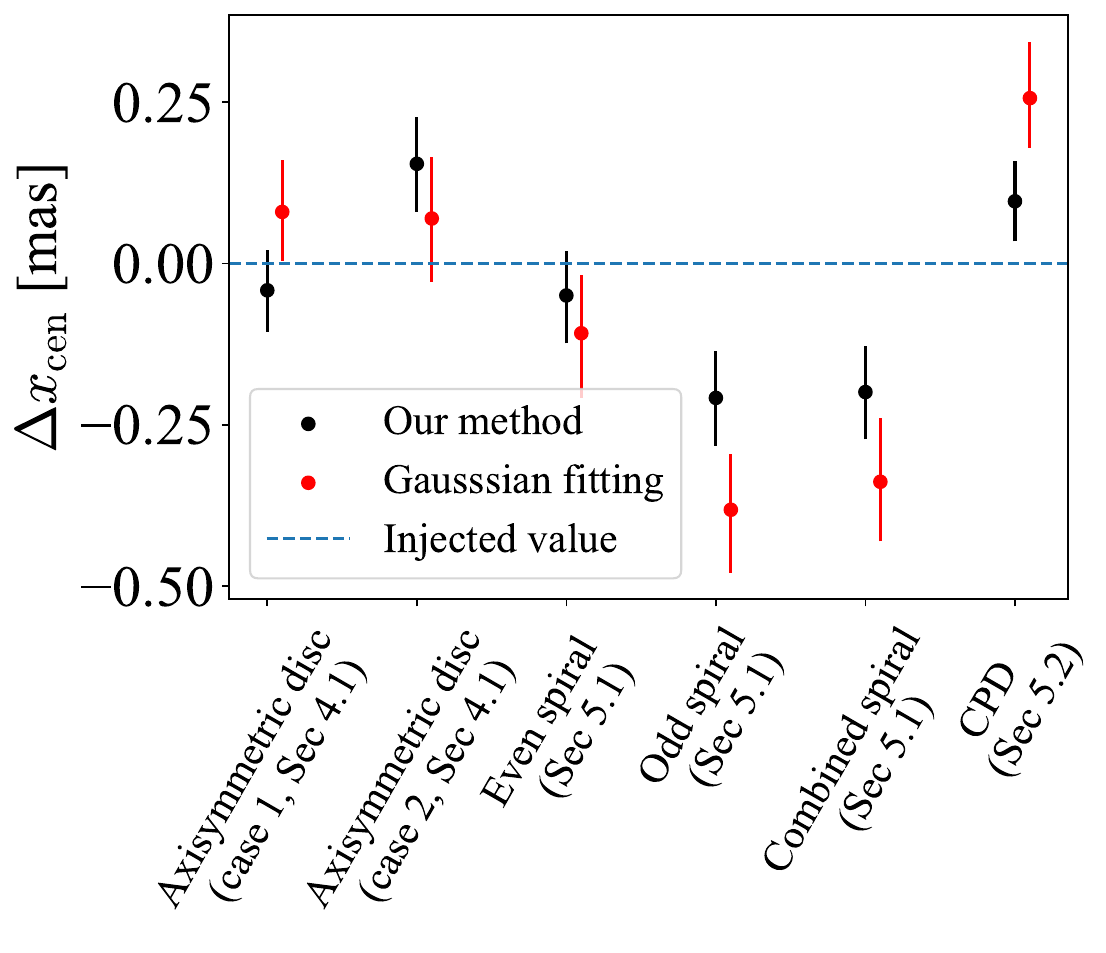}
\includegraphics[width=0.42\linewidth]{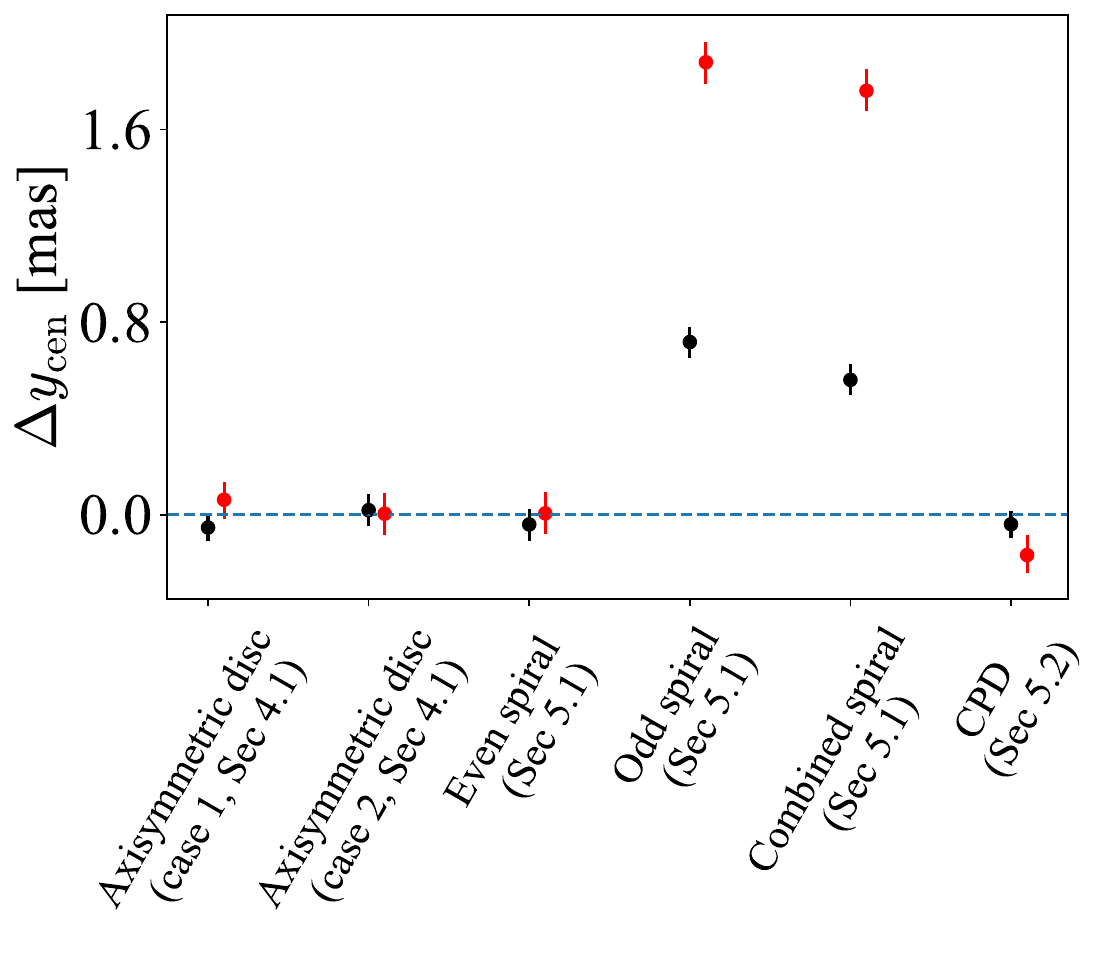}
\includegraphics[width=0.42\linewidth]{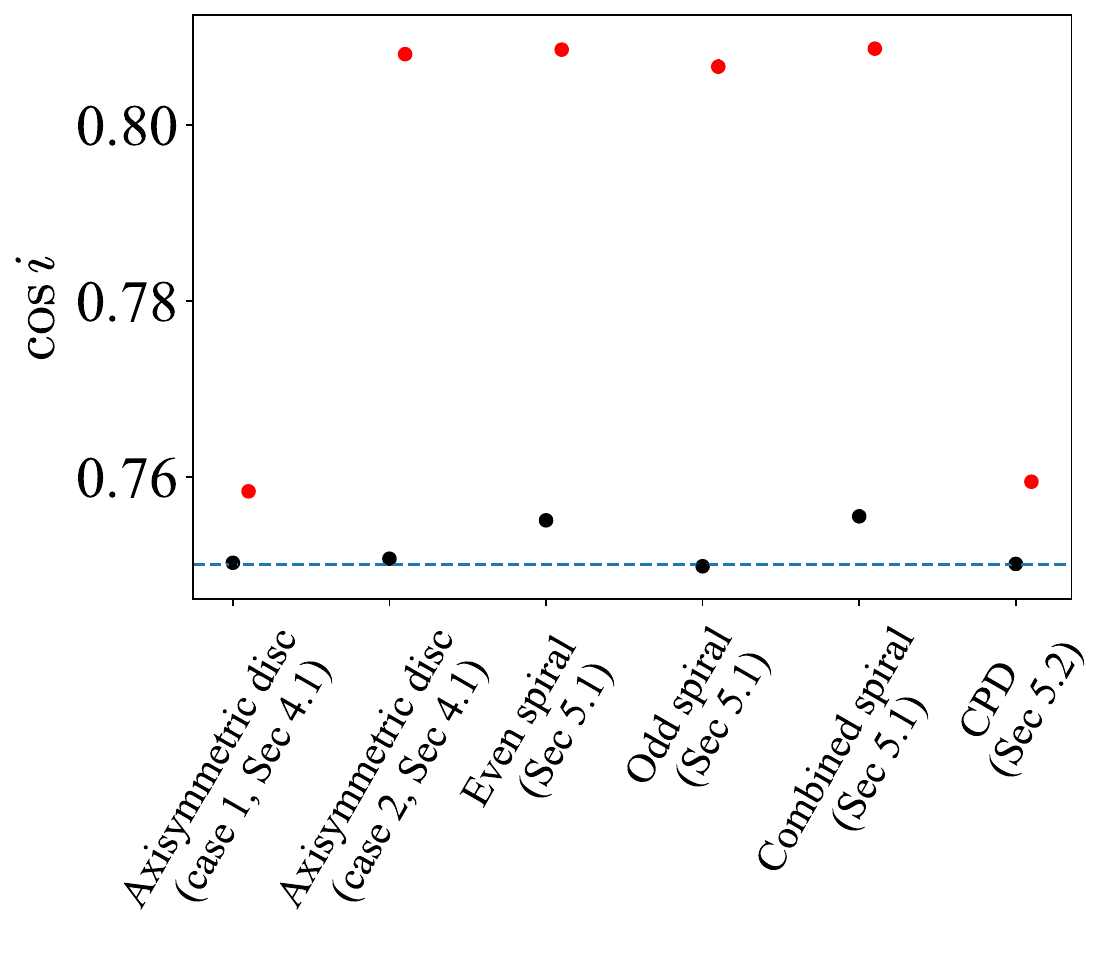}
\includegraphics[width=0.42\linewidth]{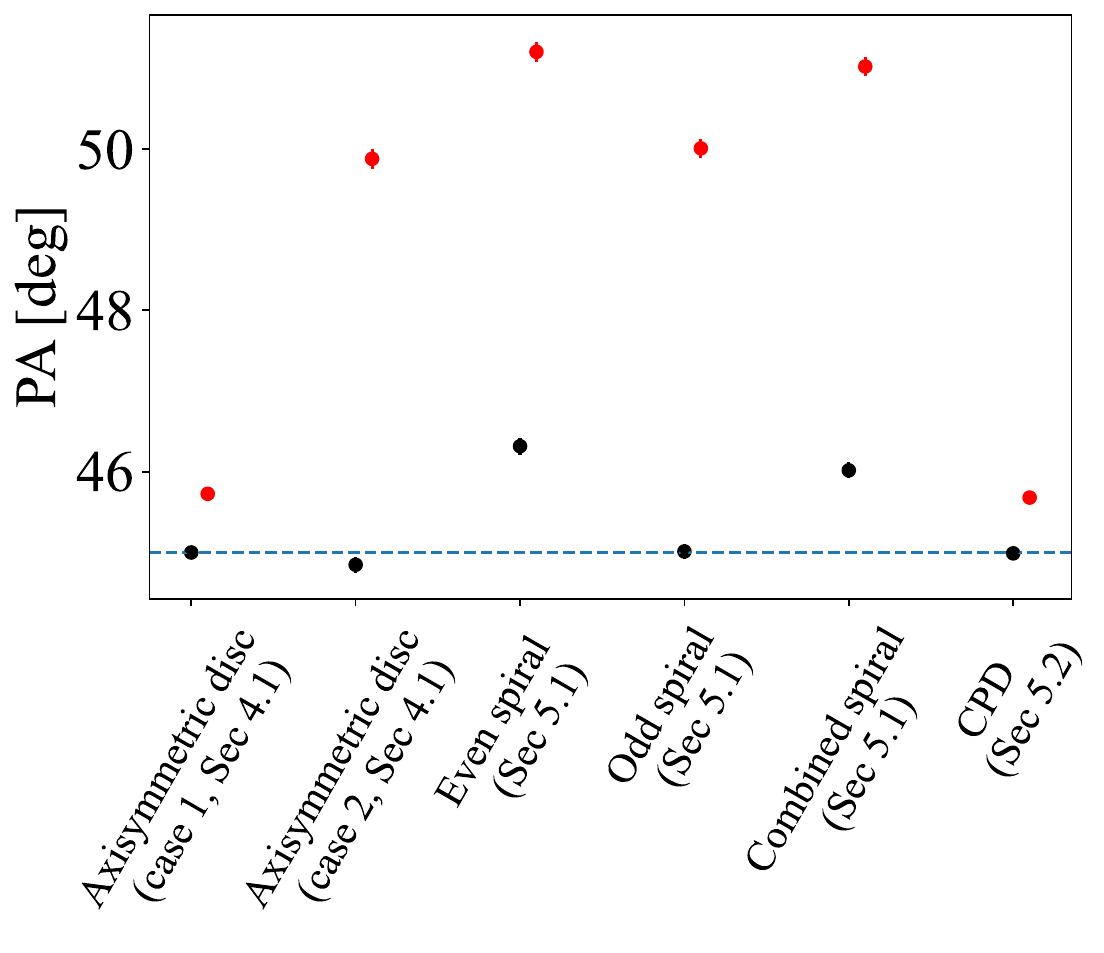}
\end{center}
\caption{Injected and recovered geometric parameters in simulations in Sec \ref{sec:sim} and \ref{sec:sim_non}. Each
panel compares the geometric parameter: the central position and the disc orientation. The geometry is estimated using the proposed method (black) and the Gaussian fitting (red). The injected parameters are indicated by the dashed blue lines. }
\label{fig:comparison_geometry}
\end{figure*}

The first and second data from the left in each panel of Fig. \ref{fig:comparison_geometry} show the injected and recovered geometric parameters obtained from the current method and the Gaussian fitting.  The central positions were well recovered in both of cases. However, the parameters $\cos i$ and PA obtained from the Gaussian fitting deviated from the injected values; $\Delta \cos i \simeq 0.01-0.05$ and $\Delta {\rm PA}\simeq 1-5$ deg. These deviations can be problematic when searching for faint signals (e.g., circumplanetary discs and spirals) because they introduce fake non-axisymmetric structures in the residual images as shown in Figs. \ref{fig:wrong_geo} and \ref{fig:wrong_geo_waoph6}. 

%-----------------------------Figure Start---------------------------
%-----------------------------Figure End------------------------------

%-----------------------------Figure Start---------------------------
%-----------------------------Figure End------------------------------

%-----------------------------Figure Start---------------------------
%-----------------------------Figure End------------------------------

%-----------------------------Figure Start---------------------------
%-----------------------------Figure End------------------------------

\section{Injection and recovery test for spirals and circumplanetary disc embedded into axisymmetric emission} \label{sec:sim_non}
As a simple extension to Sec \ref{sec:sim}, we considered an additional non-axisymmetric perturbation to the symmetric disc. We injected spirals and a point source into an axisymmetric model and simulated the visibilities incorporating noises. Subsequently, we applied our method to extract the axisymmetric model and construct residual images, which were then compared against the injected structures.

\subsection{Axisymmetric structure $+$ spirals}
\subsubsection{Injection of spirals}
We injected spiral structures into an axisymmetric structure assuming the same observational setup as the simulated case for WaOph6. The brightness profile was assumed to be the same as the case for WaOph6. In this simulation, we assume zero brightness outside $r=1.11\arcsec$, beyond which the brightness in the literature can become negative. Three different models were considered; an odd-symmetric spiral with $m=1$ component, an even-symmetric spiral with $m=2$ component, and the combined image for both spirals. Specifically, we assumed perturbations in the form of $\Delta I (m, r, \phi) = \Delta I_{m} (r) \cos (m (\phi - \phi_{m}(r)))$, where we adopted an Archimedean spiral $\phi_{m} (r) = ar + b$. We assumed $(m, a, b) = (1, 20, 1)$ for the odd-symmetric spiral and $(m, a, b) = (2, -10, -\pi/6)$ for the even-symmetric mode. The amplitudes $I_{m} (r)$ were assumed as follows: 

\begin{eqnarray}  
 I_{1}(r)  &=&
\left\{
    \begin{array}{cc}
           \displaystyle 5 \times 10^{-5} \exp (-((r-0.4 \arcsec)/0.1 \arcsec)^{2}) \\ {\rm [ Jy \;beam^{-1}]} 
    \end{array}
    \right. \label{eq: delta_I} \\
 I_{2}(r) &=&
\left\{
    \begin{array}{cc}
           \displaystyle 5\times 10^{-5} \;\; {\rm [ Jy \;beam^{-1}]}  & \;\;\;\;  (0<r< 0.6 \arcsec) \\
        0  &  \;\;\;\; (0.6 \arcsec<r) \\
    \end{array}
    \right. \label{eq: delta_I} , 
\end{eqnarray}
where the beam area was assumed to be $0.00354\; [{\rm arcsec}]^{2}$. For the calculation of model visibilities, we used functions for the non-uniform fast Fourier transform from {\tt PyNUFFT} \citep{lin2017pynufft,jimaging4030051}. We adopted the size of the image grid $N_{d}=1024$, the size of the oversampled Fourier grid $K_{d}=2048$, and the size of the interpolator $J_{d}=6$.

\begin{figure*}
\begin{center}
\includegraphics[width=0.55\linewidth]{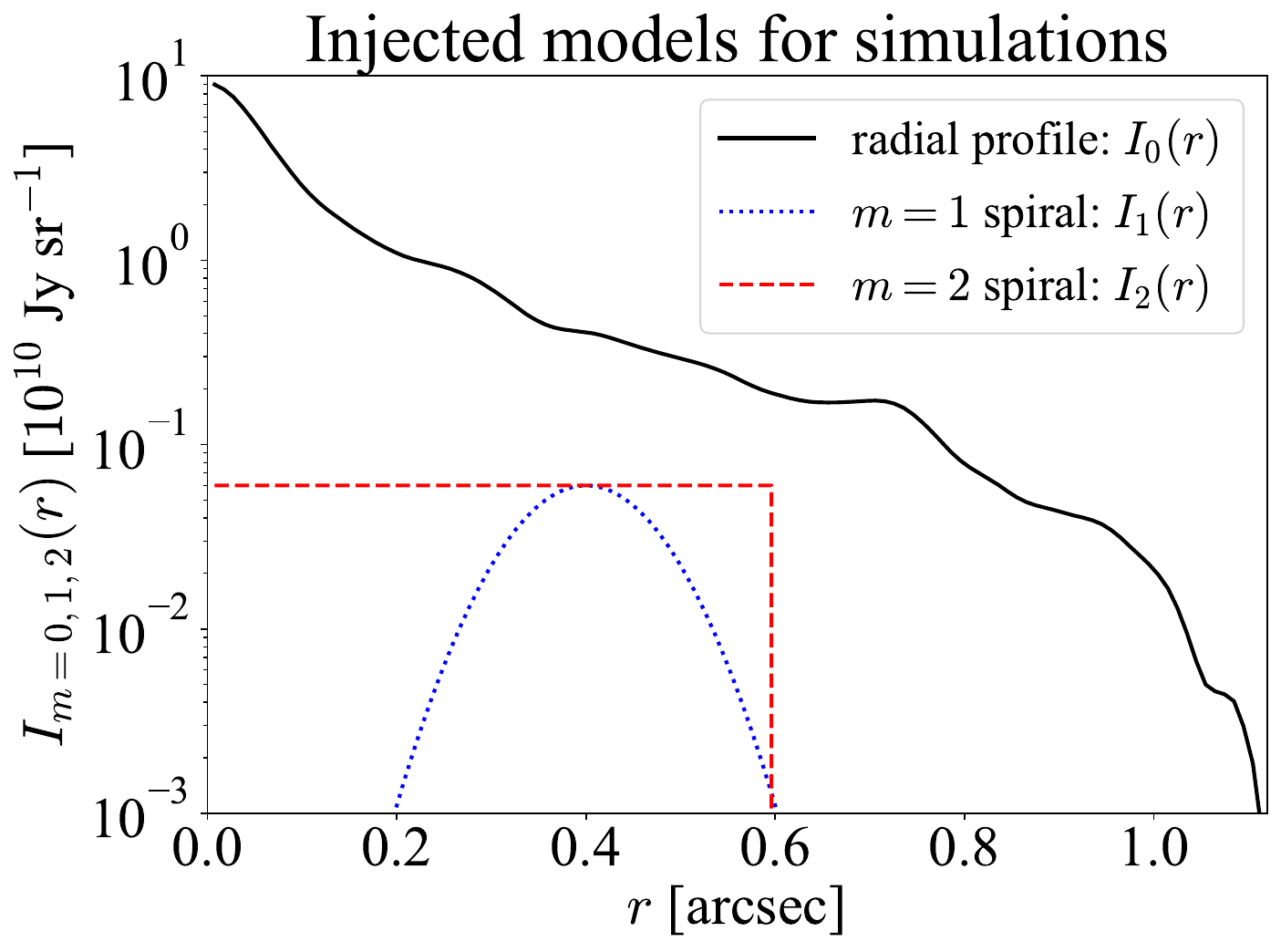}
\end{center}
\caption{Amplitudes of injected brightness profile and spirals in simulations in Sec \ref{sec:sim_non}. }
\label{fig:input_profile}
\end{figure*}

Fig. \ref{fig:input_profile} shows the amplitudes of the injected brightness profile and spirals. The amplitudes of the injected spirals were at most $\sim 30$ per cent of that of the axisymmetric part of the surface brightness, and they were comparable to the amplitudes of the observed spirals with $m=2$ mode (e.g., WaOph6 and IM Lup) such that the current simulation indeed considered the realistic situation.  The left panels in Fig. \ref{fig:spiral_recovered} show the injected images for the three cases. 

%-----------------------------Figure Start---------------------------
%-----------------------------Figure End------------------------------
\subsubsection{Comparison of injected and recovered models}
After simulating the visibilities, we applied our method to estimate the axisymmetric structures. We selected the MAP estimate on geometric parameters, drew a brightness profile, and subtracted their contributions from the observations. The residual visibilities were then used to construct the residual images using CLEAN. 

For the comparison, we also created the residual images by adopting geometric parameters estimated from the Gaussian fitting to the visibilities. Specifically, we assumed geometry with Gaussian fitting and hyperparameters from our method, and we drew a brightness profile. Subsequently, the brightness model was subtracted from the visibilities, and the residual visibilities were imaged with CLEAN. 

\begin{figure*}
\begin{center}
\includegraphics[width=0.97\linewidth]{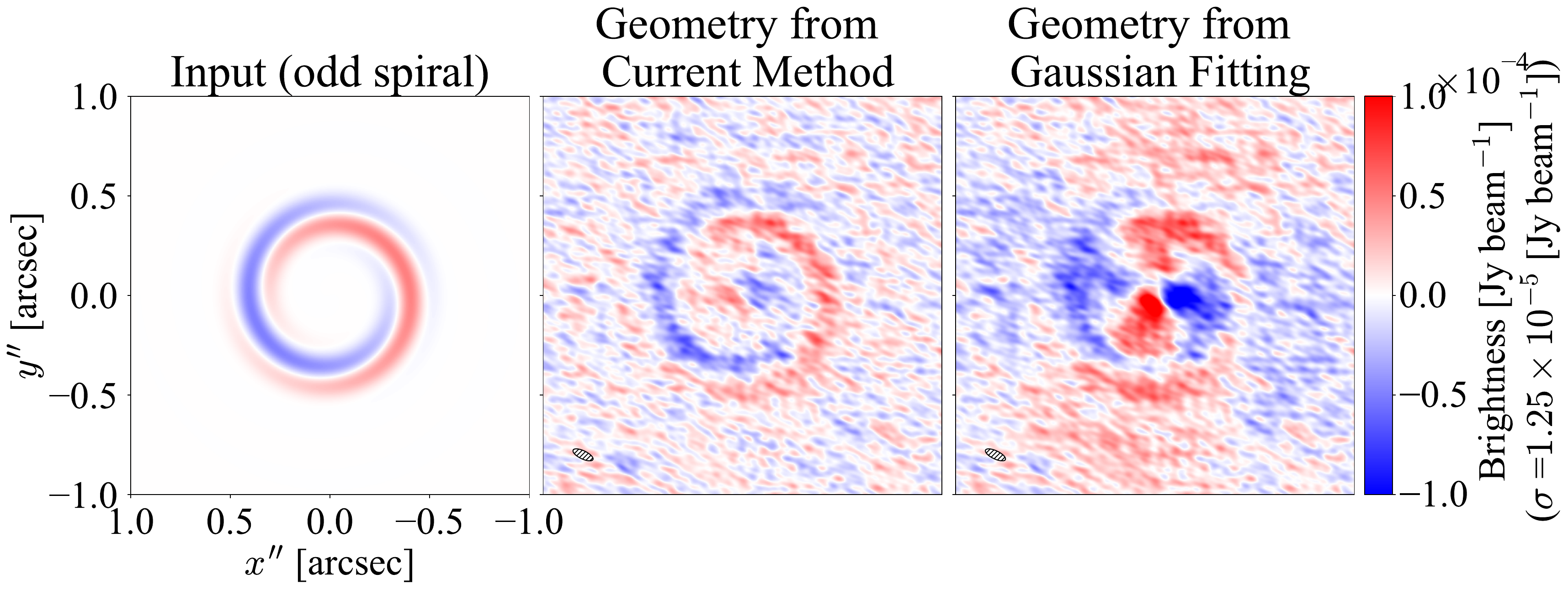}
\includegraphics[width=0.97\linewidth]{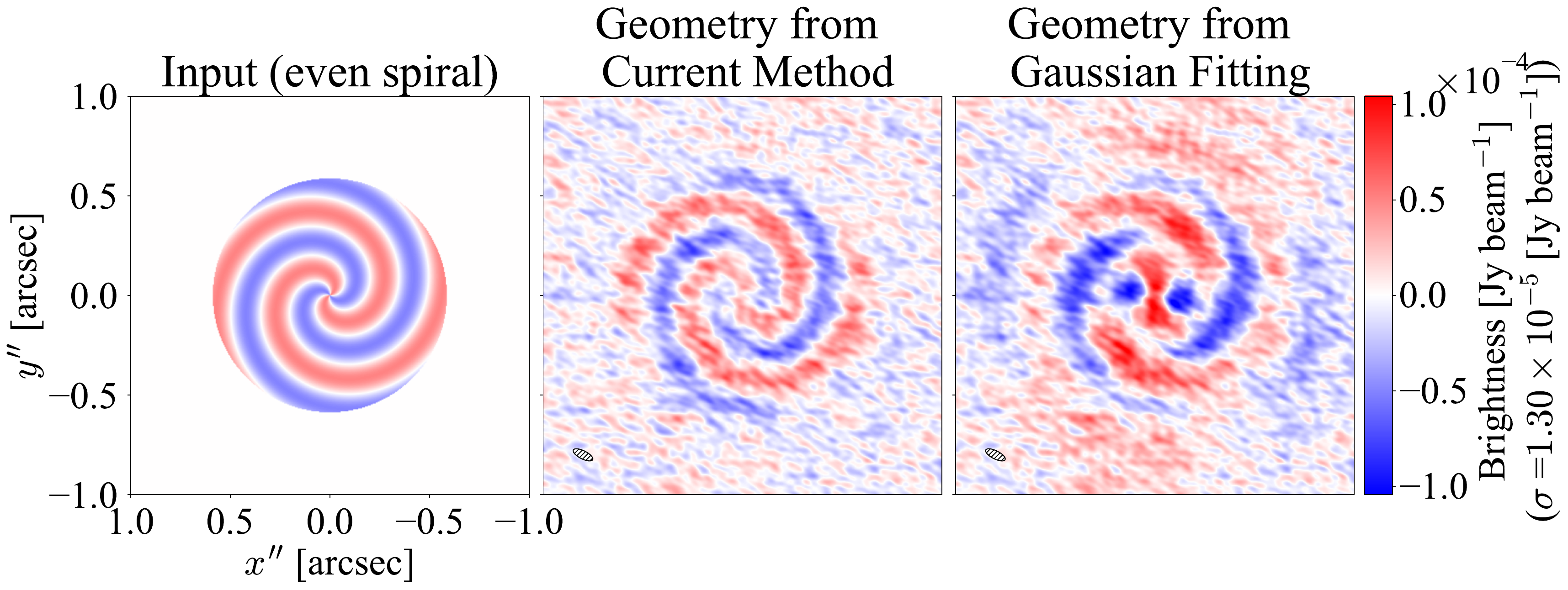}
\includegraphics[width=0.97\linewidth]{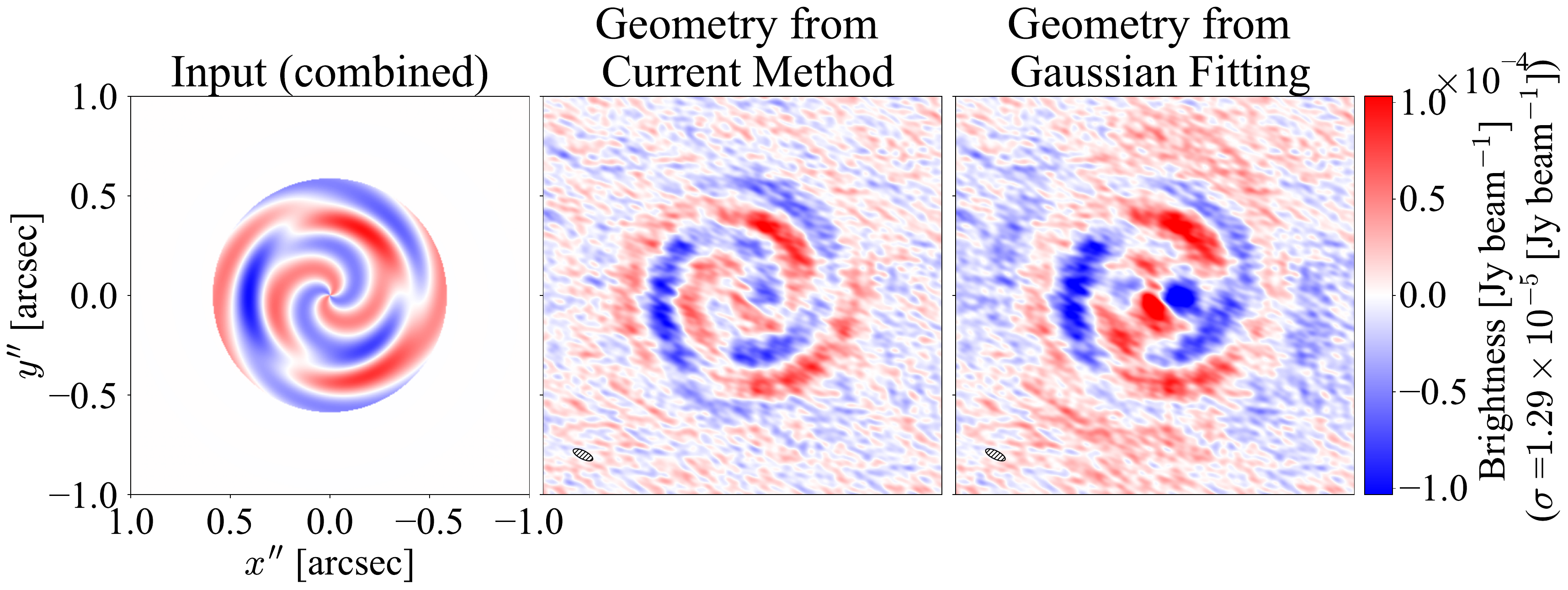}
\end{center}
\caption{Injected and recovered residual images for spirals in deprojected frames; (top rows) Even-symmetric spiral   (middle rows) Odd-symmetric spiral (bottom rows) Combined spirals. (left columns) Injected perturbations (middle columns) Recovered residual images based on the proposed method (right columns) Recovered residual images based on the geometry obtained with Gaussian fitting. The synthesized beam size for the deprojected image $(0.11\arcsec, 0.041\arcsec)$ is shown at the bottom left, and the beam size for the image before deprojection is $(0.091\arcsec, 0.037\arcsec)$. }
\label{fig:spiral_recovered}
\end{figure*}

The residual images are shown in Fig. \ref{fig:spiral_recovered}. In the second column, we show the recovered residual images obtained via the current method. The last column shows the residual images created from the geometry given by the Gaussian fitting. The residual images from our method were reasonably consistent with the injected maps, whereas those based on the Gaussian fitting exhibited larger residuals. However, the innermost structures of the discs at $r<0.2-0.3 \arcsec$ were not perfectly recovered even with our method. Specifically, the excess in the amplitude of the spiral was observed at $r<0.3\arcsec$ for the odd-symmetric spiral. Moreover, no clear spiral structure was observed at $r<0.2\arcsec$ for the even-symmetric spiral. These inconsistencies arise from the biases in the estimated geometric parameters, as discussed next. 

The comparison of geometric parameters for three cases is shown in Fig. \ref{fig:comparison_geometry}. For every parameter, larger deviations from the injected parameters were observed in the case of the Gaussian modelling than that using the proposed method. However, our method was still affected with biases; $(\Delta x_{\rm cen}, \Delta y_{\rm cen})$ for the odd-symmetric spiral, and $({\rm PA}, \cos i)$ for the even-symmetric spiral. In the case of the combined spiral, the shifts in all the geometric parameters roughly corresponded to the summation of shifts in the odd and even-symmetric spirals. These shifts are reasonable because, as discussed in Sec \ref{sec:sim}, the shifts in $(\Delta x_{\rm cen}, \Delta y_{\rm cen})$ introduce the residuals with $m=1$ mode, which can be degenerate with the injected odd-symmetric spiral. Similarly, the shifts in $( \cos i, {\rm PA})$ introduce the residuals with $m=2$ mode, which can be degenerate with the even-symmetric spiral. These degeneracies produce the biases in estimating geometric parameters. Consequently, these biases impede recovery of injected non-axisymmetric structures, as shown in Fig. \ref{fig:spiral_recovered}. 

\begin{figure*}
\begin{center}
\includegraphics[width=0.45\linewidth]{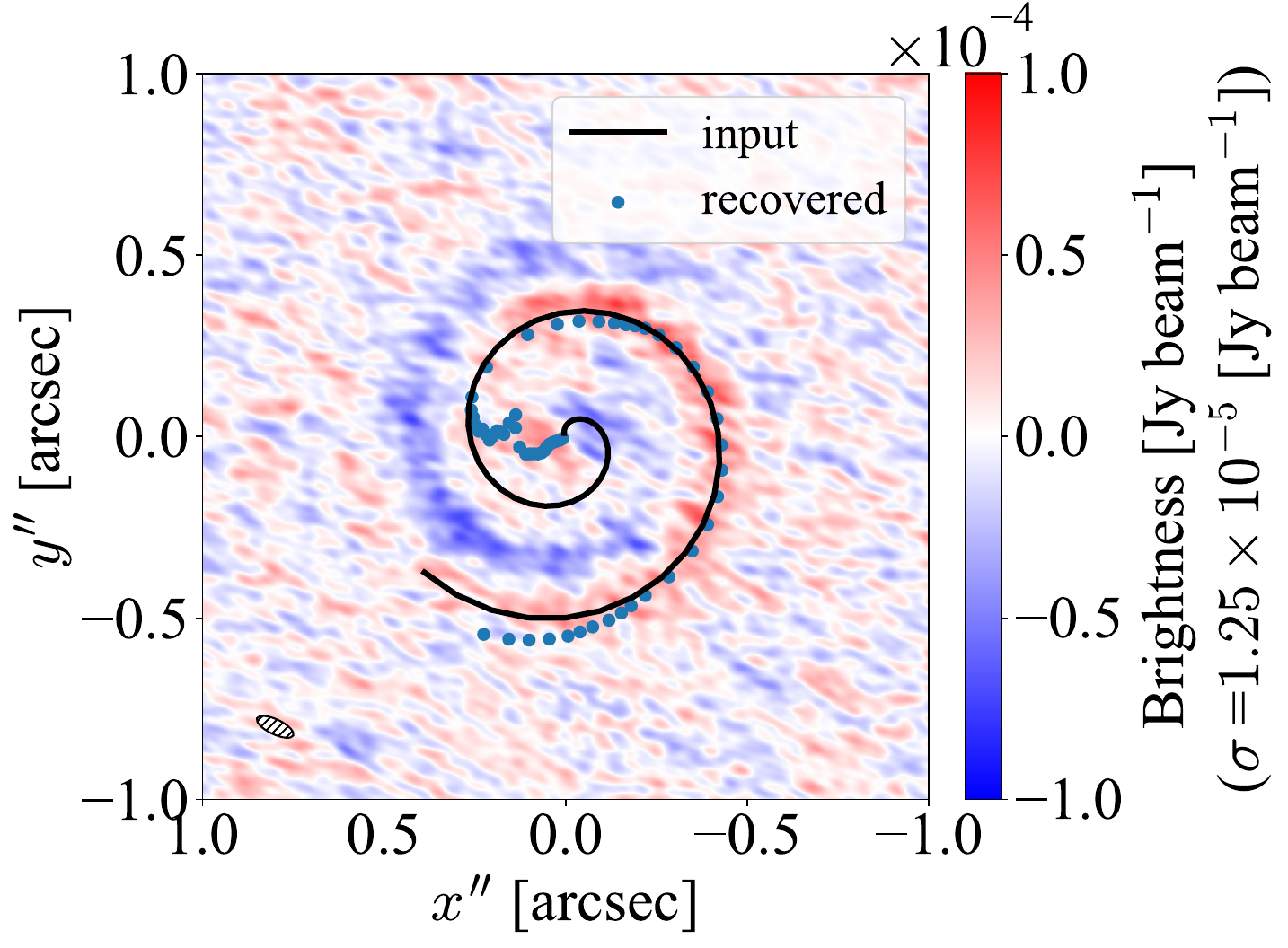}
\includegraphics[width=0.45\linewidth]{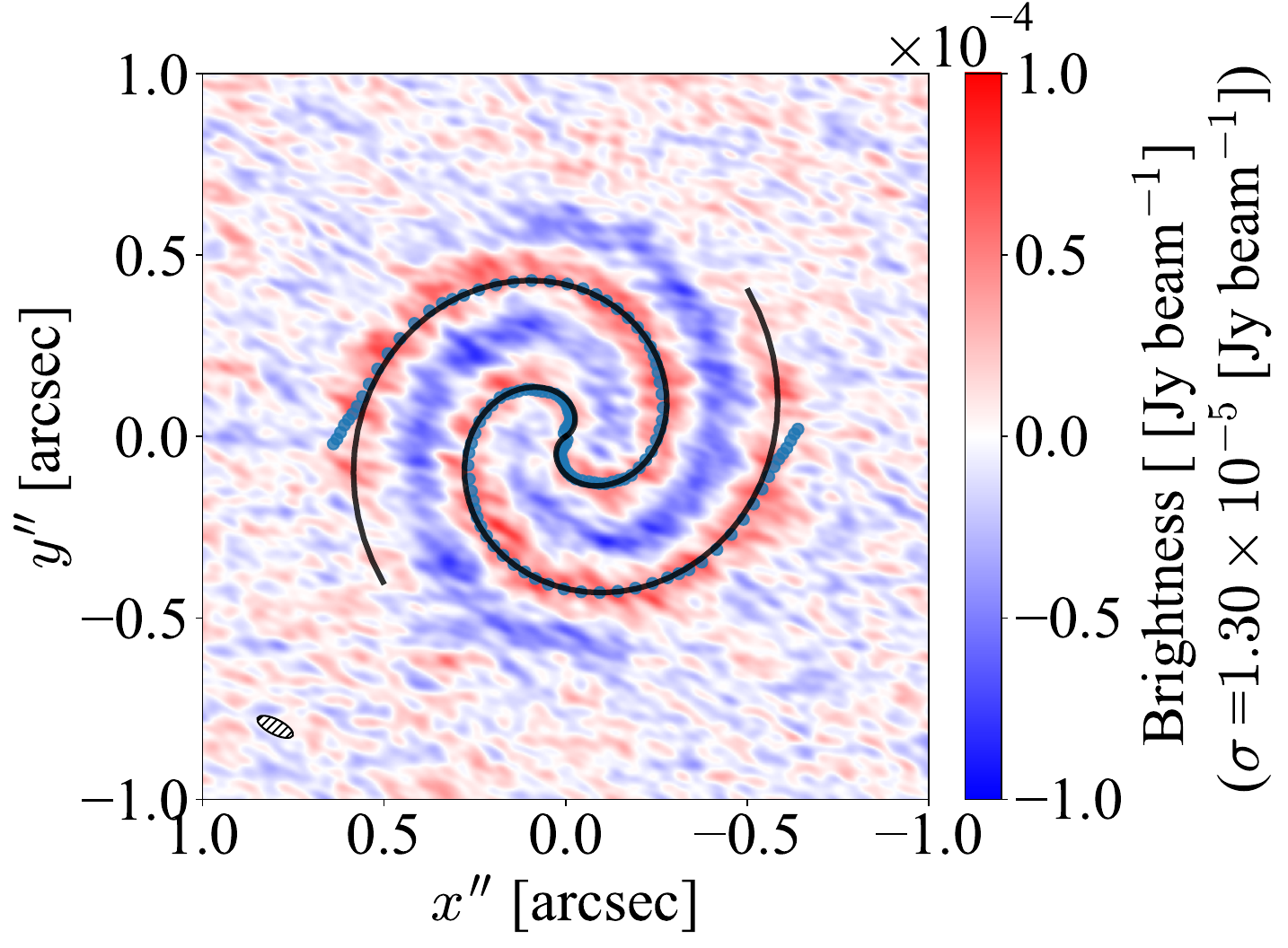}
\includegraphics[width=0.45\linewidth]{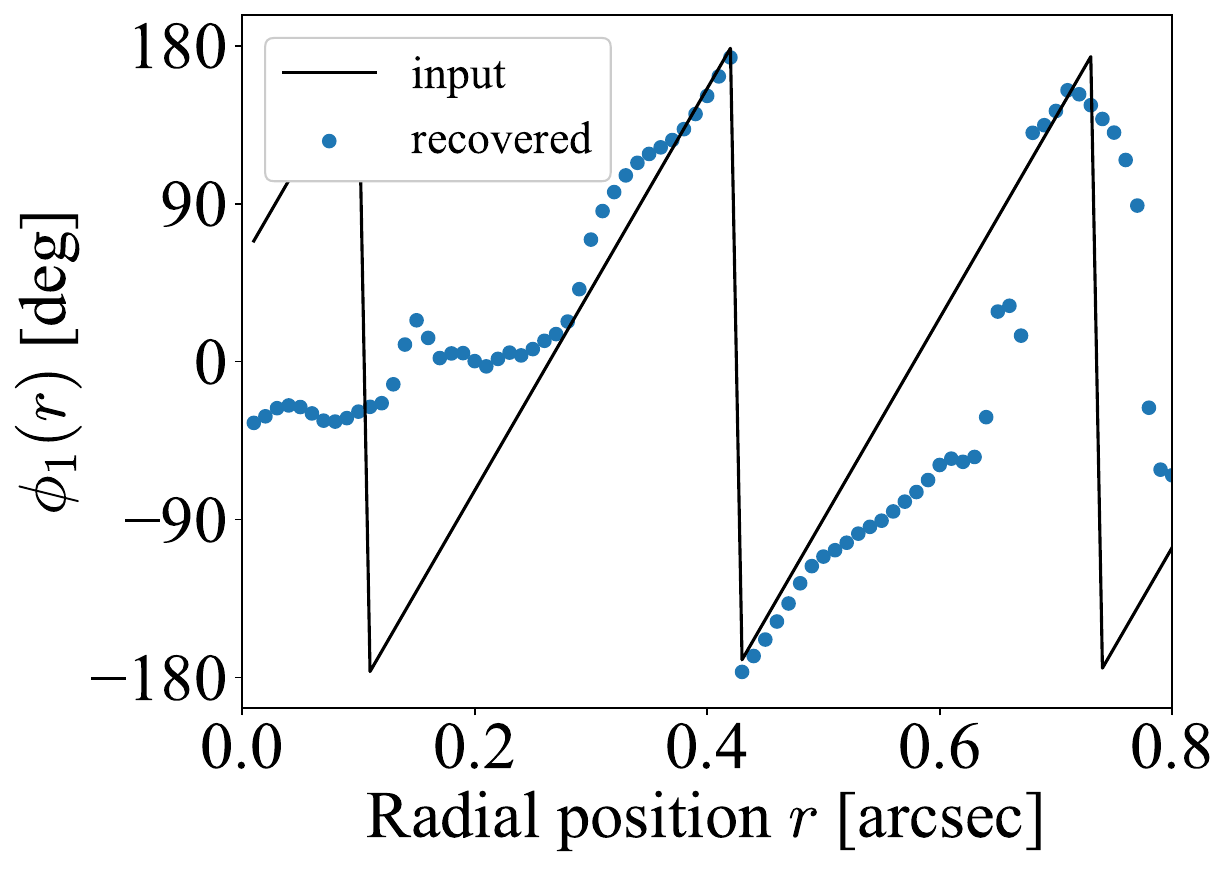}
\includegraphics[width=0.45\linewidth]{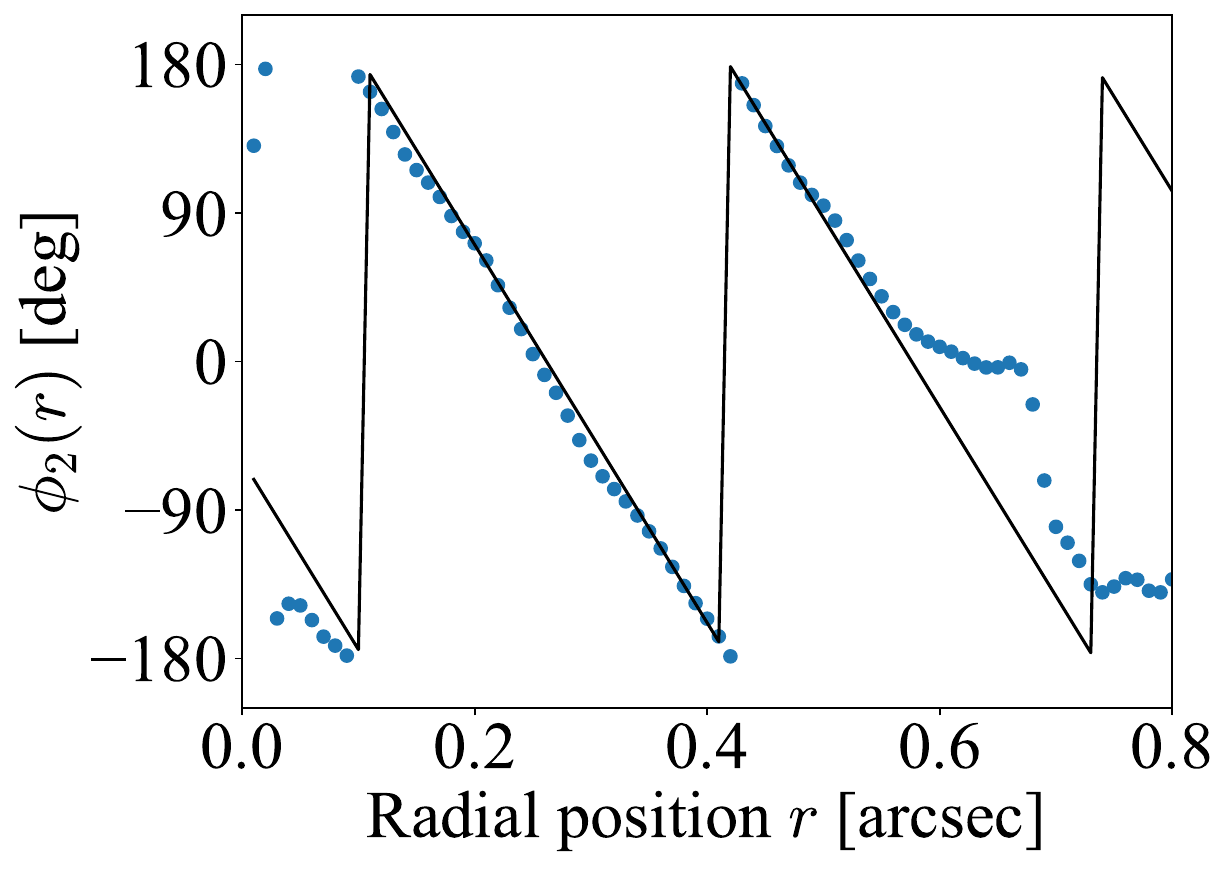}
\includegraphics[width=0.45\linewidth]{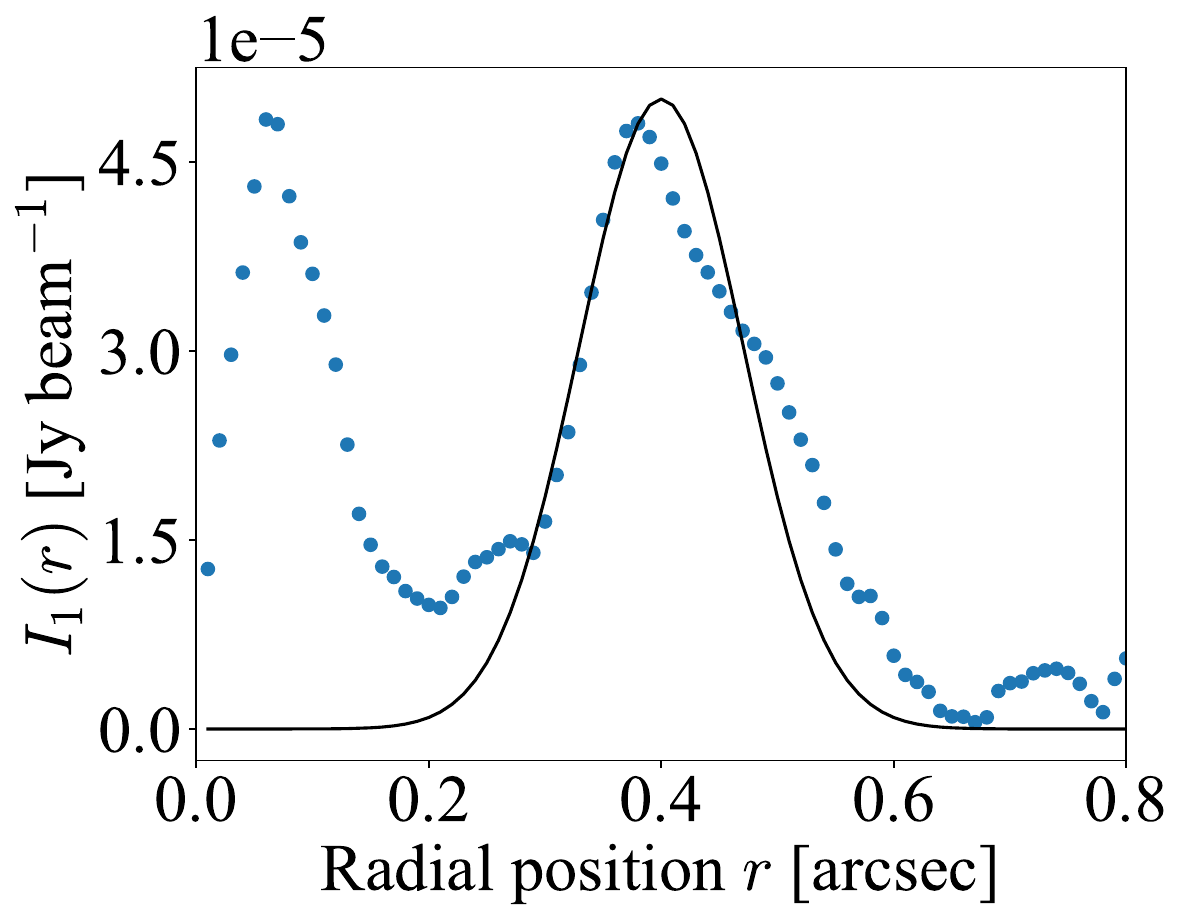}
\includegraphics[width=0.45\linewidth]{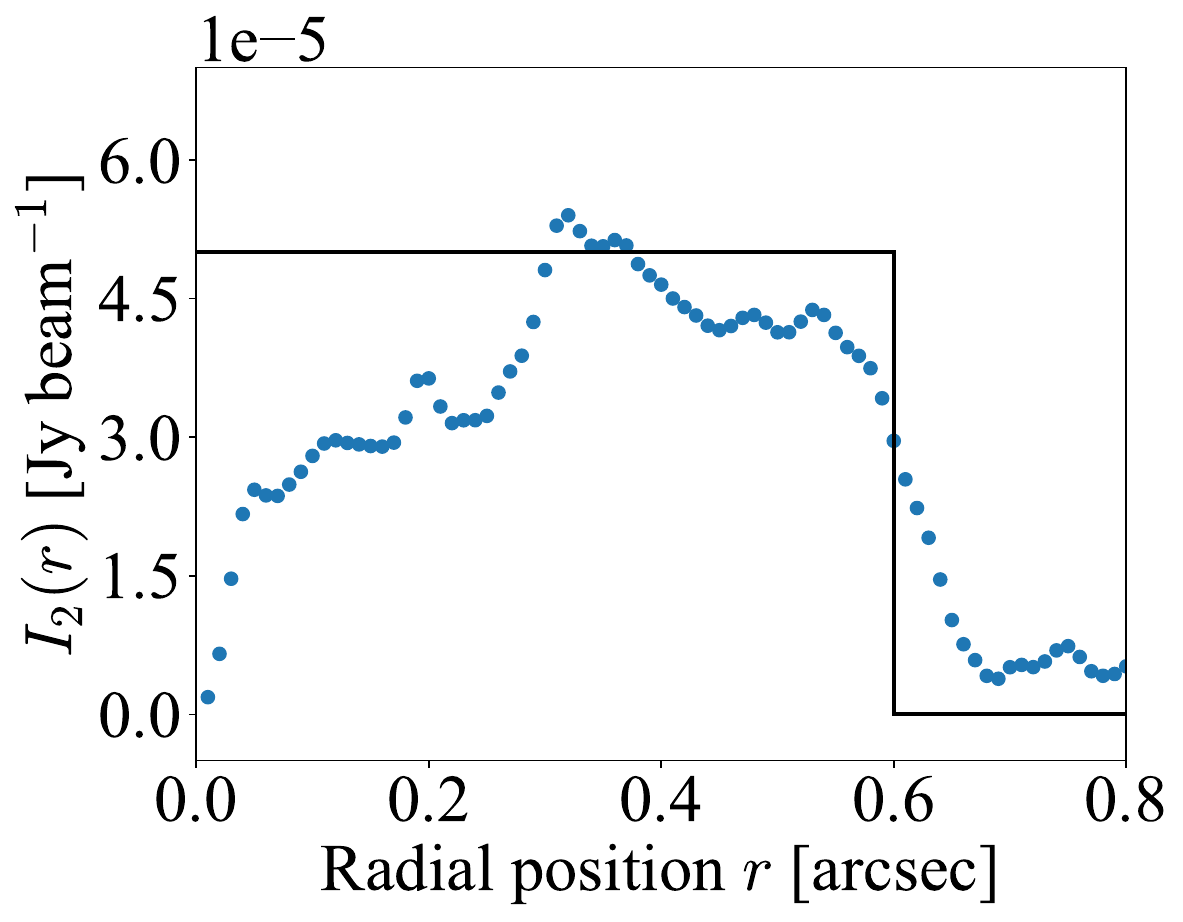}
\end{center}
\caption{Phases and amplitudes for injected and recovered odd- (left) and even- (right) symmetric spirals  in Sec \ref{sec:sim_non}. (upper panels) The residual images in deprojected frame overplotted with the locations of injected (black) and recovered phases (blue). (middle panels) Input and recovered phases in the radial direction. (lower panels) Input and recovered amplitudes of spirals. } 
\label{fig:odd_recover}
\end{figure*}

The left panels in Fig. \ref{fig:odd_recover} compare the residual image with the injected odd-symmetric spiral. The phase for the spiral was reasonably recovered in the residual image. The upper right and lower panels show the recovered radial phases $\phi_{1}(r)$ and the amplitudes $I_{1}(r)$. The amplitudes and phases were reasonably recovered in the range of $0.3\arcsec < r < 0.5\arcsec$, where the spiral amplitude was large. However, the amplitudes were clearly overestimated at inner radii $r<0.3\arcsec$. This overestimation was due to the biased central position, and is consistent with the failure of the recovery of the spiral at the innermost part in Fig. \ref{fig:spiral_recovered}. 

 The right panels in Fig. \ref{fig:odd_recover} show the comparison in the case of the even-symmetric spiral. The phases were reasonably recovered in most of the radial range; however, the deviations were large at the innermost part $r<0.1\arcsec$. The amplitudes were well recovered at $0.3\arcsec < r < 0.6\arcsec$, whereas they were underestimated in the inner regions. The deviations in the phases and amplitudes originated from the biases in the estimation of ($\cos i$, PA), which are associated with $m=2$ feature. 

\begin{figure*}
\begin{center}
\includegraphics[width=0.97\linewidth]{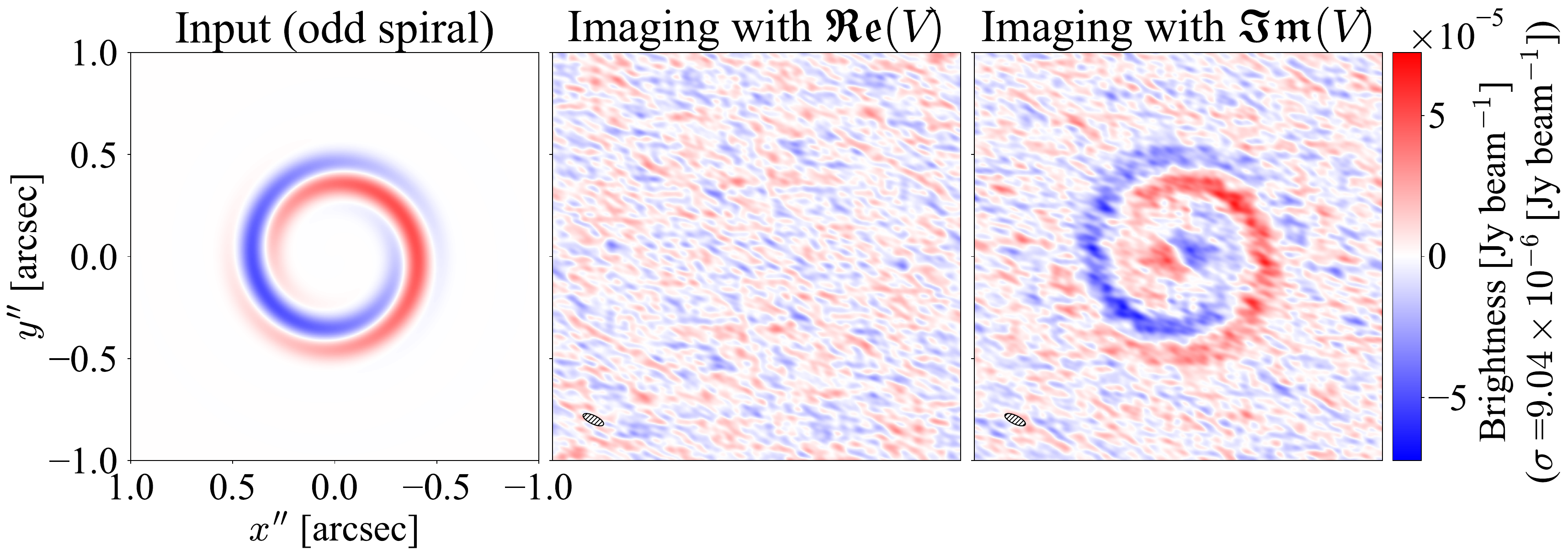}
\includegraphics[width=0.97\linewidth]{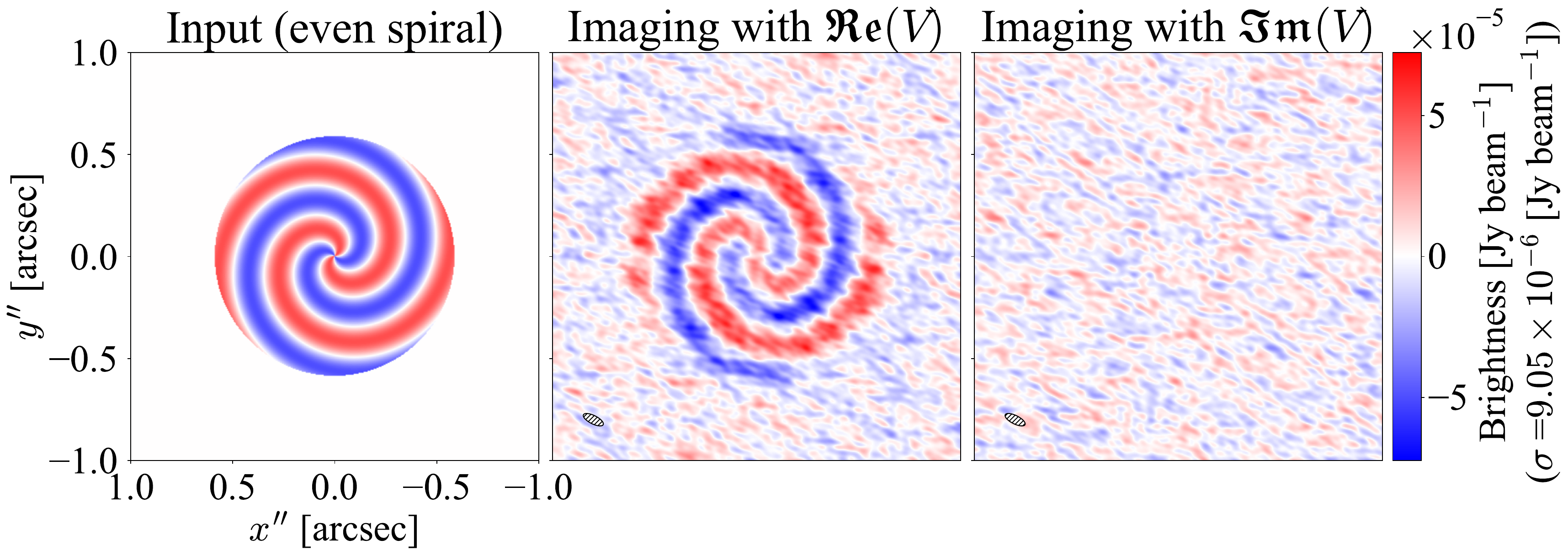}
\includegraphics[width=0.97\linewidth]{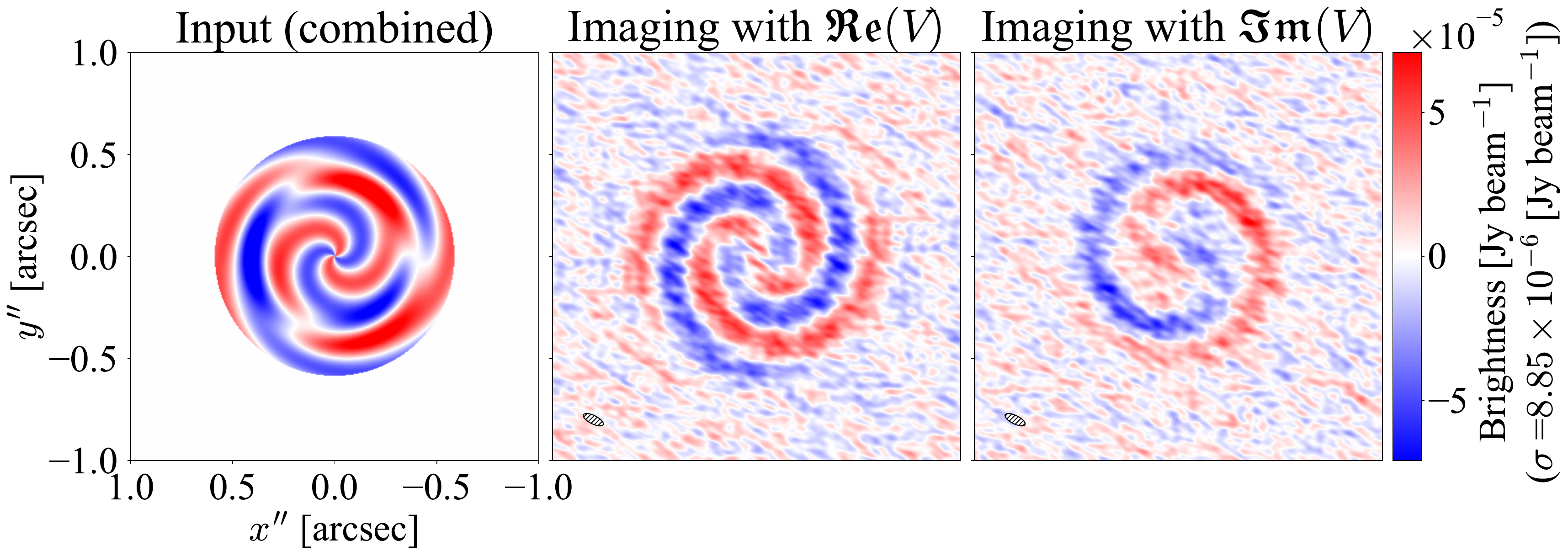}
\end{center}
\caption{Demonstration of extraction of odd- and even-symmetric spirals with real and imaginary parts of the visibilities; (top rows) Residual images for even-symmetric spiral in deprojected frames (middle rows) Residual images for odd-symmetric spiral (bottom rows) Residual images for combined spirals (left columns) Injected spirals (middle column) Image produced with only the real part of the data (right column) Image produced with only the imaginary part of the data. }
\label{fig:real_imag}
\end{figure*}

Finally, Fig. \ref{fig:real_imag} shows the residual images obtained with either real or imaginary parts of images. As evident, the imaging with only the real part successfully extracted the even-symmetric spiral, whereas that with only the imaginary part extracted the odd-symmetric spiral. This is consistent with the discussion in Sec \ref{sec:real_imag}. Notably, we successfully extracted odd- and even-symmetric spirals for the case of the combined image. Thus, in a realistic situation, where odd- and even-symmetric components are mixed, the imaging with either the real or imaginary part is indeed useful for separating them. In addition, the noise level decreased by a factor of $\sim \sqrt{2}$, or the signal-to-noise ratio increased by a factor of $\sim \sqrt{2}$ because of the assumed symmetry $I(x, y)=I(-x, -y)$ or $I(x, y)=-I(-x,-y)$. Thus, this method would be helpful in the search for faint signals.  

%-----------------------------Figure Start---------------------------
%-----------------------------Figure End------------------------------

%-----------------------------Figure Start---------------------------
%-----------------------------Figure End------------------------------

%-----------------------------Figure Start---------------------------
%-----------------------------Figure End------------------------------

\subsection{Axisymmetric structure $+$ circumplanetary disc emission \label{sec:sim_cpd}}
We injected a point source to the simulated data assuming circumplanetary disc emission and tested the recovering ability. The flux for the point source was assumed to be $0.1$ mJy, which is approximately 10 times more significant than the noise level but 10 times lower than the ambient emission at the inner disc. The planetary location was set to be at 0.5~$\arcsec$ in the deprojected frame $(x'', y'') = (0.5\arcsec,0)$, such that the source resided in a large gap of a disc. We considered the same brightness profile and geometric configuration as the simulated case for AS209 in Sec \ref{sec:sim} and computed the visibilities and added noises to the data assuming the same observational setup. 

\begin{figure*}
\begin{center}
\includegraphics[width=0.32\linewidth]{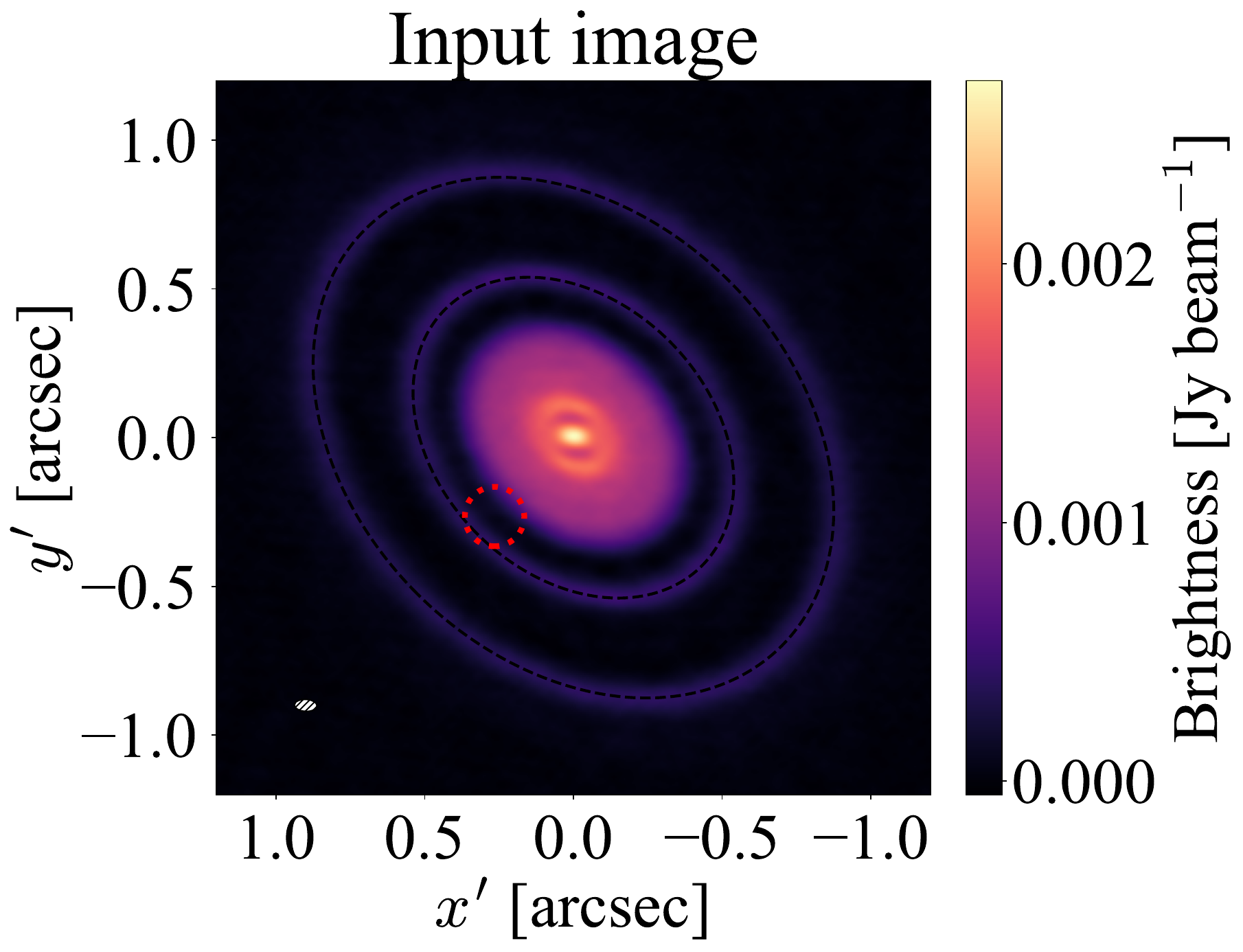}
\includegraphics[width=0.32\linewidth]{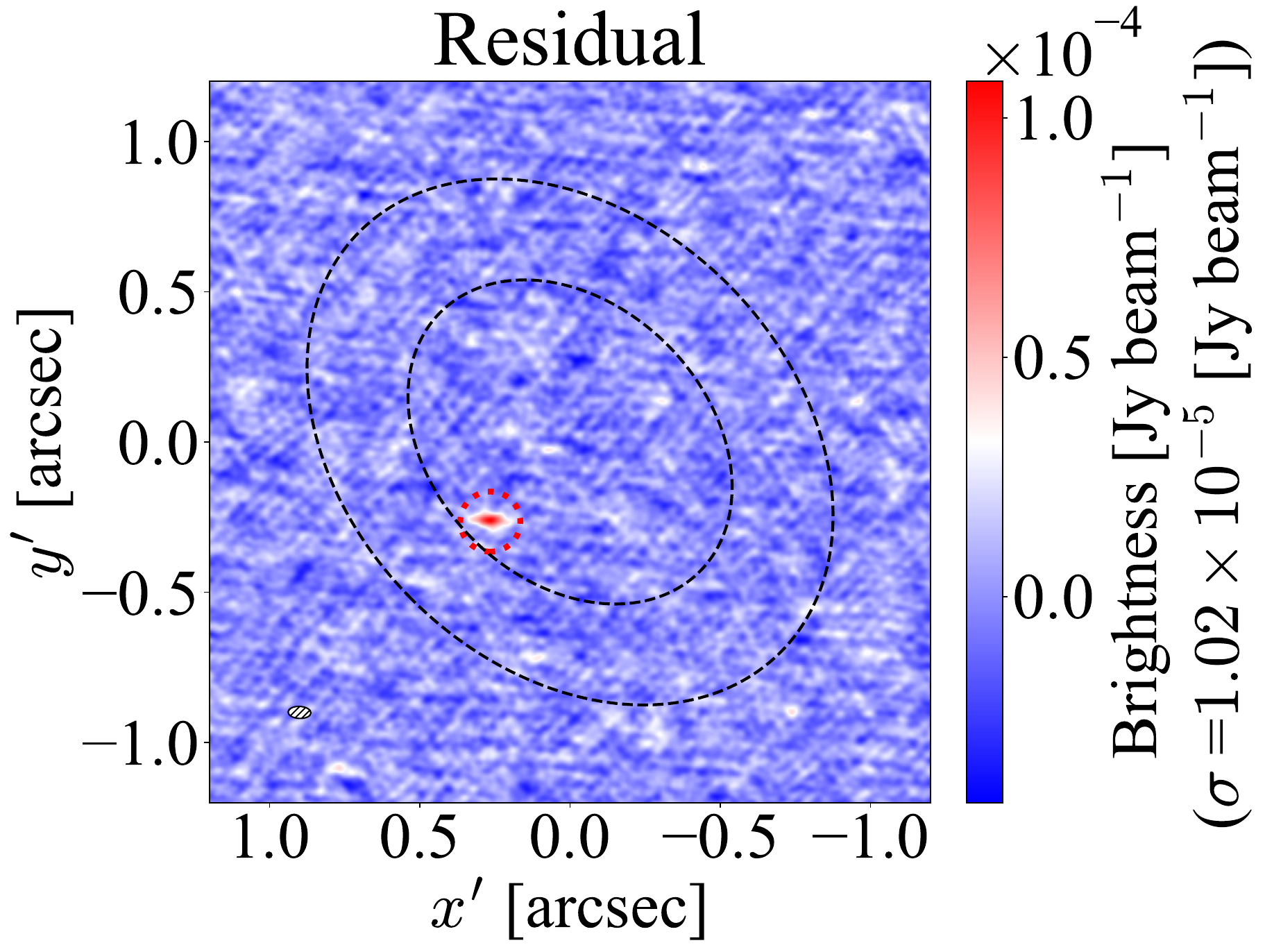}
\includegraphics[width=0.32\linewidth]{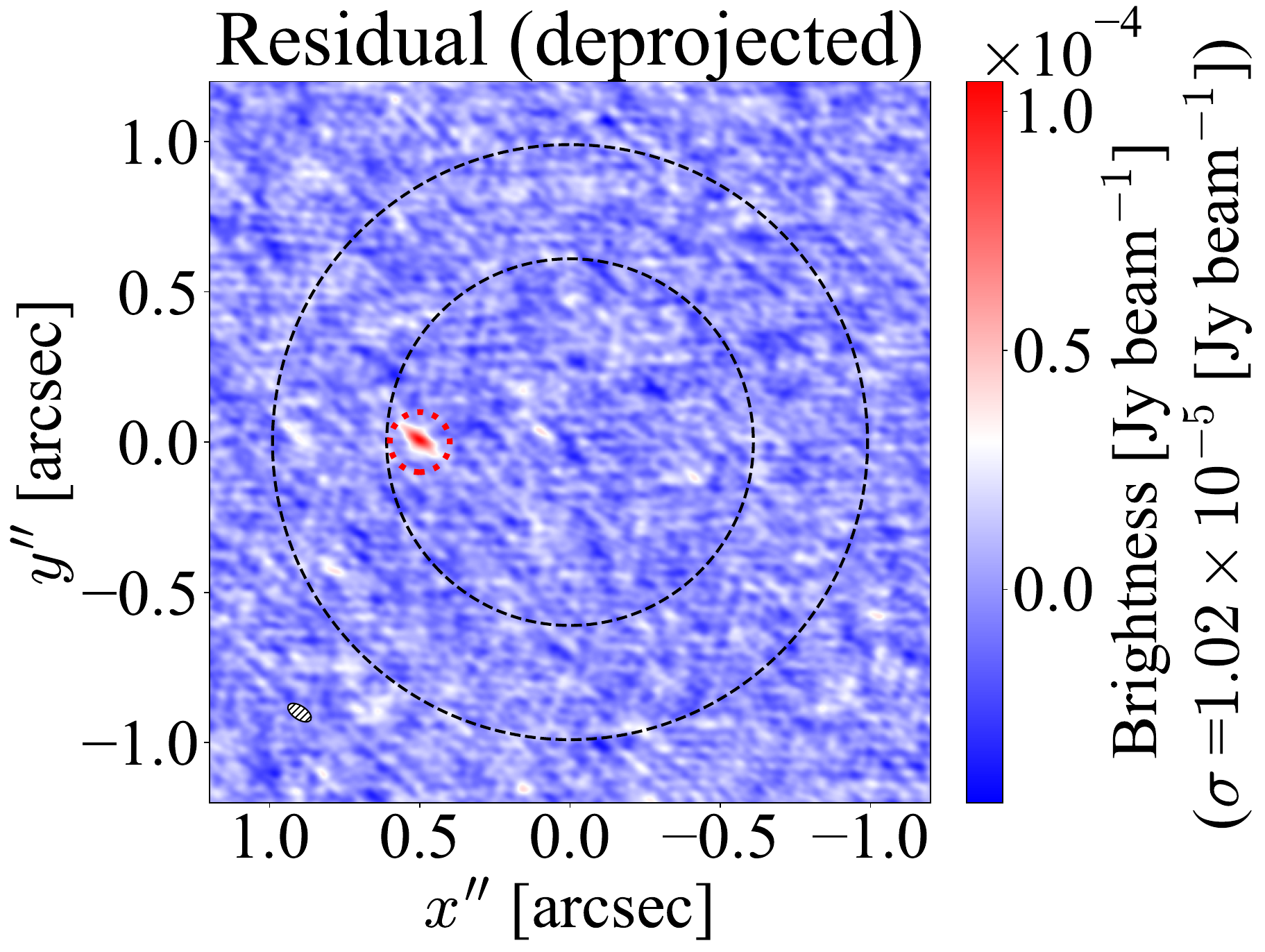}
\end{center}
\caption{Residual images for the simulated data that include a simulated circumplanetary disc emission, whose position is indicated by a red dashed circle. The central and right panels show the maps in an observed and deprojected frames, respectively. The synthesized beam sizes before and after deprojection are 
(0.076\arcsec , 0.040\arcsec ) and (0.089\arcsec , 0.045\arcsec ) are shown at the bottom left, respectively.} 
\label{fig:cpd_recover}
\end{figure*}

Using our method, we similarly recovered the brightness profile and geometric parameters from the simulated data. With the MAP estimate of geometric parameters, we randomly drew one brightness profile and subtracted the model from the simulated visibilities. Fig. \ref{fig:cpd_recover} shows the residual images, which are generated from the residual visibilities. The position of the injected point source was reasonably recovered, and the flux density was well recovered as well. 

Moreover, the central position of the disc was slightly biased, suggesting that the point source might introduce another bias. To check this possibility, we injected the brighter sources with fluxes of 1 and 2 mJy, and attempted to estimate the central positions of the discs by repeating the same analyses. Resultantly, the central positions were estimated to be  $(\Delta x_{\rm cen}, \Delta y_{\rm cen}) = (0.58^{+0.07}_{-0.07} {\rm mas}, -0.50^{+0.07}_{-0.07} {\rm mas})$ for the 1 mJy source and $(\Delta x_{\rm cen}, \Delta y_{\rm cen}) = (0.97^{+0.07}_{-0.07} {\rm mas}, -0.82^{+0.07}_{-0.07} {\rm mas})$ for the 2 mJy source. However, the positions of the flux centre with respect to the disc centre were calculated as $(0.9 {\rm mas}, -0.9 {\rm mas})$ and $(1.8 {\rm mas}, -1.8 {\rm mas})$  for the 1 and 2 mJy sources, respectively, and they were comparable to the biases in the estimation. Thus, even with our method, the central positions of the disc can be biased by approximately half of the positional difference between the disc and flux centres. This holds true not only for the point source but also for localized emission; for example, a crescent-shaped emission. In the data analysis, in case of bright localized emission, it is recommended that such an additional emission be modelled or removed because it affects the residual image creation process. Specifically, in the case of the point source, we can directly include the point source model in the axisymmetric model. However, if the emission is more complicated, we can model them on the image plane and remove their visibilities from the observations, as presented in \cite{andrews2021}.

%-----------------------------Figure Start---------------------------
%-----------------------------Figure End------------------------------

\section{Application to real data: Elias 20 and AS 209} \label{sec:apptoreal1}
We applied the proposed method to the real DSHARP data of two discs, Elias 20 and AS 209, to demonstrate its feasibility. AS 209 is one of most structured discs in the DSHARP sample, and Elias 20 shows a non-axisymmetric feature in the residual map as shown later. More systematic studies on the DSHARP discs will be presented in future studies. 

We downloaded the measurement sets of the DSHARP data for the two discs, and applied time and spectral averaging in the same manner as that of Sec \ref{sec:in_re_disc}. We manually reduced the recorded weights by 3.50 and 3.44 for Elias 20 and AS 209, respectively, to match the observations. The data were then binned with a log grid with $N_{\rm bin}=500$, and they were analyzed with the current method. After sampling the posterior distribution for the parameters, we created the residual images following the method presented in Sec \ref{sec:make_image}. To create residual images, we used two different geometries: one is obtained from our method, and the other obtained from \cite{huang_ring_2018}, who estimated the geometric parameters through ellipse fitting of annular substructures on image planes. 

Table \ref{table:disc_as209_gwlup} presents the derived geometries for Elias 20 and AS 209. For a reference, we also show the geometric parameters from \cite{huang_ring_2018}. The length scale parameters $\gamma$ for Elias 20 and AS 209 are comparable to their beam sizes $(0.076\arcsec, 0.040\arcsec)$ and $(0.048\arcsec, 0.028\arcsec)$, respectively. This result is consistent with the discussion in Sec \ref{sec:in_re_disc} and Appendix \ref{sec:scaling}; $\gamma$ was  determined by the combination of the intrinsic brightness profile and the UV-coverage. 

The derived geometries from our method were mostly consistent with the previous estimates from \cite{huang_ring_2018}, although there are slight or moderate discrepancies. The central positional estimates for AS209 are offset by approximately $1$ mas for both directions. Similarly, in the case of Elias 20, the positions also show a difference of 1-2 mas, and notably, there is a 0.07 difference in $\cos i$.

\begin{figure*}
\begin{center}
\includegraphics[width=0.45\linewidth]{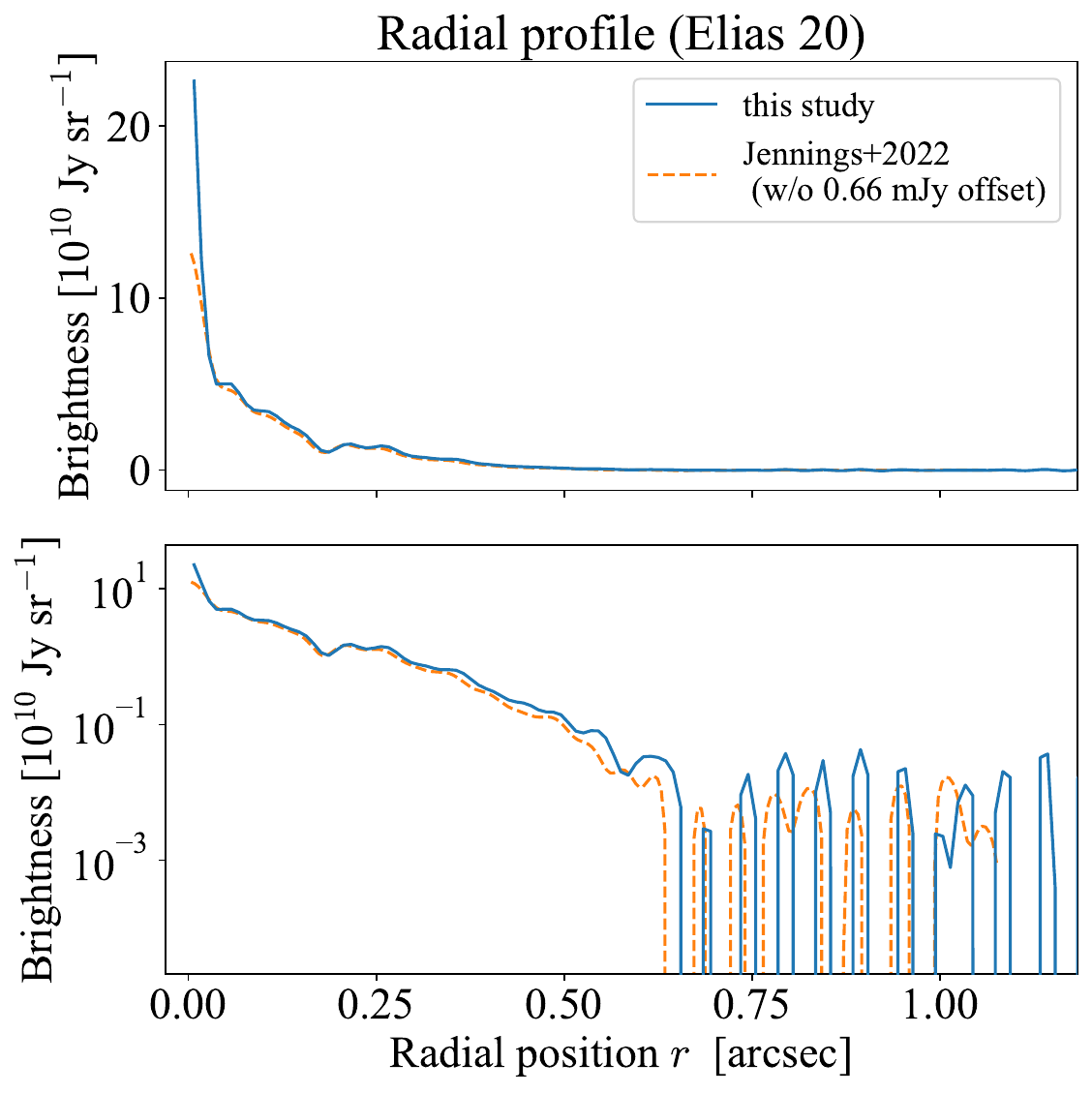}
\includegraphics[width=0.45\linewidth]{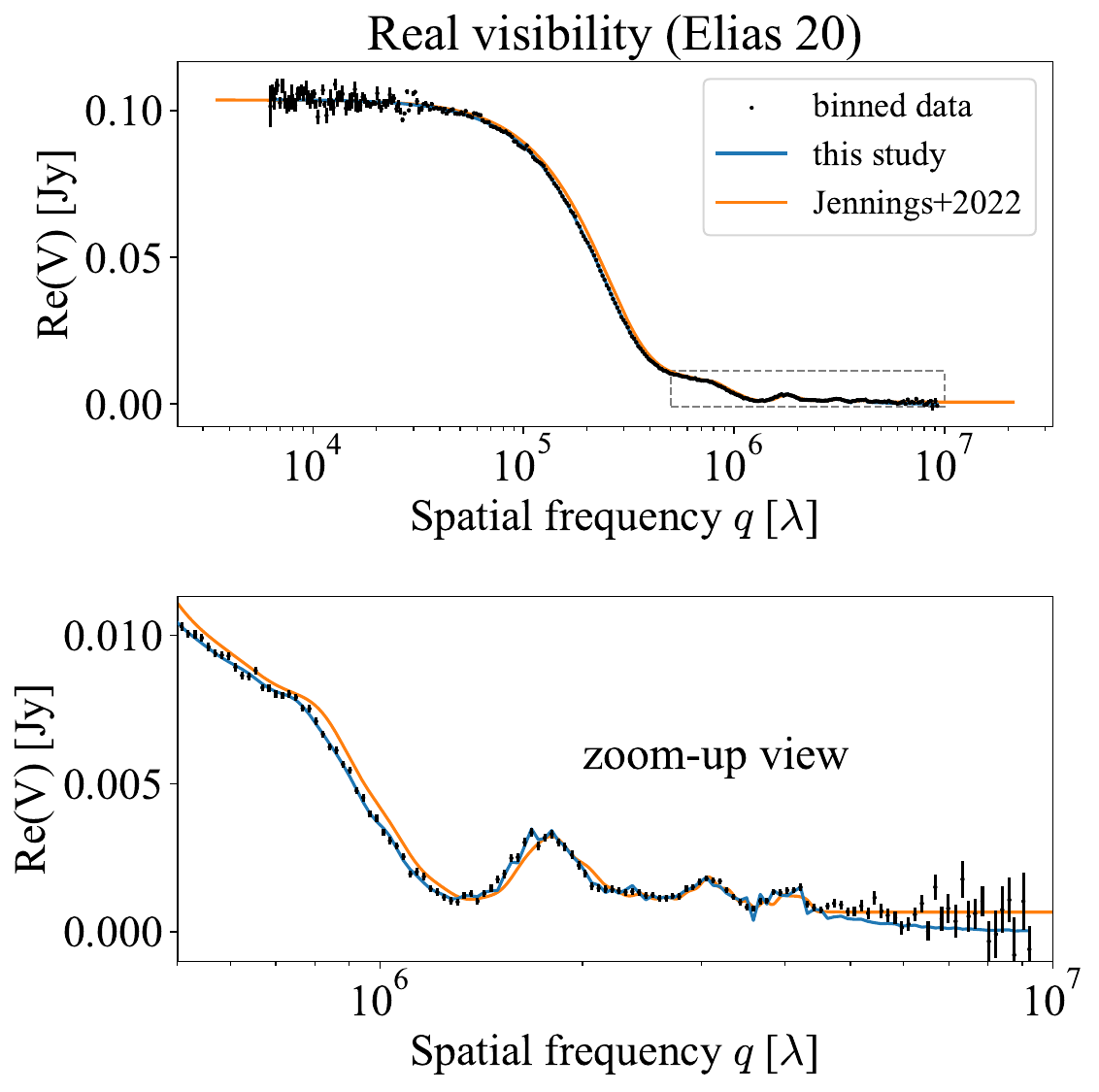}
\includegraphics[width=0.98\linewidth]{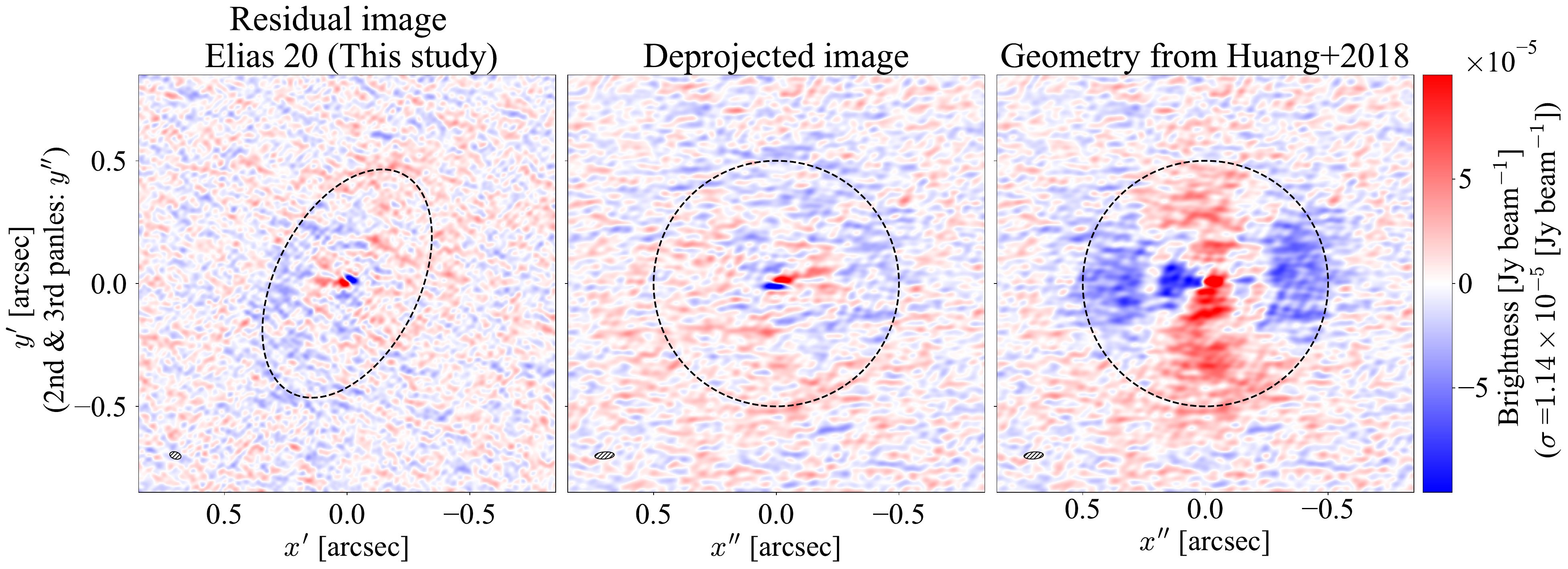}
\end{center}
\caption{Analysis of DSHARP data of Elias 20. (upper-left panels) Brightness profiles recovered by this study and \protect\cite{jennings2022}. The profiles are shown in linear and log scales in the upper and lower panels, respectively. (upper-right panels) Model and observed real visibilities and deprojected spatial frequencies. The model by \protect \cite{jennings2022} is also shown for comparison. (lower panels) Residual image made with the geometry from our method in the observational frame, that in the deprojected frame, and the image made with that of \protect \cite{huang_ring_2018} in the deprojected frame. In the residual images, the reference lines at $r=0.5\arcsec$ are also shown. As treference, the ellipse and circles are shown. The synthesized beam sizes before and after deprojection are $(0.048\arcsec, 0.028\arcsec)$ and $(0.079\arcsec, 0.029\arcsec)$ are shown at the bottom left, respectively.}
\label{fig:elias20}
\end{figure*}

\begin{figure*}
\begin{center}
\includegraphics[width=0.45\linewidth]{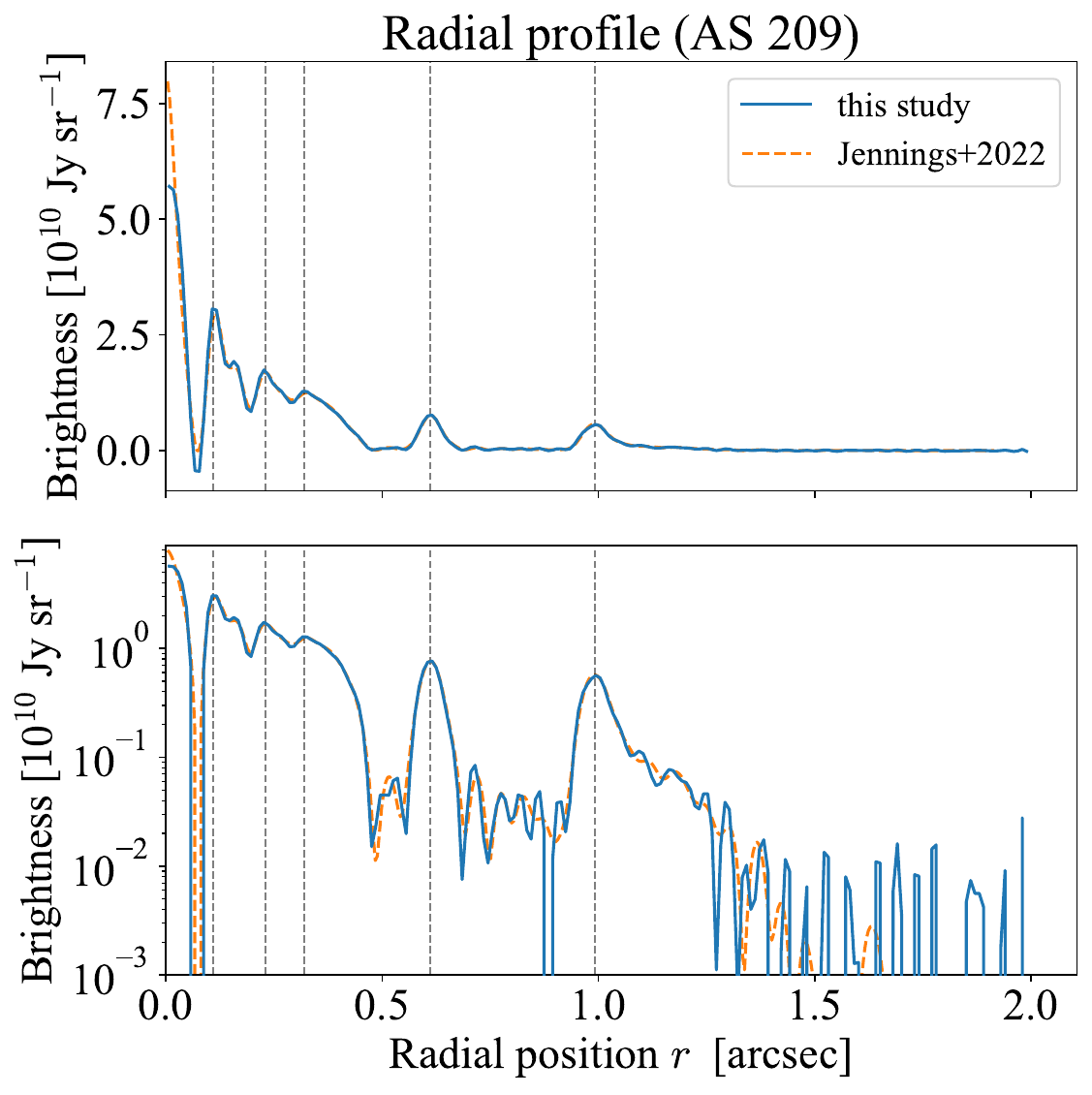}
\includegraphics[width=0.45\linewidth]{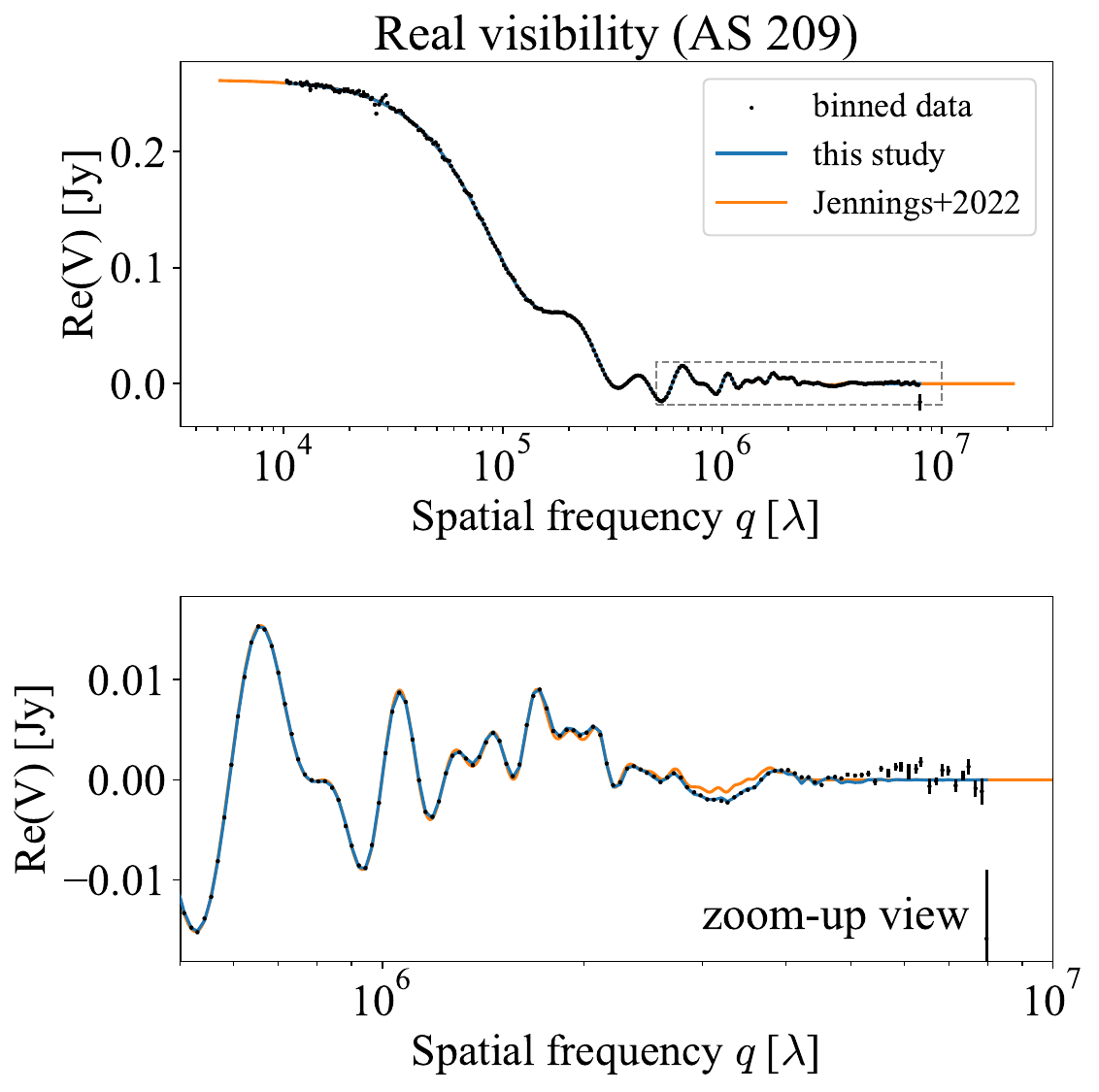}
\includegraphics[width=0.98\linewidth]{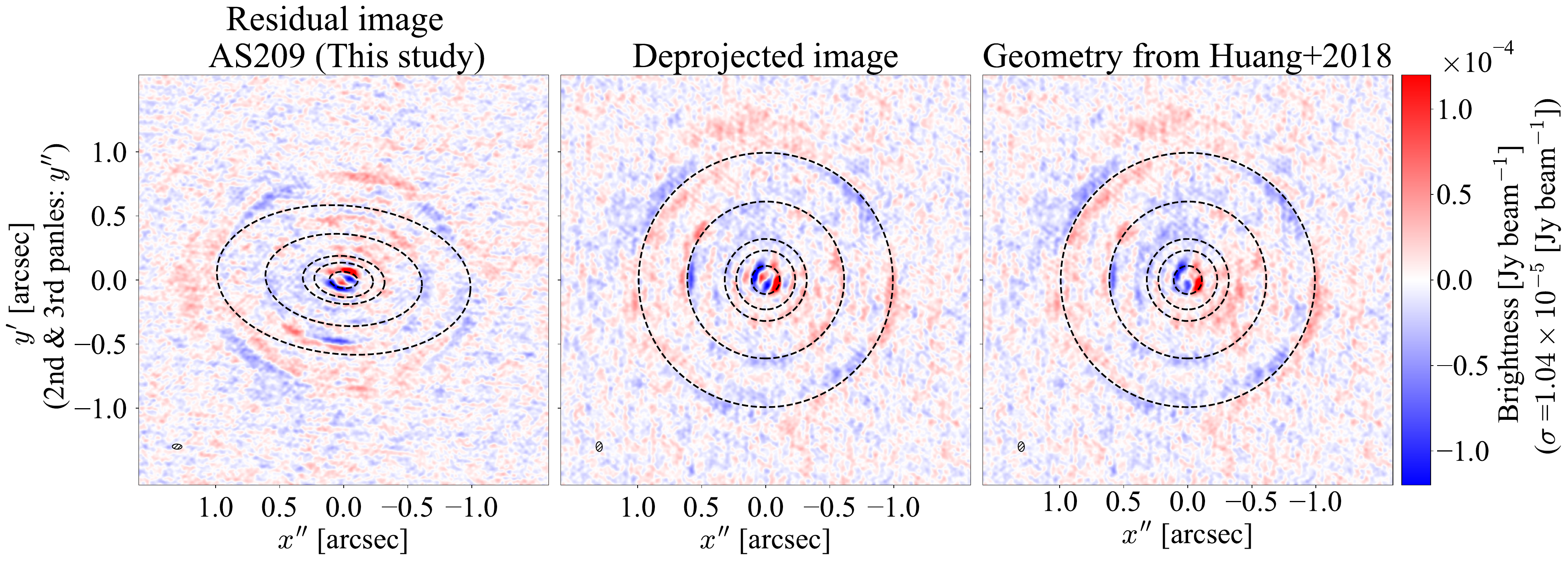}
\end{center}
\caption{Analysis of DSHARP data of AS 209. The format of the figure is the same as that of Fig. \ref{fig:elias20}. The reference lines corresponding to gaps and rings are shown in brightness profiles and the residual images. The synthesized beam sizes before and after deprojection are $(0.076\arcsec, 0.040\arcsec)$ and $(0.076\arcsec, 0.049\arcsec)$, respectively.} 
\label{fig:as209}
\end{figure*}

These small discrepancies are important for creating the residual images. Figs. \ref{fig:elias20} and \ref{fig:as209} show the brightness profiles, visibilities, and residual images. For comparison, we also show the brightness profiles and model visibilities obtained by \cite{jennings2022}, who systematically analysed the DSHARP data. Here, they used {\tt frank} \citep{jennings2020}, which reconstructs the radial brightness profile by fitting the real part of the deprojected visibilities. Note that \cite{jennings2022} adopted the non-negativity condition on the brightness profile in case of AS 209, and assumed additional point-source emission with flux of 0.66 mJy at the disc centre to suppress the artificial oscillating features for Elias 20 (detailed discussion on point-source correction is presented in Appendix A in their paper). Their model visibilities, as shown in Fig. \ref{fig:elias20}, for Elias 20 are thus converged to 0.66 mJy, which corresponds to the flux of this point-source emission. 

The brightness profiles and model visibilities of the proposed method and that of {\tt frank} were mostly consistent; however, there existed slight or moderate differences. 

In the case of Elias 20, the model visibilities are horizontably offset, and this is due to the difference in the $\cos i$. In addition, the models using the proposed method gradually converged to zero at high spatial frequencies, whereas those from {\tt frank} sharply converged to the flux of the point source around $5$ M$\lambda$. The locations of the tipping points for the convergence were determined by $\gamma$ in our method or hyperparameters in {\tt frank}, and the differences in the way of convergence were due to the different choices for the regularization. 

The notable difference was observed in the peak brightness values near $r=0\arcsec$.The peak brightness is generally hard to estimate accurately because of the small flux at the small radii. This is mainly due to the point-source correction with 0.66 mJy adopted in \cite{jennings2022}, which was however not included in their brightness profile. The innermost brightness $I(r_{1})$ in our model corresponding to the flux 0.66 mJy  is $\sim 6\times 10^{10}$Jy sr$^{-1}$, which can largely explain the observed difference of about $10 \times 10^{10}$ Jy sr$^{-1}$ between this study and \cite{jennings2022}.  Another potential explanation could be the difference in the strength of the regularization for smoothing, although identifying the superior model remains challenging at this stage. 

In the case of AS 209, the brightness profiles and model visibilities were mostly consistent for our method and that of {\tt frank}. The peak brightness values, on the other hand, showed the difference of about $2.5 \times 10^{10}$ Jy sr$^{-1}$, which is yet smaller than that of Elias 20. The discrepancy potentially arises from the difference in regularization strengths or the non-negative condition adopted in \cite{jennings2020}. Overall, our method exhibited the same capability at recovering the brightness profiles as that of {\tt frank}. 

The estimation of the residual images was also improved using our updated geometry. In case of Elias 20, the residual image derived with the geometry from \cite{huang_ring_2018} exhibits the $m=2$ pattern, which is mainly attributed to the inclination offset, as shown in Figs. \ref{fig:wrong_geo} and \ref{fig:wrong_geo_waoph6}. This coherent pattern, also seen in \cite{jennings2020}, mostly disappears in our update image, suggesting that our method suppresses the artificial pattern owing to wrong geometric parameters. In case of AS 209, the previous literature identified significant residual patterns \citep{guzman2018, jennings2020}. The residual image with the updated geometry is less noisy than the previous estimate; however, structured patterns were still present. We thus conclude that the residual pattern for AS 209 cannot be removed by solely optimizing the geometric parameters for a geometrically-thin axisymmetric model. 

There can be multiple possible explanations for the residual pattern in AS 209. It may be due to the real non-axisymmetry in the physical parameters, or because of the geometric effect. The individual rings may have different geometric parameters, that is, misalignment or positional offsets, where the latter case was reported for HL Tau \citep{2015ApJ...808L...3A}. Further, the vertical structure may also render the residual pattern complicated, although the observed residual image is not perfectly consistent with this hypothesis, which predicts the symmetric pattern with respect to the $x$ axis (minor axis). To unveil the origin of the residual pattern for AS 209, we require a more complicated model with multiple geometric parameters or vertical structures; however this is beyond the scope of this paper. 
\bgroup
\def\arraystretch{1.2}

\begin{center}  
\begin{table*} 
\caption{Hyperparameters for Gaussian Process and geometric parameters of Elias 20, AS 209, and PDS 70. Geometric parameters assumed in previous studies are also shown for comparison. }  \begin{tabular}{ccccccc} 
\hline 
Name& $\gamma$ ["]& $\log_{10} \alpha$& $\Delta x_{\rm cen}$ [mas]& $\Delta y_{\rm cen}$ [mas] & $\cos i$& PA [deg]\\ 
\hline 
Elias 20 (this study) & $0.038 ^{+0.001}_{-0.001}$& $0.800 ^{+0.161}_{-0.150}$& $-53.919 ^{+0.064}_{-0.065}$& $-488.849 ^{+0.071}_{-0.073}$& $0.584 ^{+0.001}_{-0.001}$& $153.930 ^{+0.060}_{-0.060}$ \\  
Elias 20 (\cite{huang_ring_2018}) & & &-54.5 & -491.0 & 0.656 & 153.2   \\ \hline 
AS 209 (this study) & $0.028 ^{+0.001}_{-0.001}$& $-1.202 ^{+0.092}_{-0.087}$& $0.767 ^{+0.068}_{-0.066}$& $-1.225 ^{+0.048}_{-0.050}$& $0.821 ^{<0.001}_{>-0.001}$& $85.766 ^{+0.036}_{-0.038}$ \\ 
AS 209 (\cite{huang_ring_2018}) & ... & ... & 1.9 &-2.5 & 0.819 & 85.8   \\ \hline 
PDS 70 (this study) & $0.064^{+0.003}_{-0.003}$& $-2.396^{+0.112}_{-0.110}$& $10.827^{+0.079}_{-0.078}$& $14.953^{+0.101}_{-0.105}$& $0.6427^{+0.0002}_{-0.0002}$& $160.003^{+0.020}_{-0.022}$  \\
PDS 70 \cite{benisty2021} & ... & ... & 12 & 15 & 0.6494 & 161  \\ 
\hline  \end{tabular}  
\label{table:disc_as209_gwlup} 
\end{table*} \end{center} 
\egroup

%-----------------------------Figure Start---------------------------
%-----------------------------Figure End------------------------------

%-----------------------------Figure Start---------------------------
%-----------------------------Figure End------------------------------

\section{Application to real data: PDS 70}\label{sec:pds70}
\subsection{Analysis with axisymmetric disc model} \label{sec:current_ana_pds70}

As a practical application to recovering a circumplanetary emission, we applied our method to the data of PDS 70 in \cite{benisty2021}. The same data were already analyzed by {\tt frank} in \cite{benisty2021}. We used the combined dataset, including long, medium, and short baseline data, for continuum emission in Band 7 as used in \cite{benisty2021}. The data were then averaged in a similar manner to Sec \ref{sec:sim} to reduce the data size. 

In the continuum emission, there is a notable crescent feature in the North-West direction. \\cite{benisty2021} removed the asymmetric feature by following the method described in \cite{andrews2021}. Specifically, using the CLEAN model image, they defined the asymmetry model by isolating the emission of the crescent feature, and subtracted the mean radial profile outside the area from the model, leaving only the asymmetric contribution. The constructed model was then Fourier transformed, and the model visibilities are subtracted from the observed visibilities, which were analyzed using {\tt frank}. As demonstrated in \cite{andrews2021}, the method is effective for extracting strong asymmetric features that hinder the detection of weak signals, such as emission from CPD (see Appendix B of their paper for details).

In this paper, we applied our method to the original data without any such prior subtraction, aiming to minimize the manual adjustment. A major concern with this simpler approach was that the localized crescent feature might bias the estimates of the geometric parameters, especially the position estimate $(\Delta x_{\rm cen}, \Delta y_{\rm cen})$, similar to that in Sec. \ref{sec:sim_cpd}. However, as will be shown later, the derived parameters agree well with those estimated by \cite{benisty2021}, with slight differences, such as $1 \;{\rm mas}$ in $\Delta x_{\rm cen}$ and $1\; {\rm deg}$ in PA. The effect of the positional difference in the residual image is negligible in the current analysis (see Fig. \ref{fig:pds_ben_comp} in Appendix \ref{sec:comp_residuals}). Therefore, we simply present the result of the analysis with our method, without implementing the prior subtraction of the asymmetric features.  

We employed a logarithmically spaced grid with $N_{\rm bin}=500$ and $(x_{\rm min}, x_{\rm max}) = (10^{2} \lambda, 1.1 \times 10^{7} \lambda)$ to bin the data. We drew samples from the posterior of parameters using {\tt emcee} with 16 walkers and 10,000 samples, and we ensured the convergence of MCMC. Table \ref{table:disc_as209_gwlup} lists the parameters from this study and \cite{benisty2021} (specifically, Appendix B in their paper). The two studies yielded consistent values, while there are slight differences: approximately $1$ mas in $\Delta x$, $0.007$ in $\cos i$, and $1^{\circ}$ in PA. 

We constructed a visibility model using the MAP solution of parameters, and subtracted it from the observed visibilities.  Subsequently, the residuals were processed with CLEAN, with a threshold of nsigma$=3.5$. Note that we did not adopt JvM correction adopted in the previous literature \citep{czekala2021, benisty2021,balsalobre2023} to prevent potential exaggeration of signal-to-noise ratios in an image \citep{Casassus2022}. In the imaging process, we experimented with various robust parameters, specifically setting them to be 0.0, 0.5, 1.0, and 1.5. We found that a setting of 1.0 offers a small standard deviation image while maintaining relatively high angular resolution. For comparison, we also generated a residual image using the geometric parameters in \cite{benisty2021} while adopting the hyperparameters in the MAP solution from our modeling.

\begin{figure*}
\begin{center}
\includegraphics[width=0.48\linewidth]{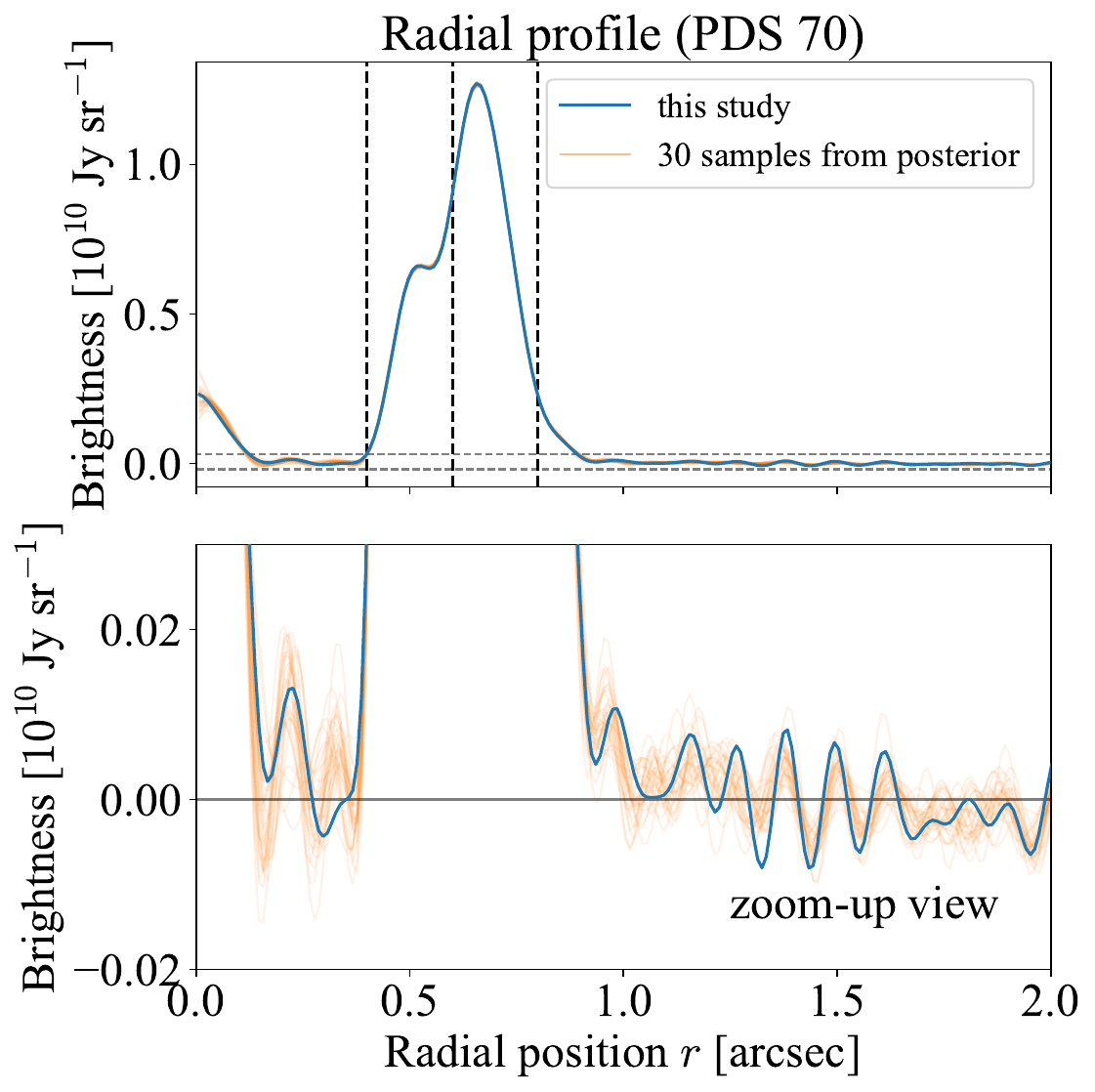}
\includegraphics[width=0.48\linewidth]{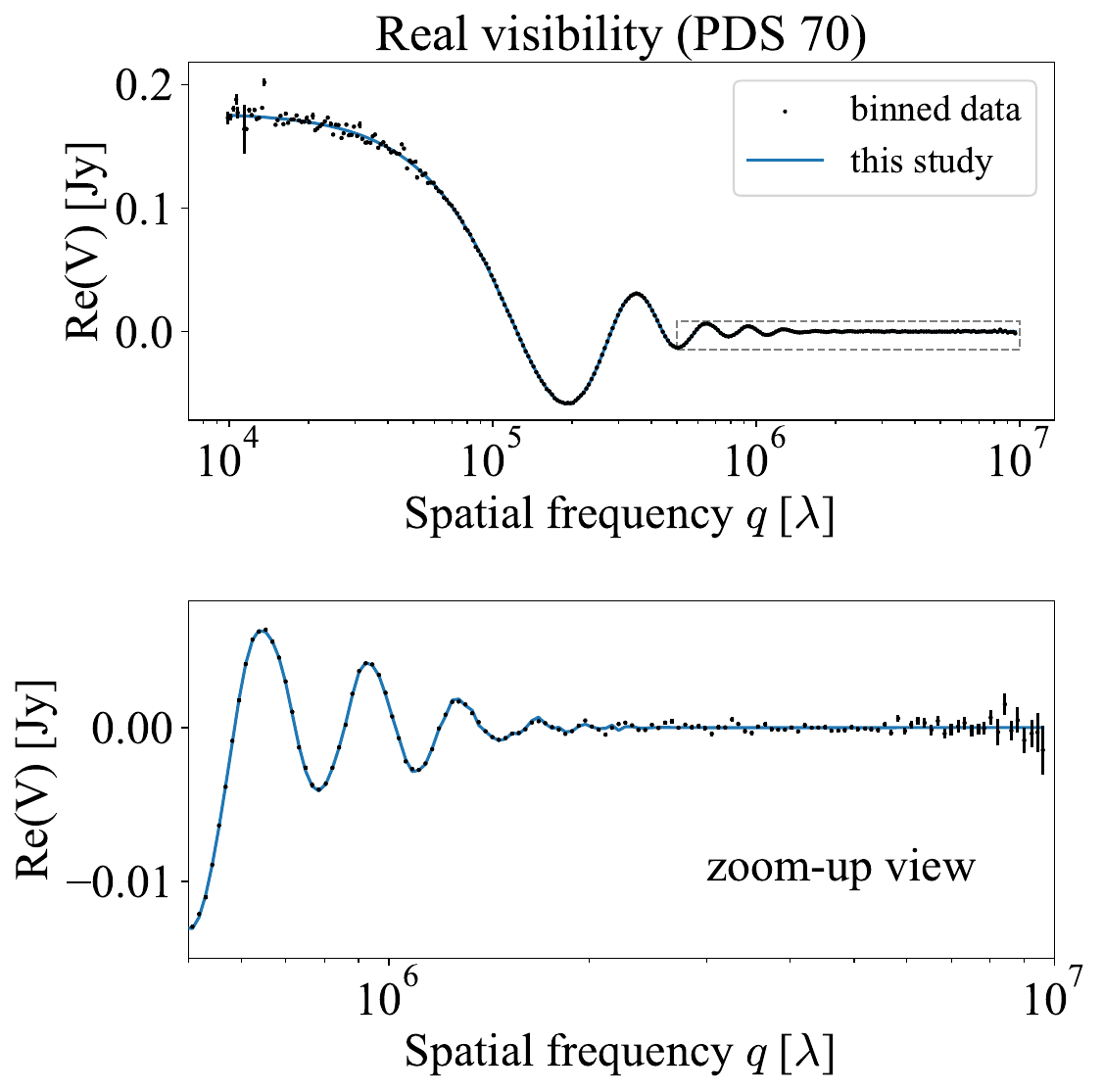}
\includegraphics[width=0.98\linewidth]{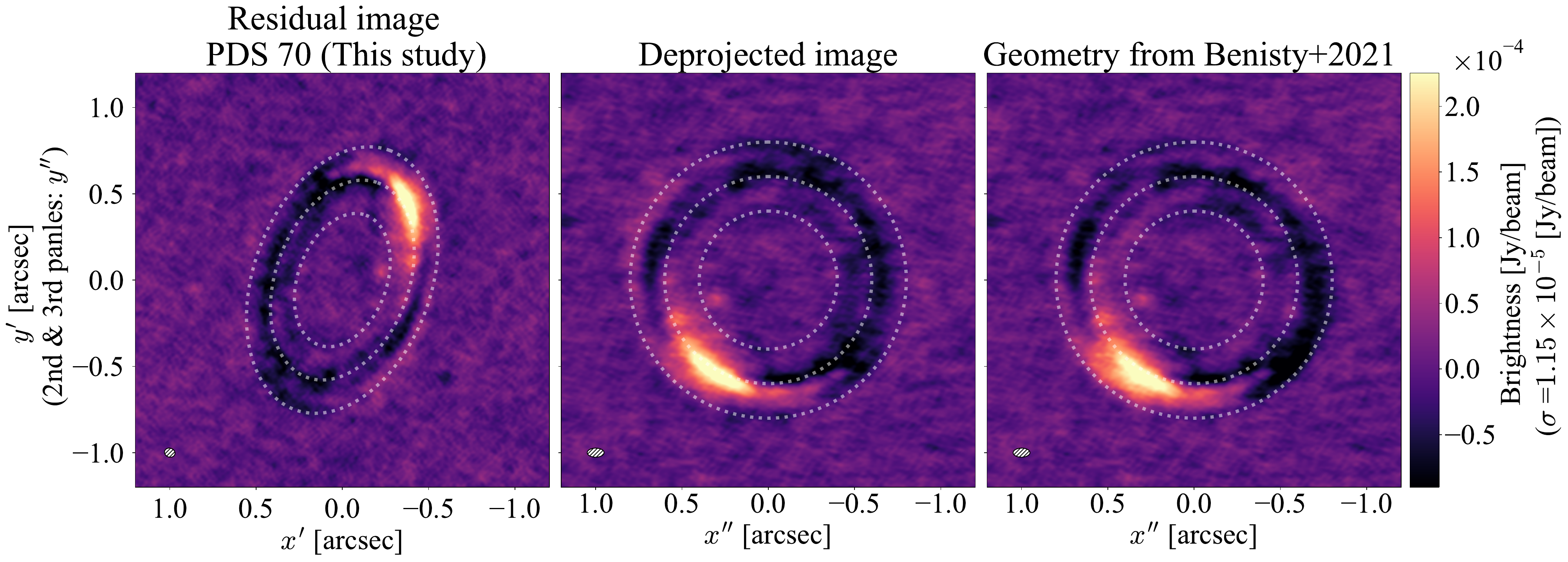}
\end{center}
\caption{Analysis for PDS 70. The format of the figure is the same as that of Fig. 13 except that the zoom-up view of the radial profile is shown in the linear scale. The reference lines are shown at $r=(0.4\arcsec, 0.6\arcsec, 0.8\arcsec)$ in the panels of the brightness profile and the residual maps. The synthesized beam sizes before and after deprojection are $(0.063\arcsec$ , $0.053\arcsec$) and $(0.098\arcsec$ , $0.053\arcsec)$, respectively.}
\label{fig:pds70_model_obs}
\end{figure*}

Figure \ref{fig:pds70_model_obs} illustrates the brightness profiles and visibilities derived from our method. For the brightness profiles, we show 30 random samples, indicated by the light orange lines. The brightness profile was characterized by the presence of the inner disc as well as the outer disc with two local maxima, consistently with previous studies \citep{keppler2019, benisty2021}. At the outer disc, there is another shoulder at $0.85\arcsec$, as also reported in \cite{benisty2021}. A slight positive excess was also observed in the cavity region, between the inner and outer discs. The flux would arise from emission around PDS 70 b\&c and their Lagrange points \citep{benisty2021,balsalobre2023}.

Figure \ref{fig:pds70_model_obs} shows the residual images for PDS 70; the left figure in the lower panels shows the residual map based on our study in the observational frame, the middle one is shown in the deprojected frame, and the right one is based on the geometry in \cite{benisty2021}. The residual images from both geometries are consistent, but there is a slight discrepancy due to difference in geometry. We investigated which parameter made the large difference, and found that the difference in PA largely accounts for the discrepancy (also see Appendix \ref{sec:comp_residuals}).

The upper panels in Figure \ref{fig:pds70_cre_sub} shows the residual images in deprojected coordinate and the polar coordinate. Our methods successfully recovered the circumplanetary emission around PDS 70 c, in addition to the crescent feature. The emission of the circumplanetary disc remains unresolved, consistently with the analysis in \cite{benisty2021}. We measured the brightness of the circumplanetary emission around PDS 70 c. The estimated peak brightness of the emission were $95.2 \pm 17.7$  $\mu$Jy beam$^{-1}$, $89.8 \pm 13.1$ $\mu$Jy beam$^{-1}$, and $91.2 \pm 11.9$ $\mu$Jye beam$^{-1}$ for ${\rm robust} =0, 0.5, 1.0$, respectively. Our estimates for $\rm{robust}=0$ and $0.5$ were consistent with the result from \cite{benisty2021}, which reported $86 \pm 10 $ $\mu$Jy beam$^{-1}$ and $79 \pm 7 $ $\mu$Jy beam$^{-1}$. It should be noted that in our study, the minor/major beam sizes exhibit a slight deviation from those reported in \cite{benisty2021} (see Table 4) with discrepancies ranging between 0.01 and 0.03$\arcsec$. This minor variation is likely due to the time and spectral averaging processes applied to our measurement sets, but the discrepancies give negligible impacts on the measurements of intensities for PDS 70 c. On the other hand, our estimate for ${\rm robust}=1.0$ significantly differed from $170 \pm 4$  $\mu$Jy beam$^{-1}$ reported in \cite{benisty2021}. The discrepancy likely arises from the contamination from the main disc emission, which is however mostly mitigated by our method.

An extended emission around PDS 70 b was reported in \cite{benisty2021}. Our analyses similarly identified a positive excess around it, and the enclosed flux for each robust parameter was around 50 $\mu$Jy, comparable to 38 $\mu$Jy reported by \cite{benisty2021}, although our measured flux was influenced by the cumulative noise within the region. Additionally, there was a claim on the tentative co-orbital emission near PDS 70 b's L5 Lagrangian point \citep{balsalobre2023}. We found the signal to be visually insignificant in the images. Indeed, the peak emission in this region showed significance of 2.0-2.7$\sigma$ for $\rm{robust}=(0,0.5,1)$, slightly lower than the significance around 3.3-3.4$\sigma$ reported in \cite{balsalobre2023} for the robust parameters that we considered. The slight discrepancy might be due to the subtraction of the annular brightness in our study. On the other hand, \cite{balsalobre2023} observed higher significance at 5-6$\sigma$ if they employed ${\rm robust}\geq1.5$ and JvM correction, which were not investigated in this study. The further observations and analyses of these marginal signals will be needed.

%-----------------------------Figure Start---------------------------
%-----------------------------Figure End------------------------------

\subsection{Additional analysis on residual images by subtracting crescent model} \label{sec:sub_cres}
Apart from the crescent feature and circumplanetary emission, residual features still persist in the image. \cite{benisty2021} also identified the residual features in the image after subtracting the axisymmetric component and asymmetric feature including the crescent feature (rightmost end of Fig. 8 in their paper), although the detailed discussion was not presented. Here, we revisited the residual feature in more detail using our updated residual image.

For the further investigation, the high contrast of the bright crescent feature hinders the search of faint features in the images. In addition, the bright crescent unnecessarily introduces the negative flux, because the residual image is supposed to have zero flux when averaged over the polar direction. To avoid the latter problem, \cite{benisty2021} removed the asymmetric feature in the image-based analysis before the visibility analysis. To mitigate these effects, we instead attempted to subtract the bright crescent feature by employing a provisional analytical model. Specifically, we considered a model comprising of the super Gaussian function in the radial direction and the von Mises distribution in the polar direction as follows: 

\begin{align}
I_{\rm crescent} (r, \phi) &= I_{{\rm crescent, amp} } \exp\left( - \left(\frac{(r - r_{0})^{2}}{2\sigma_{r}^{2}} \right) ^{\alpha} \right) \nonumber \\ &\times \left( \frac{\exp(\kappa \cos(\phi - \mu))}{2\pi I_{n=0, {\rm modfied\; bessel}}(\beta)} - \frac{1}{2 \pi}  \right), \label{eq:model_polar}
\end{align}

where $I_{n, {\rm modfied\; bessel}}$ is the modified Bessel function of the first kind, $I_{{\rm crescent, amp} }$ determines the amplitude of the crescent model, $\alpha$ is the exponent for the super Gaussiaan function, $r_{0}$ denotes the radial position for the crescent peak, $\sigma_{r}$ indicates the radial width, $\kappa$ specifies the azimuthal concentration, and $\mu$ determines the azimuthal location for the peak. The integral of the equation (\ref{eq:model_polar}) along $\phi$ is zero, consistent with the construction of the residual image. The adopted model does not perfectly explain the crescent feature, but it is still useful for removing the bright component of the crescent. We fitted the model to the residual image in Sec \ref{sec:current_ana_pds70}, and the optimization was performed using {\tt curve\_fit} in  {\tt scipy}, yielding $(I_{{\rm crescent, amp}}, \alpha, r_{0}, \sigma_{r}, \kappa, \mu)=(0.339\;{\rm mJy/beam}, 0.793, 0.623\arcsec, 0.0723\arcsec, 5.88, -0.960 \;{\rm rad})$.

\begin{figure*}
\begin{center}
\includegraphics[width=0.98\linewidth]{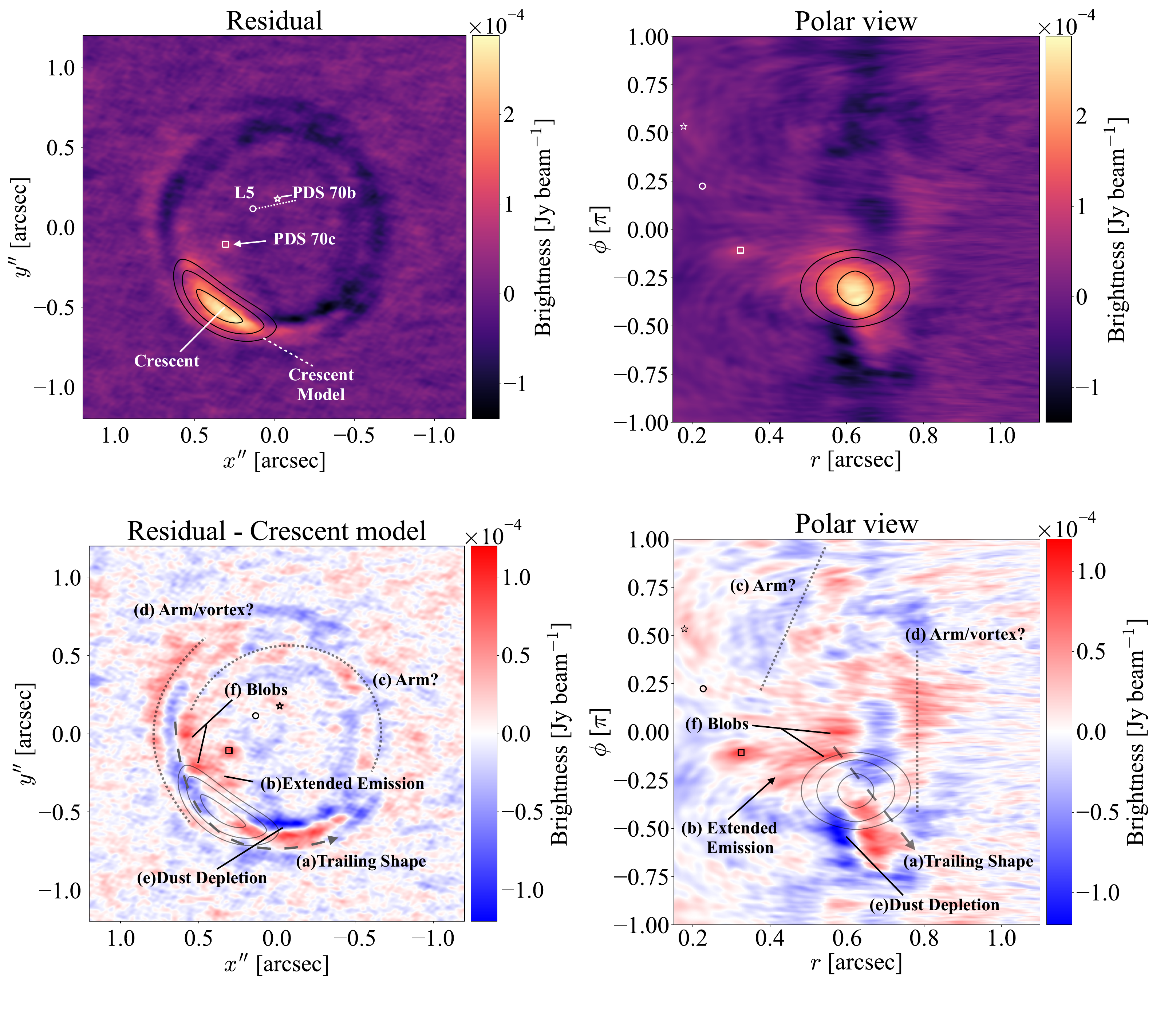}
\end{center}
\caption{ The residual images for PDS 70 in the deprojected coordinate (left panels) and in the polar coordinate (right panels). The upper and lower panels show the residual images without and with the subtraction of the crescent models, as discussed in Sec \ref{sec:sub_cres}. Contours of the crescent models are also overlaid as black lines with levels of $(50, 100, 200)$ $\mu$Jy beam$^{-1}$. Annotations in the upper panels show the locations of the protoplanets and the L$_5$ point of PDS 70 b, as provided by \protect \cite{balsalobre2023}. Annotations in the lower panels specify the possible structures identified in this system, and the further explanations for the features (a)-(f) are provided in Sec \ref{sec:sub_cres}. }
\label{fig:pds70_cre_sub}
\end{figure*}

The lower panels in Figure \ref{fig:pds70_cre_sub} shows the residual images with the crescent models being subtracted in deprojected coordinate and the polar coordinate. We confirmed the presence of emission from PDS 70 c, while no significant detection for point-source emission was observed from PDS 70 b and its L5 point. 

The subtraction process facilitated the identification of faint features in the residual images. We here briefly summarize their characteristics in the following: 
\begin{itemize}
\item[(a)] {\it Crescent is trailing }: As evident in the residual images with and without the subtraction in a polar coordinate, the observed crescent feature appears to exhibit a trailing pattern rather than circular shape. Here the disc rotation is clockwise, as we deduced from the velocity gradient map and the emission surface of the gas \citep{keppler2019}. One interesting possibility is that a planetary gravity perturbs the crescent, resulting in the trailing shape. 
\item[(b)] {\it Extended emission from crescent to PDS 70 c?}: We also observed that the flux excess at the crescent appears to extend toward the vicinity of PDS 70 c. One interesting possibility is that this feature implies a dust accretion flow to PDS 70 c from the outer disc. 
\item[(c)] {\it Faint arm stemming from inner end of crescent}: A faint leading arm is observed to stem from the inner end of the crescent. This feature is also faintly discernible in the residual images without the subtraction. 
\item[(d)] {\it Arm/Vortex-like structure around $0.75\arcsec$ stemming from crescent}: Another arc or vortex stemming from the crescent was observed, and it is radially concentrated around $r=0.75\arcsec$. This feature is more pronounced in our updated residual image than the residual image based on the geometry from \cite{benisty2021}, due to the difference in the adopted geometries, especially in PA (also see Appendix \ref{sec:comp_residuals}). The base of the arm is overlapped with the crescent model in the lower panels, potentially resulting in over-subtraction of the flux of the arm near the region. 
\item[(e)] {\it Depletion near crescent}: We observed a strong negative feature near the crescent feature. It is visible on both residual maps, with and without subtraction of crescent models. This might be attributed to the counter effect of the dust concentration at the crescent. 
\item[(f)] {\it Blobs in crescent}: We observed blobs inside the crescent, suggesting that the crescent is not unimodal but rather multimodal. There are at least two visible blobs in the crescent feature, located at $(x'', y'') = (0.53\arcsec, -0.24\arcsec)$ and $(x'', y'') = (0.58\arcsec, 0.02\arcsec)$ in the deprojected frames. The blobs exhibited an excess of 3-5$\sigma$ with respect to the background level of the the crescent feature. 
\end{itemize}
All of the features appear to be associated with the crescent feature. In Appendix \ref{sec:comp_residuals}, we presented the difference in the residual images derived with geometries from our study and \cite{benisty2021}. Apart from the possible arm (d), all of the features were similarly identified in both of the residual images, reinforcing the coherence of the signals within the current data set. 

Substructures in the PDS 70 disc have also been reported in different bands. The arm-like structure was found in the northwest direction in the near-infrared bands, near the crescent observed in the radio band \citep{muller2018,juillard2022,Christiaens2024}. \cite{juillard2022} investigated the motion of the arm over six years. However, they did not find the anticipated rotation expected if the arm were comoving with  PDS 70 c. This suggests that the arm might be a vortex rather than a spiral. Indeed, as demonstrated in Fig 4 of \cite{marr2022}, such a circular arc/vortex may be misunderstood as one-armed spiral in the near-infrared band, where the disc thickness becomes significant. On the other hand, the crescent observed in the radio band in this study displays a trailing pattern, which is unlikely to result from a purely geometric effect. One alternative explanation for the arm might be the presence of an undetected companion in the outer disc. \citet{Christiaens2024} set an upper limit on the mass of such a potential planet, obtaining limits  1-4 Jupiter masses from the JWST observation in the near-infrared band. 

Additionally, \cite{Christiaens2024} identified a spiral accretion stream to the vicinity of PDS 70 c in the near infrared band by the JWST observation. It might be related to the extended emission excess near PDS 70 in the radio band (feature b in this study), but the connection between the features is not clear at this point. 

The present findings rely on the assumption that the disc structure is well approximated by the thin axisymmetric disc with a single set of geometric parameters. This underlying assumption may be inadequate if the inner and outer discs are misaligned, or the disc thickness is substantial. In addition, there is still room for improvement for the modeling of the crescent feature, and the model might be better to be incorporated into the framework of the axisymmetric modelling. A comprehensive analysis of these points is essential for the robust detection the features, and we leave this issues for the future study.

\section{Summary}  \label{sec:summ}
This study proposed a scheme for estimating the geometry, hyperparameters, central position, 
and brightness profiles assuming a geometrically thin disc in radio interferometry. Our approach is less susceptible to human biases due to manual tuning of parameters in contrast to {\tt frank}, where the non-linear parameters need to be determined a priori. 

Simulating observations for an axisymmetric disc, we demonstrated that the proposed method can successfully retrieve geometric parameters in a more precise manner than that using Gaussian fitting. Additionally, we performed injection and recovery tests for the non-axisymmetric perturbations to the simulated data, and showed that the proposed method can reasonably recover the injected structures. However, the estimated geometric parameters were slightly shifted from the assumed values. This is attributed to the degeneracy between the non-axisymmetric perturbations and the residuals caused by biases in the geometric parameters. 

The model was then applied to the real data for Elias 20 and AS 209, and the ability of the method to determine the disc geometry and brightness profile was demonstrated especially for Elias 20. Moreover, the data for the continuum emission of the PDS 70 were reanalyzed with our method. Our methods successfully identified the circumplanetary emission from PDS 70 c and the crescent feature. Furthermore, we tentatively identified several new features, including trailing nature of crescent, extended emission near PDS 70 c, arm-like structures, dust depletion near crescent, and blobs. The origins of these features are unclear, and a further modeling is needed. 

The current methodology can be applied to any type of continuum data in radio interferometric observations, and future studies will analyse the DSHARP data \citep{andrews2018} to explore their non-axisymmetric structures. 

One of the notable strengths of this research is its capacity to adapt to more complex problems involving a multitude of geometric or hyper parameters, which are difficult to adjust manually. Below, we present the potential directions for extending our study: 
\begin{itemize}
\item The model can be extended to cover multiple rings/gaps with different central positions, 
inclinations, or position angles as observed in GW Ori \citep{bi2020}. 
This can be realized by considering multiple sets of geometric parameters, 
each of which is applied to one of the separate ranges. 
\item The current model can be extended to incorporate frequency-dependent radial 
profiles for multi-frequency data. We can straightforwardly expand our formulation 
to include the frequency dependence in a linearized formulation by Taylor 
expansions in a manner similar to multi-frequency CLEAN \citep{rau2011}:
\begin{eqnarray}
I(r; \nu)= \sum_{t} \left(\frac{ \nu - \nu_{0} }{\nu_{0}}\right)^{t} I_{t, \nu_{0}} (r), \label{eq:taylor_mfeq}
\end{eqnarray}
where $I(r; \nu)$ is the radial brightness at the frequency $\nu$, $\nu_{0}$ is 
the reference frequency, and $I_{t, \nu_{0}} (r)$ is the coefficient of $t$-th Taylor expansions. With this expression, we can analytically derive the model parameters in the same manner as that presented in the current study. 
\item The current study focuses on the axisymmetric component $I(r)$; however, we may be able to include non-axisymmetric components $(I_{m}(r), \phi_{m}(r))$ in the model, and directly solve them. Such modelling efforts might ease the degeneracy between the geometric parameters and non-axisymmetric structures. 
\item Although we created residual images with CLEAN, we can also use 
other imaging techniques, for example, sparse modelling, to produce images alternatively \citep{honma2014,nakazato2020,yamaguchi2020,aizawa2020}. 
In particular, sparse modelling favours solutions with many zeros, and it can achieve better angular resolutions; thus it will be helpful for resolving or identifying the new substructures, including spirals and circumplanetary emission. 

\item A more comprehensive analyse for PDS 70 will be essential for the reliable detection of the discovered structures. Incorporating the crescent model into our model will allow us to separate the residual signal accurately. Exploring more complicated models that consider a disc thickness or misalignment between inner and outer disc planes will be rewarding. The multi-wavelength data will be also useful for assessing the consistency of the signals across different wavelengths. 

\end{itemize} 
These will be considered in future studies. 

\section*{Acknowledgments}
We thank Kazuhiro Kanagawa, Hongping Deng, and Ruobing Dong for their insightful discussions. We also thank the anonymous
referee for constructive comments, which significantly improved the quality of the paper. Furthermore, we would like to extend our gratitude to Myriam Benisty for sharing the data for PDS 70 used in this research.

This work was supported by JSPS KAKENHI grant numbers 17H01103, 18H05441, and 23K03463. M.A. is supported by Special Postdoctoral Researcher Program at RIKEN. T.M. is supported by Yamada Science Foundation Overseas Research Support Program. 

Data analysis was in part carried out on the Multi-wavelength Data Analysis System operated by the Astronomy Data Center (ADC), National Astronomical Observatory of Japan.
This research is based on the following ALMA data \#2013.1.00226.S, \#2015.1.00486.S, and \#2016.1.00484.L. ALMA is a partnership of ESO (representing its member states), NSF (USA) and NINS (Japan), together with NRC (Canada), MOST and ASIAA (Taiwan), and KASI (Republic of Korea), in cooperation with the Republic of Chile. The Joint ALMA Observatory is operated by ESO, AUI/NRAO and NAOJ. 

{\it Software}: {\tt Astropy} \citep{astropy2013}, {\tt CASA} \citep{mcmullin2007}, {\tt corner} \citep{foreman2016}, {\tt emcee} \citep{foreman2013}, {\tt Jupyter Notebook} \citep{kluyver2016}, {\tt Matplotlib} \citep{hunter2007}, {\tt NumPy} \citep{walt2011}, {\tt Pandas} \citep{mckinney-proc-scipy-2010}, {\tt PyNUFFT} \citep{jimaging4030051,lin2017pynufft}, {\tt SciPy}  \citep{virtanen2020}. 

%%%%%%%%%%%%%%%%%%%%%%%%%%%%%%%%%%%%%%%%%%%%%%%%%%
\section*{Data Availability}
The DSHARP data used in this article are available in the DSHARP Data Release at https://bulk.cv.nrao.edu/almadata/lp/DSHARP. The data underlying this article will be shared on reasonable request to the corresponding author.

\bibliographystyle{mnras}
\bibliography{export-bibtex}

\appendix

\section{Efficient computation of evidence given geometry} \label{sec:efficient_comp}
The direct computation of $\mathcal{N}(\bm{d}|0,\bar{\bm{\Sigma}}_{d} + \bar{H} \bar{\bm{\Sigma}}_{a} \bar{H}^{T})$ 
in equation (\ref{eq:d_b_theta_post}) requires the inverse matrix 
of $\bar{\bm{\Sigma}}_{d} +   \bar{H} \bar{\bm{\Sigma}}_{a} \bar{\bm{H}}^{T}$, whose computational 
cost can be $\mathcal{O}(M^{3})$ with $M$ the number of data points. 
As $\mathcal{O}(M^{3})$ can be very large for the usual case in interferometric 
observations $M>10^{5-6}$, we aimed at reducing this computation through the formula transformation. 

Using the Woodbury identity 
$(A+ UWV)^{-1}= A^{-1} - A^{-1} U (W^{-1} + VA^{-1}U)^{-1}VA^{-1}$, 
we obtain
\begin{eqnarray}
(\bar{\bm{\Sigma}}_{d} +   \bar{\bm{H}} \bar{\bm{\Sigma}}_{a} \bar{\bm{H}}^{T})^{-1} = 
\bar{\bm{\Sigma}}_{d}^{-1} - \bar{\bm{\Sigma}}_{d}^{-1} \bar{\bm{H}} \bar{\bm{\Sigma}}_{a|d} \bar{\bm{H}}^{T} \bar{\bm{\Sigma}}_{d}^{-1}.
\end{eqnarray}
With this equation, we can compute the log probability for $p(\bm{d}|  \bm{g}, \bm{\theta} )$ as follows: 
\begin{eqnarray}
 -2 \log(p(\bm{d}|  \bm{g} , \bm{\theta} ) )&=& \log|{\rm det}
 ((\bar{\bm{\Sigma}}_{d} +   \bar{\bm{H}} \bar{\bm{\Sigma}}_{a} \bar{\bm{H}}^{T}))|    
 + \bm{d}^{T} \bar{\bm{\Sigma}}_{d}^{-1} \bm{d}  \nonumber\\
  &-& \bm{d}^{T}  \bar{\bm{\Sigma}}_{d}^{-1} \bar{\bm{H}} \bar{\bm{\Sigma}}_{a|d} \bar{\bm{H}}^{T} \bar{\bm{\Sigma}}_{d}^{-1}  \bm{d}.
\end{eqnarray}
Using another identity 
${\rm det} (A + UWV) = {\rm det} (W^{-1} + VA^{-1}U) {\rm det}(W) {\rm det} (A)$, 
we can compute the first term as follows: 
\begin{eqnarray}
&& \log |{\rm det}((\bar{\bm{\Sigma}}_{d} + \bar{\bm{H}} \bar{\bm{\Sigma}}_{a} \bar{\bm{H}}^{T}))|  
=\log|{\rm det}(\bar{\bm{\Sigma}}_{a|d}^{-1})| \nonumber \\ &+& \log| {\rm det}(\bar{\bm{\Sigma}}_{d})| + \log| {\rm det} (\bar{\bm{\Sigma}}_{a})| .
\end{eqnarray}
Consequently, we find the following expression:
\begin{eqnarray}
 -2 \log(p(\bm{d}|  \bm{g} , \bm{\theta} ) ) &=& \log|{\rm det}(\bar{\bm{\Sigma}}_{a|d}^{-1})|  + \log| {\rm det} (\bar{\bm{\Sigma}}_{a})|  \nonumber\\
 &-&  \bm{d}^{T}  \bar{\bm{\Sigma}}_{d}^{-1} \bar{\bm{H}} \bar{\bm{\Sigma}}_{a|d} \bar{\bm{H}}^{T} \bar{\bm{\Sigma}}_{d}^{-1} \bm{d} + c,  \label{eq:post_compute}
 \end{eqnarray}
where $c$ corresponds to the terms independent on $(\bm{g} , \bm{\theta})$.  
In the case of $M \gg N$, this incurs a computational cost that is much lesser than that using equation (\ref{eq:d_b_theta_post}).

\section{Revisiting visibility binning with uniform and log gridding} \label{sec:data_bin}
The computational cost strongly depends on the number of visibilities. Here, we discuss the data binning in an interferometric observation assuming linear and log grids.  

\subsection{Formulation}
We consider the data $\bm{d}_{\rm obs}$ with the data weights $\bm{w}_{\rm obs}$. 
We binned them on a grid specified by bin edges  $\{(E_{\rm i}, E_{\rm j})\}$ with $i= -(N_{\rm bin}+1), -N_{\rm bin}, ..., N_{\rm bin}, N_{\rm bin}+1$, wherein $N_{\rm bin}$ determines the number of grids. For a cell with $E_{i-1}<u_{k}< E_{i}$ and $E_{j-1}<v_{k}< E_{j}$, we define the
summing operation for a vector $\bm{x}$ with the same length as $\bm{d}$ as follows:
\begin{eqnarray}
{\rm Bin}(\bm{x}, i, j) = \sum_{E_{i-1}<u_{k}<E_{i} \atop
E_{j-1}<v_{k}<E_{j}} x_{k}, 
\end{eqnarray}
where $E_{i}$ denotes either $E_{{\rm uniform}, i}$ or $E_{{\rm log}, i}$. 
Bin edges for a uniform gridding are defined as follows: 
\begin{eqnarray}  
E_{{\rm uniform}, i} =
\left\{
    \begin{array}{cc}
         x_{\rm min} +(i-1)\left( \frac{ x_{\rm max}- x_{\rm min} } {N_{\rm bin}} \right) &   i=1, 2, ..., N_{\rm bin}+1  \\
        0        &   i=0 \\
        - x_{\rm min} + (i+1) \left( \frac{ x_{\rm max}- x_{\rm min} } {N_{\rm bin}}   \right)  &   i=-1, -2, ..., -(N_{\rm bin}+1),  \\
    \end{array}
    \right. \nonumber
\end{eqnarray}
\begin{eqnarray}
    \label{eq: linear_bin} 
\end{eqnarray}
where ($x_{\rm min}$, $x_{\rm max}$) are values that determine the limits of the grid.  

Similarly, we define bin edges for a log grid as follows: 
\begin{eqnarray} 
E_{{\rm log} , i} =
\left\{
    \begin{array}{cc}
         x_{\rm min} \times 10^{ (i-1)\Delta_{\rm log}} &  i=1, 2, ..., N_{\rm bin}+1  \\
        0        &   i=0 \\
        - x_{\rm min}  \times 10^{- (i+1)\Delta_{\rm log}}    &   i=-1, -2, ..., -(N_{\rm bin}+1)
    \end{array}
    \right. \label{eq: log_bin} 
\end{eqnarray}
where we define a spacing $\Delta_{\rm log}$ for the log grid as follows: 
\begin{eqnarray}
\Delta_{\rm log}= \frac{\log_{10} (x_{\rm max} ) - \log_{10} (x_{\rm min} )}{N_{\rm bin}}.  \label{eq:bin_end}
\end{eqnarray}

In each cell, we computed a weighted average $\bm{d}_{{\rm bin}}$, 
a total sum of weights $\bm{w}_{{\rm bin}}$, standard deviations for the noise $\sigma_{{\rm bin}, i, j}$, and weighted average of spatial frequencies $( \bm{u}_{{\rm bin}}, \bm{v}_{{\rm bin}})$ as follows:

\begin{eqnarray}
w_{{\rm bin}, i, j} &=&  {\rm Bin}(\bm{w}, i, j) \\
\sigma_{{\rm bin}, i, j} &=&  \frac{1}{\sqrt{w_{{\rm bin}, i, j}}}\\
d_{{\rm bin}, i, j} &=& \frac{ {\rm Bin}(\bm{d} \circ \bm{w}, i, j)}   {w_{{\rm bin}, i, j}}\\
u_{{\rm bin}, i, j} &=& \frac{ {\rm Bin}(\bm{u} \circ \bm{w}, i, j)}  {w_{{\rm bin}, i, j}}\\
v_{{\rm bin}, i, j} &=& \frac{ {\rm Bin}(\bm{v} \circ \bm{w}, i, j)}  {w_{{\rm bin}, i, j}}, 
\end{eqnarray}
where $ \circ $ denotes the Hadamard product for 
vectors $\bm{x}$ and $\bm{y}$: $(\bm{x} \circ \bm{y})_{i} = x_{i} y_{i}$. 

\subsection{Quantifying binning errors with uniform and log grids} \label{sec:q_binning}
We quantified binning errors by simulating an observation of a proto-planetary disc. The simulation setup was same as simulated case of AS 209 in Sec \ref{sec:sim}.  

The data were binned with $N_{\rm bin}=(125, 250, 500, 1000, 2000, 4000)$, $x_{\rm min}=10^{2}$ $\lambda$, and $x_{\rm max}=10^{7}$ $\lambda$. 
At the binned spatial frequencies $(\bm{u}_{\rm bin}, \bm{v}_{\rm bin})$, 
we computed model visibilities $\bm{d}_{\rm bin, model}$, and quantified a binning error relative to an observational error as follows: 
\begin{eqnarray}
\chi_{{\rm bin}, i,j} = \frac{ d_{{\rm bin}, i, j} - d_{{\rm model}, i, j}}{\sigma_{{\rm bin}, i, j} }, \label{eq:binning_err}
\end{eqnarray}
which is the ratio of the binning error with respect to the noise amplitude. Additionally, we also computed the mean squared binning errors $\chi_{{\rm bin error, mean}}^{2}$: 
\begin{eqnarray}
\chi_{{\rm bin, mean}}^{2} = \frac{1}{N_{\rm d, bin}}\sum \chi_{{\rm bin}, i,j} ^{2}, 
\end{eqnarray}
where $M_{\rm d, bin}$ is the number of data after binning. 

\begin{figure*}
\begin{center}
\includegraphics[width=0.42\linewidth]{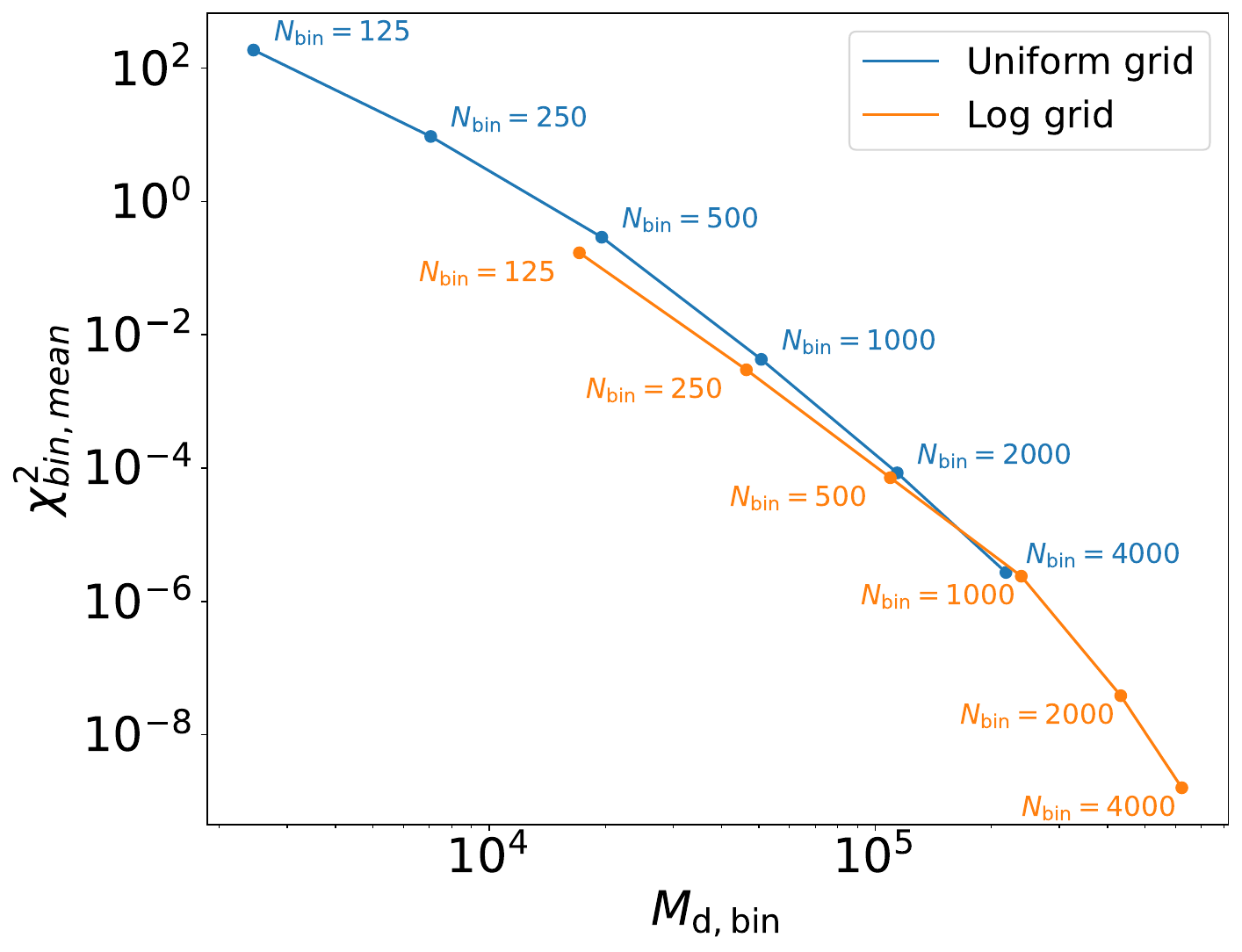}
\includegraphics[width=0.42\linewidth]{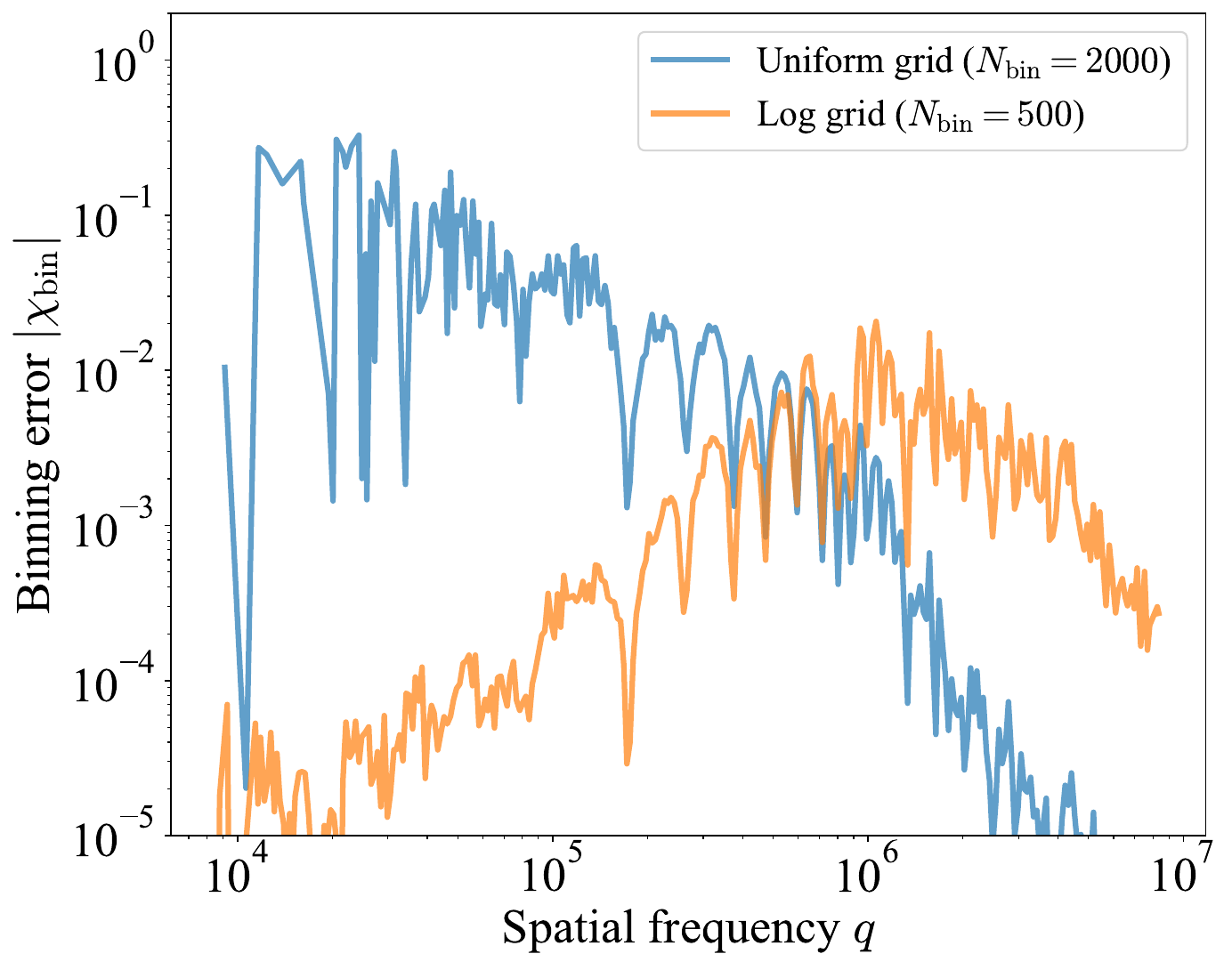}
\end{center}
\caption{Comparison of binning errors for log (orange line) and linear (blue line) grids. 
(left) The average binning error $\chi_{{\rm bin, mean}}^{2}$ and the number of binned data 
$M_{\rm d, bin}$.  We vary $N_{\rm bin}$ in the range of $125$-$4000$. (right) Individual 
binning error $|\chi_{\rm bin}|$ and deprojected spatial frequency $q$. We assume 
$N_{\rm bin}=2000$ for the uniform grid and $N_{\rm bin}=500$ for the log grid.}
\label{fig:bin_2d}
\end{figure*}

We computed the binning errors for the linear and log grids by adopting $N_{\rm bin}=(125, 250, 500, 1000, 2000, 4000)$. 
The left panel in Fig. \ref{fig:bin_2d} shows the relation between the mean squared binning errors $\chi_{\rm bin, mean}^{2}$ and the number of binned data $M_{\rm d, bin}$, which determines the computational time. As $N_{\rm bin}$ increases, the binning errors became small, and $M_{\rm d, bin}$ became large as expected, thereby increasing the computational cost. On average, uniform and log grids yielded similar binning errors with  $M_{\rm d, bin}$ being fixed as shown in Fig. \ref{fig:bin_2d}.

The right panel in Fig. \ref{fig:bin_2d} shows the binning errors 
$|\chi_{\rm bin}|$ at deprojected spatial frequencies $q$. To obtain a similar number of data points $M_{\rm d, bin}  \simeq 10^{5}$, we assumed $N_{\rm bin}=2000$ for the uniform grid and $N_{\rm bin}=500$ for the log grid (see left panel in Fig. \ref{fig:bin_2d}). 
The binning errors in the uniform grid increased at low spatial frequencies, and they reduced at higher frequencies. With $N_{\rm bin}=2000$ for the uniform grid, the binning error can reach $40$ per cent at most, and such errors might be problematic to estimations on large-scale emissions. The binning errors increase at small spatial frequencies because the binning errors in the uniform grid are proportional to the first derivative of the radial visibility profile, which tends to be large at smaller scales. This is supported by the small binning errors at large spatial frequencies. In the case of the log grid, the binning errors are suppressed at small spatial frequencies, in contrast to the uniform grid. This accords with the narrow bin widths at small spatial frequencies in the log grid; the grid is sufficiently fine to resolve the visibility profile. At the middle to high spatial frequencies, the binning errors increased and decreased, with a peak at $10^{6}$ M$\lambda$. At the higher spatial frequencies, the grids become coarse, while the first derivative of the radial visibility profile becomes small. These two different trends result in this dependency. 

Comparing the two grids, the log grid tends to yield moderate binning errors, whereas the uniform grid yields errors in a broad range. Considering the robustness, we adopted the log grid in the analysis of the main text, because the large binning errors at small spatial frequencies for the uniform grid can degrade the estimated accuracy of the flux on a large scale. 

\subsection{Dependence of geometric parameters and hyperparameters on number of grids} \label{sec:q_binning}
\begin{figure*}
\begin{center}
\includegraphics[width=0.32\linewidth]{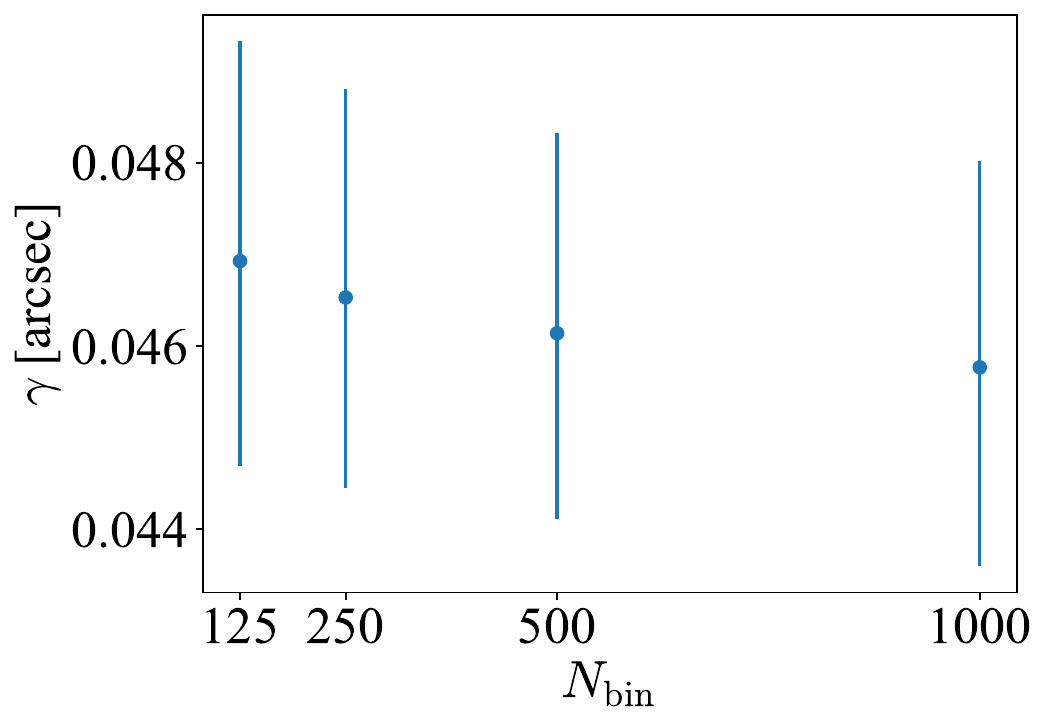}
\includegraphics[width=0.32\linewidth]{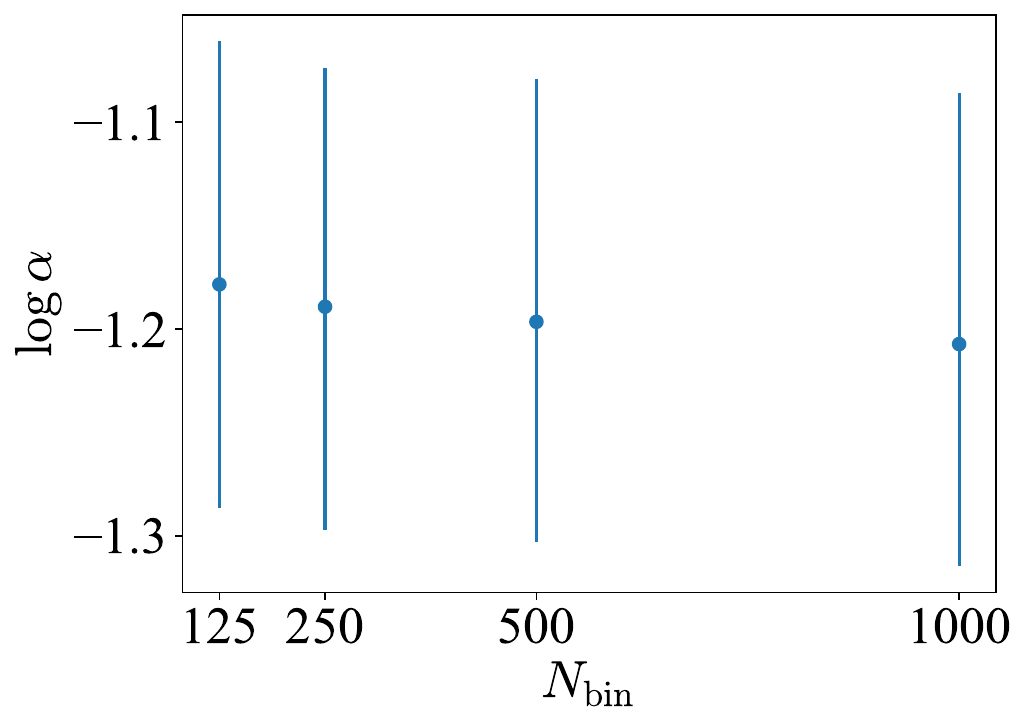}
\includegraphics[width=0.32\linewidth]{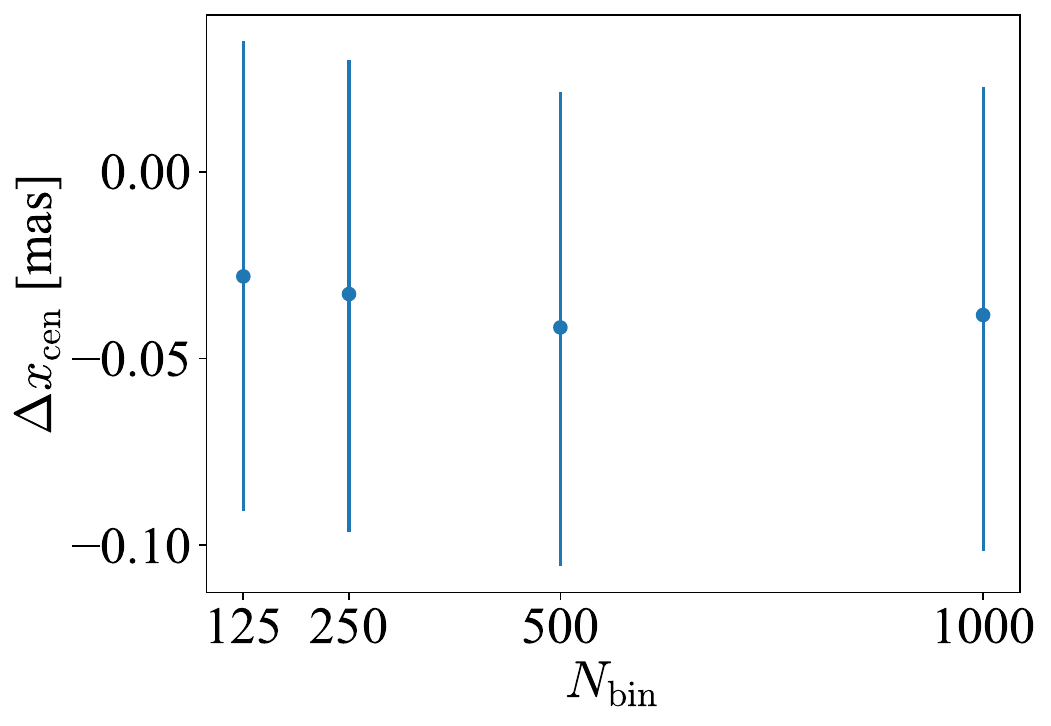}
\includegraphics[width=0.32\linewidth]{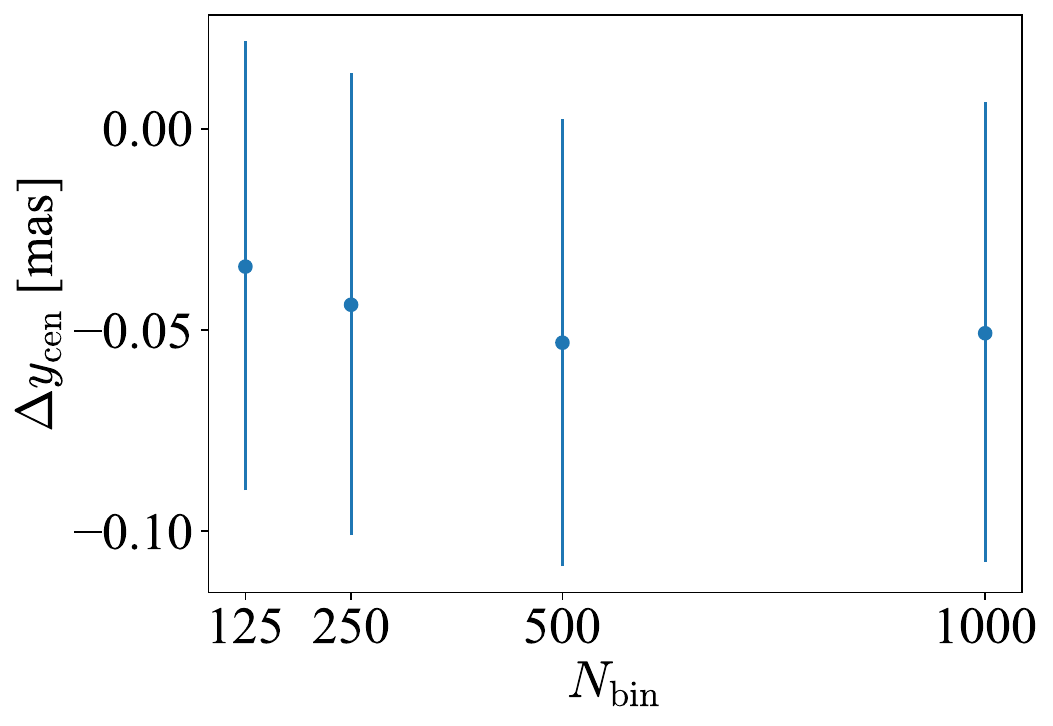}
\includegraphics[width=0.32\linewidth]{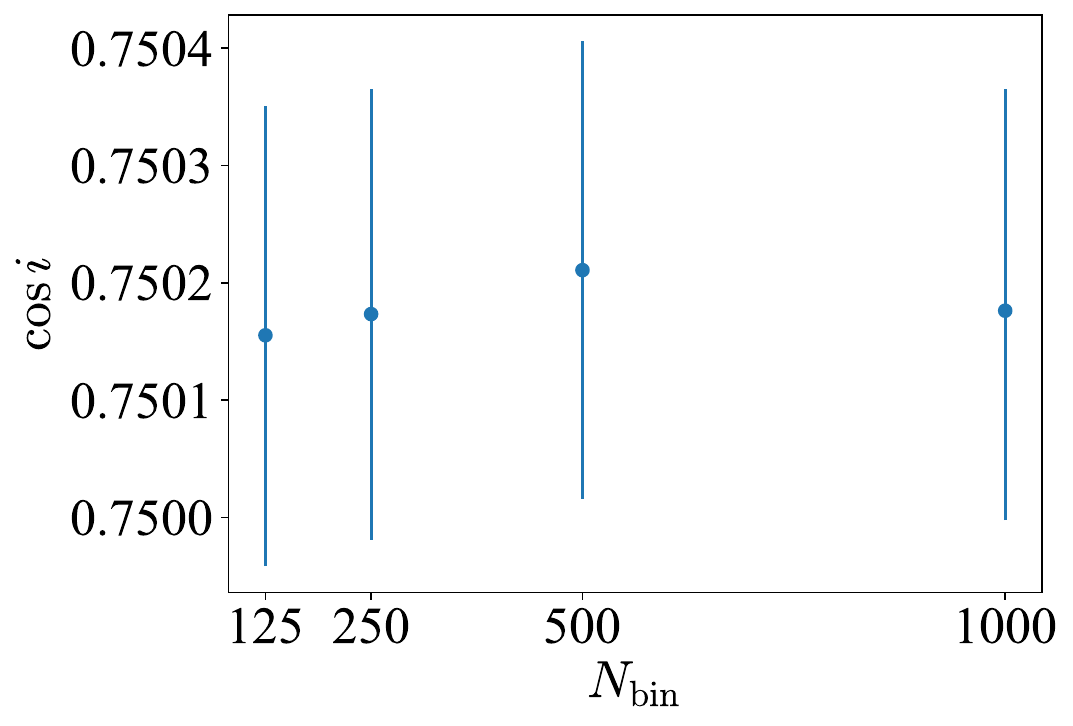}
\includegraphics[width=0.32\linewidth]{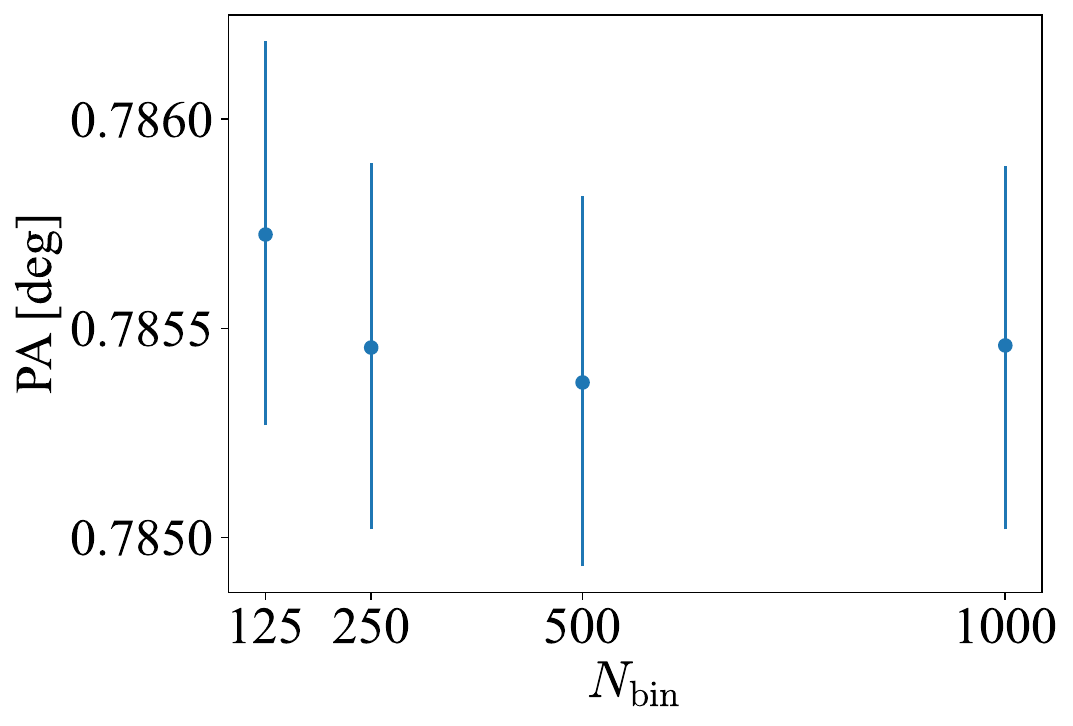}
\end{center}
\caption{Recovered geometric and hyperparameters with different numbers of bins $N_{\rm bin}$. }
\label{fig:bin_change}
\end{figure*}

In the main text, we adopted the log grid with $N_{\rm bin}=500$. Here, we show that the choice of $N_{\rm bin}$ does not affect the estimation on non-linear parameters. Simulating the same observational setup and the brightness profile as in the simulated case for AS 209 in Sec \ref{sec:sim}, we derived the geometric parameters by varying the number of grids, $N_{\rm bin}=$ 125, 250, 500, and 1000. We injected $(\Delta x_{\rm cen}, \Delta y_{\rm cen},\cos i, {\rm PA})=( 0\arcsec, 0\arcsec, 0.75, 45^{\circ})$. Fig. \ref{fig:bin_change} shows the results for recovered geometric parameters and hyperparameters. The results appear to be unaffected by the choice of $N_{\rm bin}$. 
%-----------------------------Figure Start---------------------------
%-----------------------------Figure End------------------------------

\section{Dependence of residual images on robust parameters} \label{sec:robust_vary}
\begin{figure*}
\begin{center}
\includegraphics[width=0.45\linewidth]{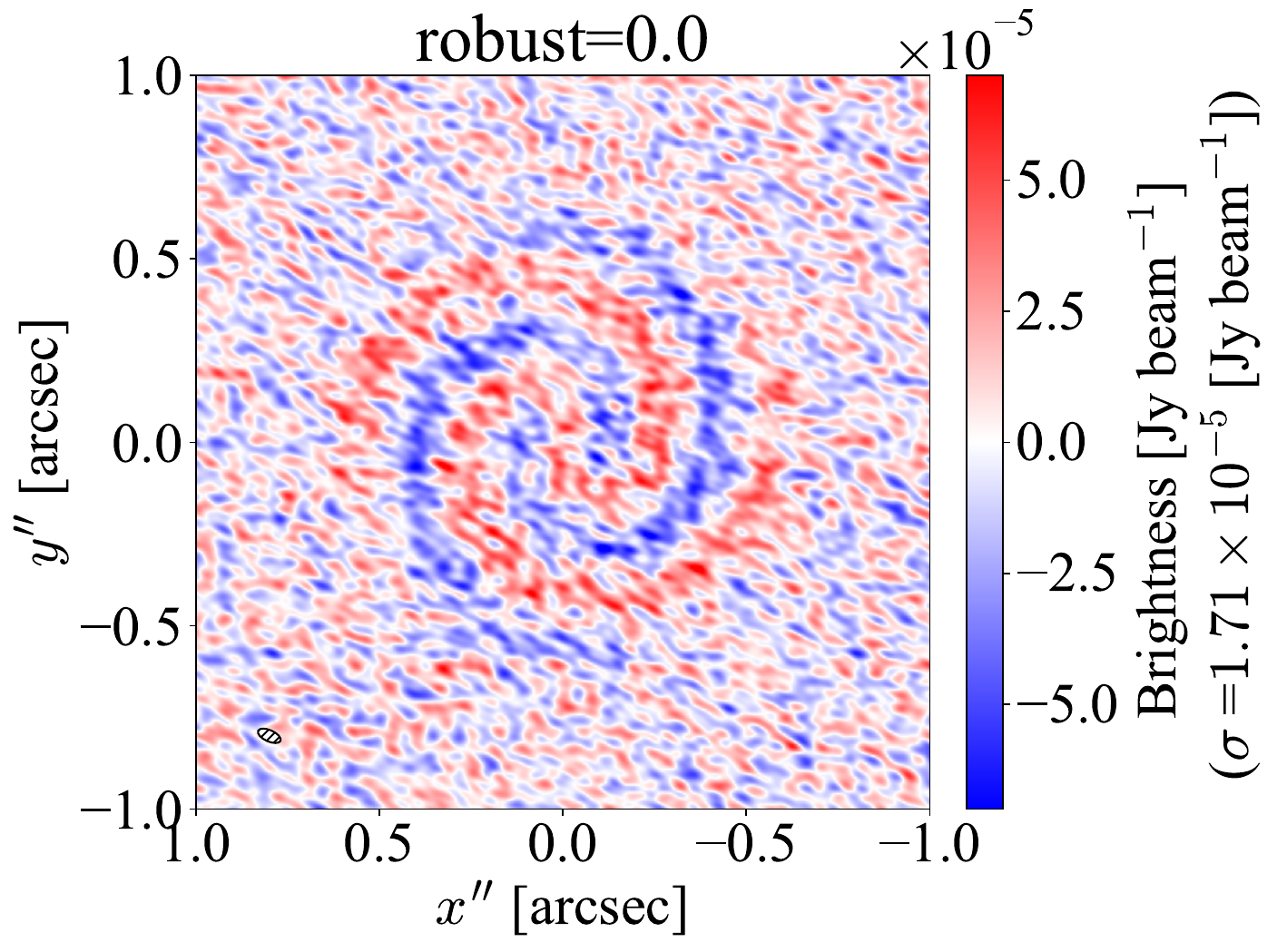}
\includegraphics[width=0.45\linewidth]{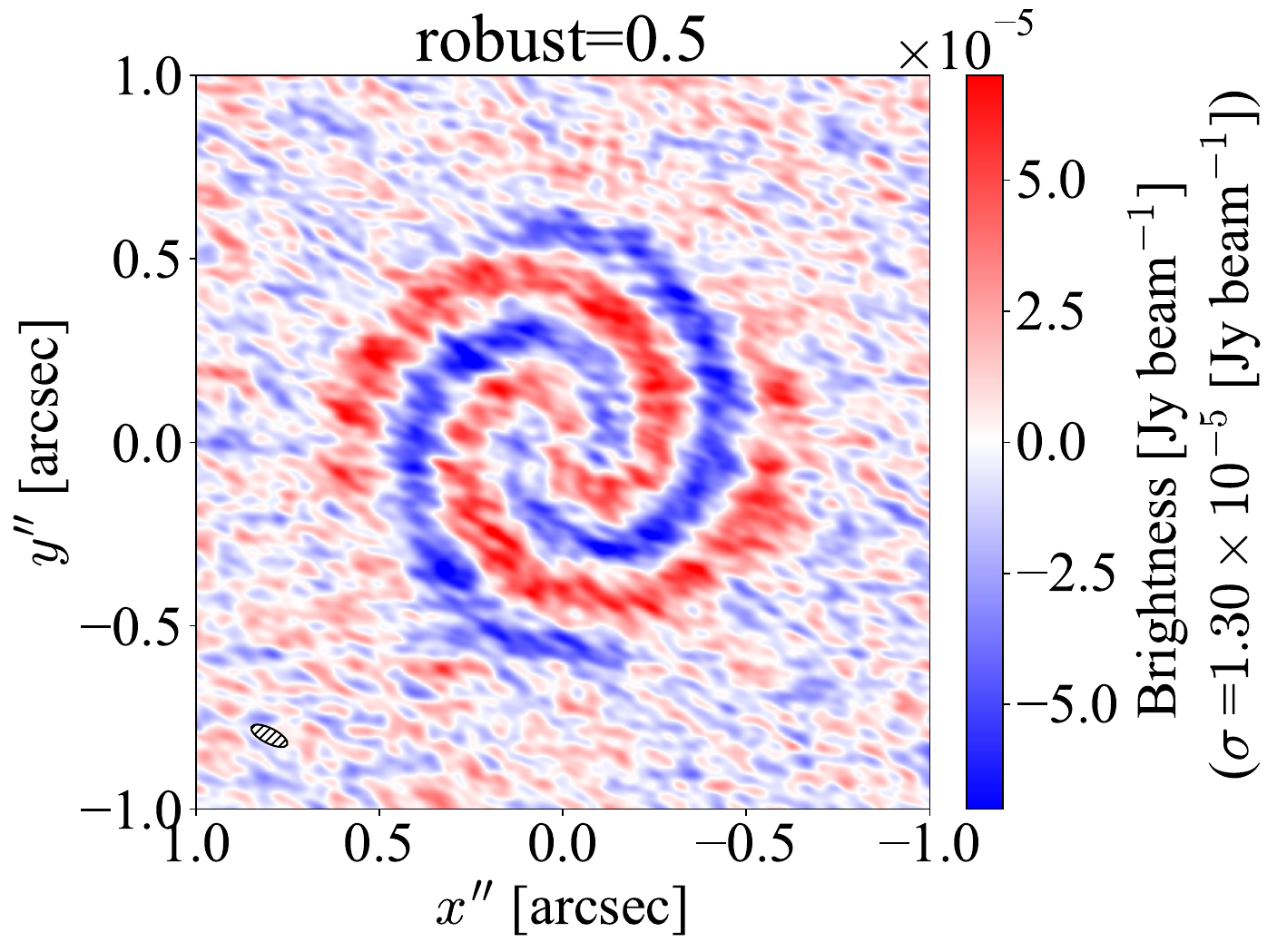}
\includegraphics[width=0.45\linewidth]{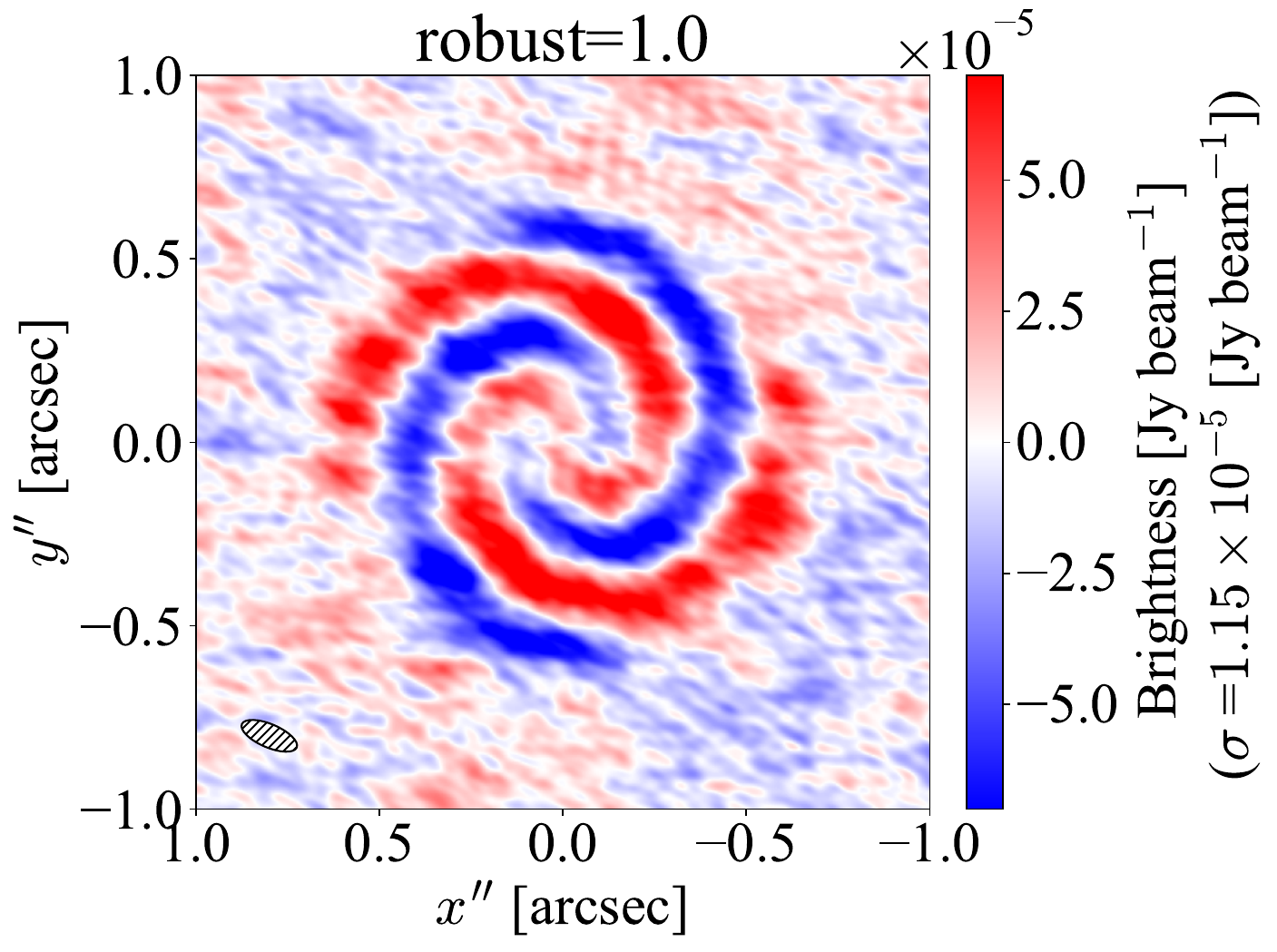}
\end{center}
\caption{Dependence of residual images on robust parameters [0, 0.5, 1.0].  The synthesized beam sizes $(0.067\arcsec, 0.031\arcsec)$, $(0.11\arcsec, 0.041\arcsec)$, and $(0.16\arcsec, 0.063\arcsec)$ are shown at the bottom left of panels. }
\label{fig:robust_change}
\end{figure*}

We varied the robust parameters as [0, 0.5, 1] in Briggs gridding, and investigated the dependence of the resultant residual images. Fig. \ref{fig:robust_change} shows the deprojected residual images adopting three different robust parameters. They were constructed from the same residual visibilities as those used in the case of the even-symmetric spiral in Fig. \ref{fig:odd_recover}. Visual inspections indicated that the robust parameter of 0.5 yielded the most balanced image in terms of both sensitivity and resolution. 

%-----------------------------Figure Start---------------------------
%-----------------------------Figure End------------------------------

\section{Recovered spatial scales for different angular resolutions and brightness profiles} \label{sec:scaling}
\begin{figure*}
\begin{center}
\includegraphics[width=0.48\linewidth]{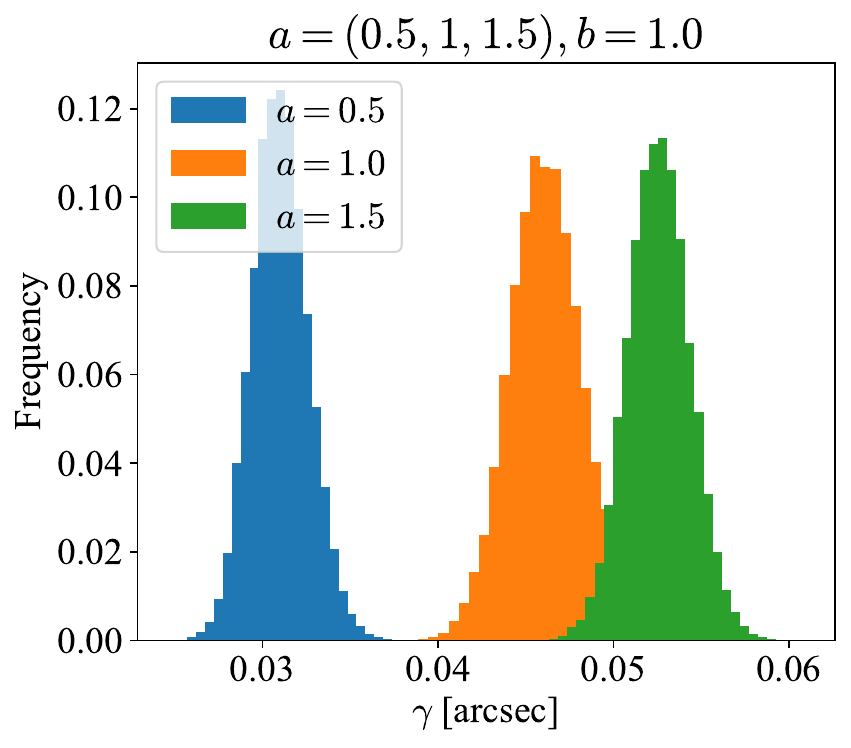}
\includegraphics[width=0.45\linewidth]{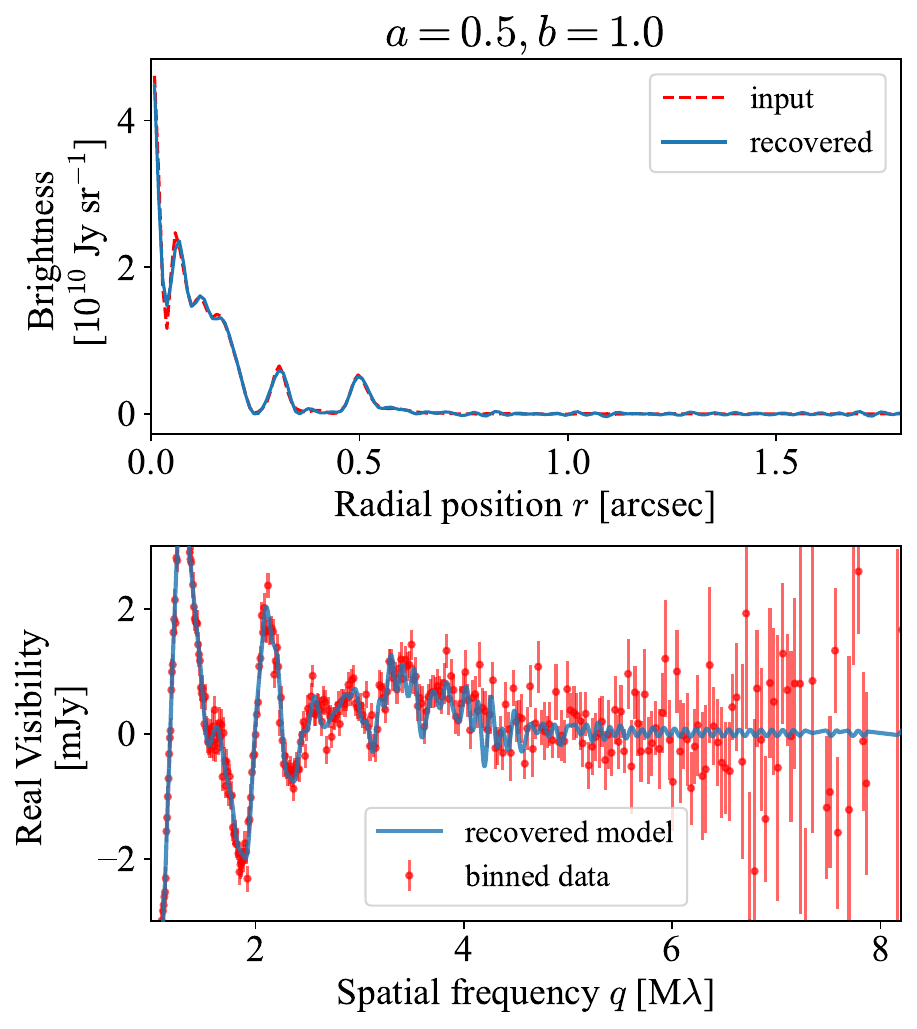}
\includegraphics[width=0.45\linewidth]{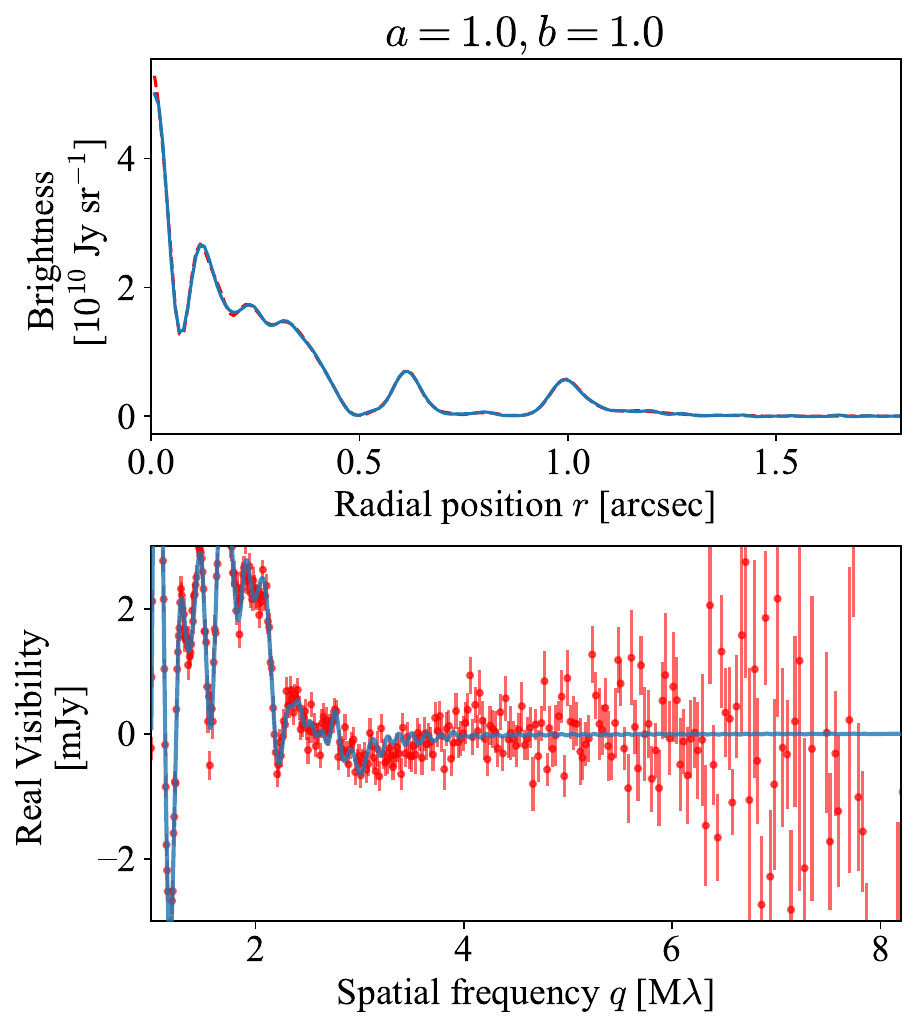}
\includegraphics[width=0.45\linewidth]{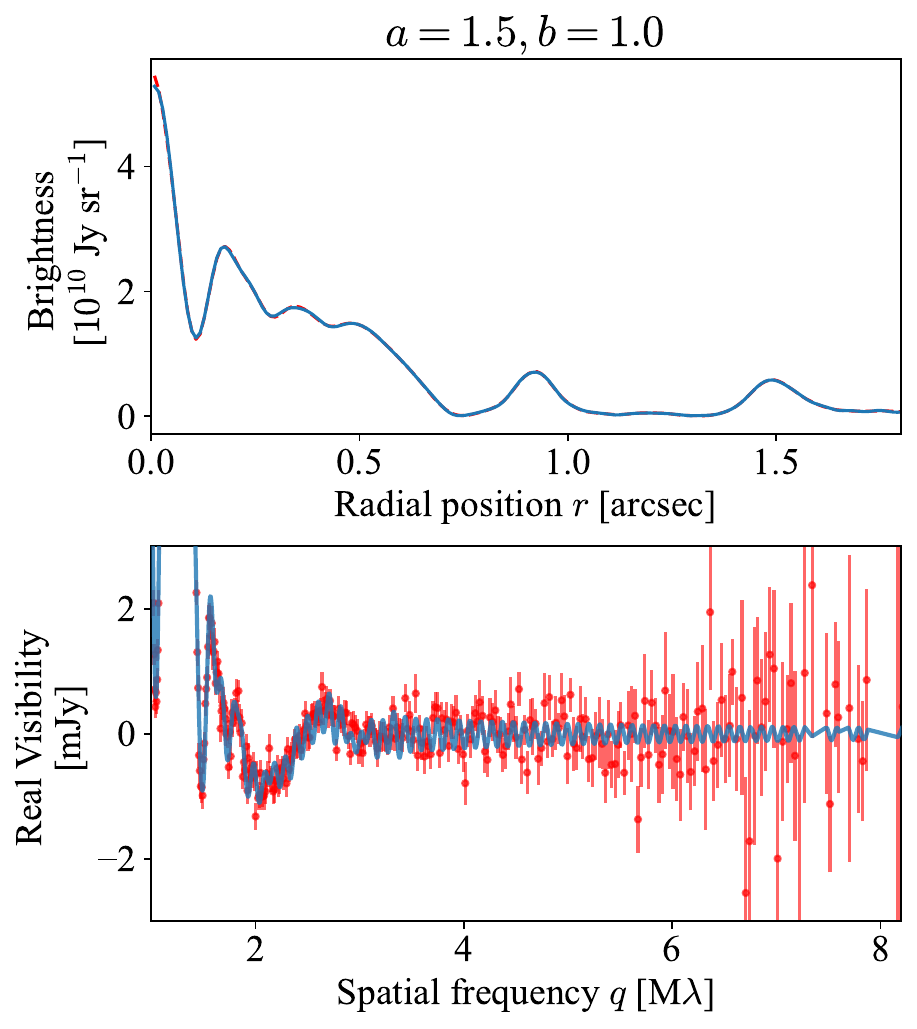}
\end{center}
\caption{Simulation result with varying length scales of injected brightness profile $f_{a}(r)=f(ar)$ with $a=(0.5, 1, 1.5)$. (left upper) Posterior distribution of recovered length scale $\gamma$ for $a=(0.5, 1, 1.5)$. (right upper) Result with $a=0.5$. Injected and recovered brightness profile in the upper panel, and model and observed visibilities in the lower panel. The observed visibilities are binned with a log grid with $N_{\rm bin}=2000$. (lower panels) Result with $a=(1,1.5)$. }
\label{fig:a_vary_b}
\end{figure*}

\begin{figure*}
\begin{center}
\includegraphics[width=0.48\linewidth]{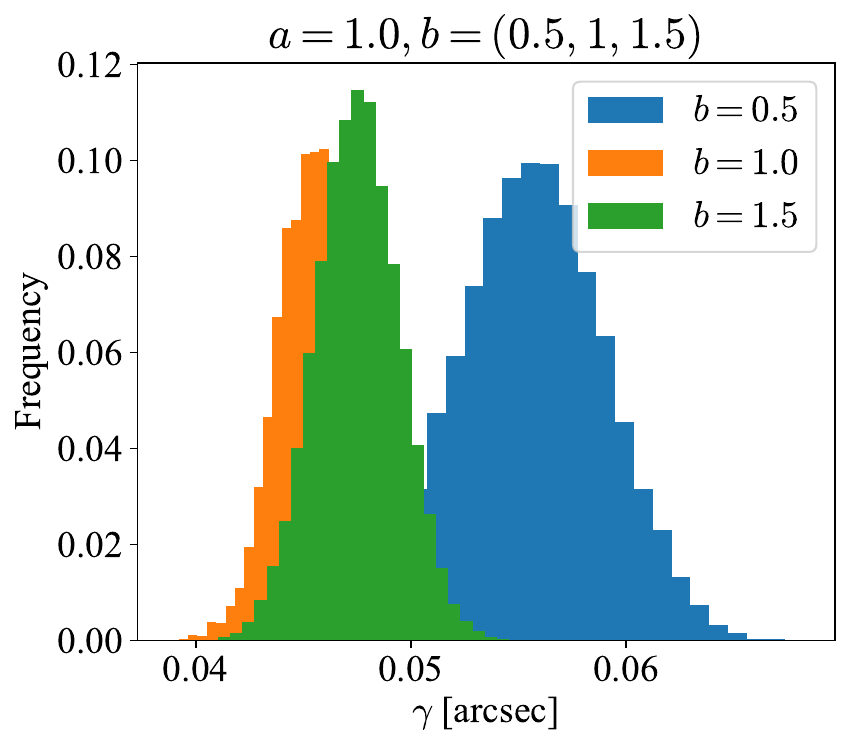}
\includegraphics[width=0.45\linewidth]{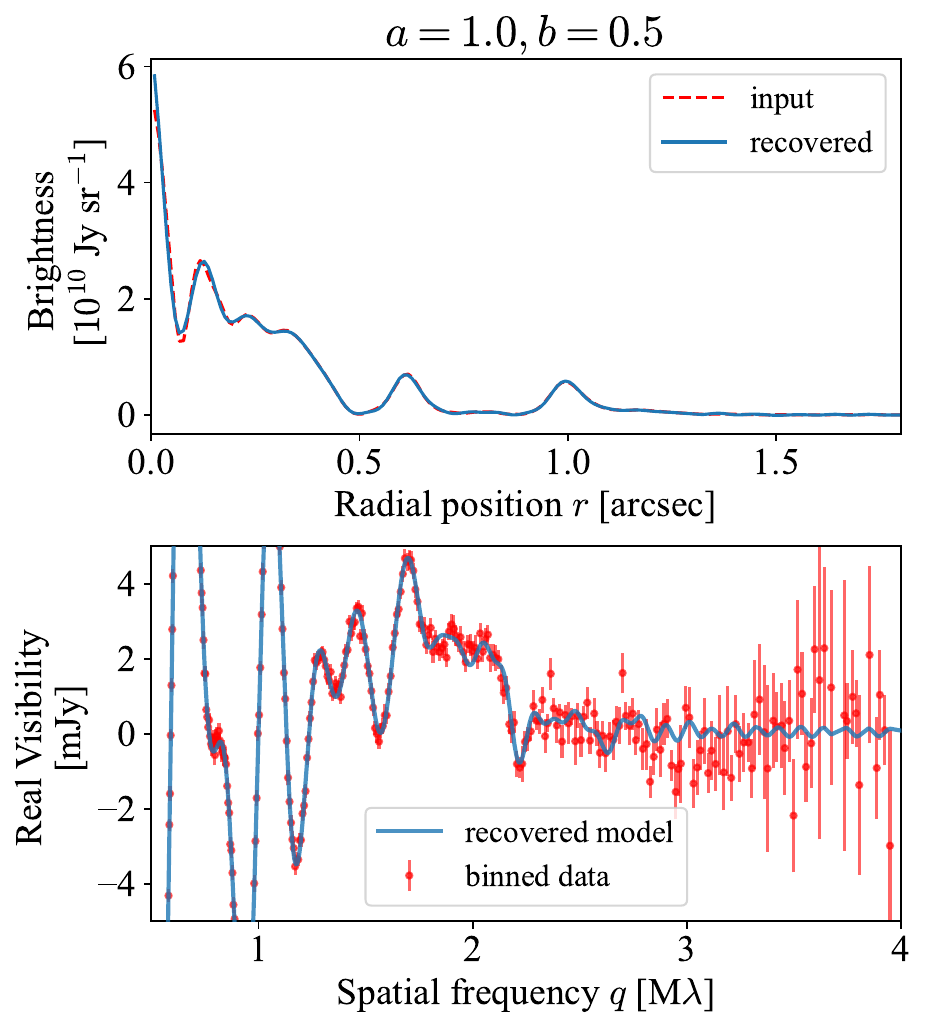}
\includegraphics[width=0.45\linewidth]{./fig/varying_scale/a1.0_b1.0.pdf}
\includegraphics[width=0.45\linewidth]{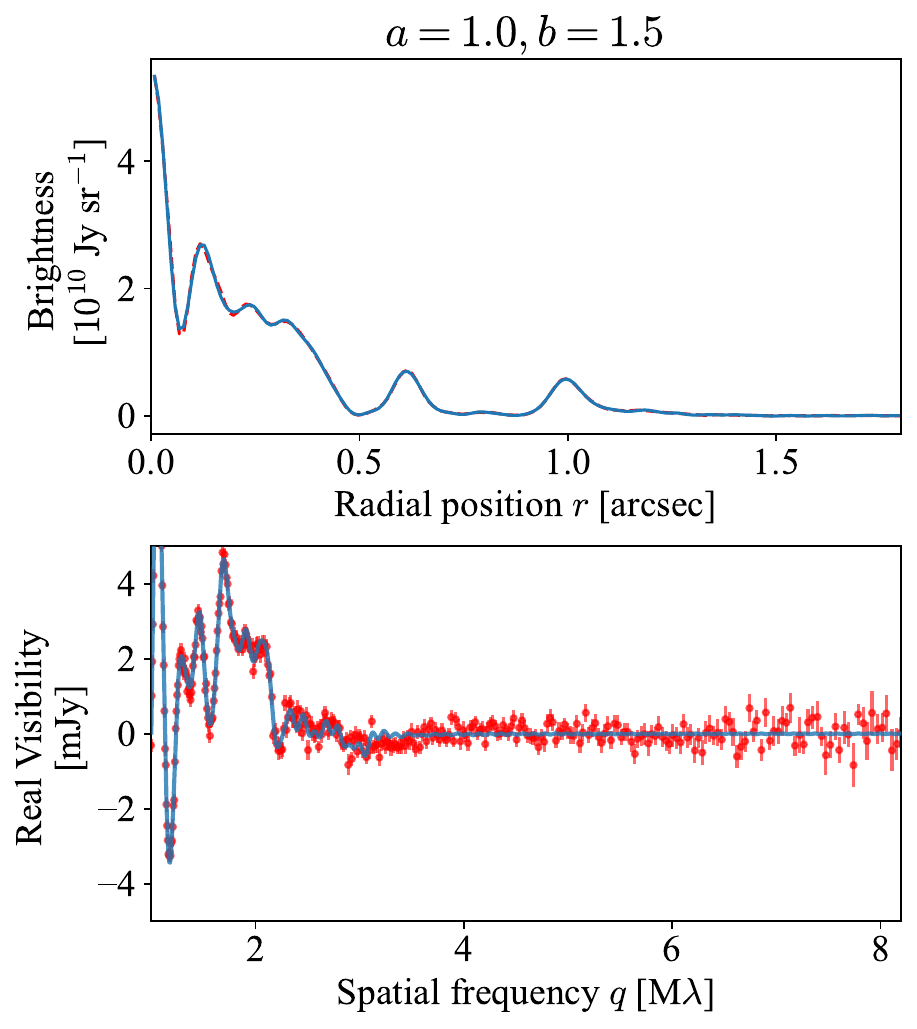}
\end{center}
\caption{Same figure as Fig. \ref{fig:a_vary_b} except that we vary $b=(0.5, 1, 1.5)$ and fix $a=1$.  }
\label{fig:b_vary_a}
\end{figure*}

We changed the length scales for the input profile $I(r)$ and observational spatial frequencies $\{u_{j},v_{j}\}$ to investigate the variations of recovered length parameters $\gamma$. Specifically, we considered a modified brightness profile $I_{a}(r) = I(ar)$ and modified spatial frequencies $\{u_{b, j},v_{b, j}\} = \{b u_{j},b v_{j}\}$, where we introduced scaling parameters $(a, b)$. We considered the same observational setup and brightness profile as that of AS 209 in Sec \ref{sec:sim}, simulated the data, and derived the posterior distribution for parameters including $\gamma$. Fig. \ref{fig:a_vary_b} shows the result with varying scale of brightness profile $a=(0.5, 1, 1.5)$ and fixed observed spatial frequency $b=1$. The optimized length scale $\gamma$ positively scaled with $a$. Beyond the optimized length scale, the power of model visibilities was suppressed and exhibited damped oscillations.   Fig. \ref{fig:b_vary_a} shows the result with varying scale for spatial frequency for the data $b=(0.5, 1, 1.5)$ and the same brightness profile $a=1$. The optimized length scale was unchanged for $b=1, 1.5$; however, it reduced for $b=0.5$. This indicates that if the structure is already well resolved, the improvement in the angular resolution does not change the optimized length scale $\gamma$. However, if we adopt the worse angular resolution, the resolution cannot be sufficient to resolve the structure. This results in the larger value of $\gamma$. From this discussion, we can conclude that the optimized length scale $\gamma$ 
is determined by both the brightness profile and the UV-coverage.
%-----------------------------Figure Start---------------------------
%-----------------------------Figure End------------------------------

%-----------------------------Figure Start---------------------------
%-----------------------------Figure End------------------------------
\section{Mean and variance of residual images } \label{sec:effect_of_mcmc}
\begin{figure*}
\begin{center}
\includegraphics[width=0.44\linewidth]{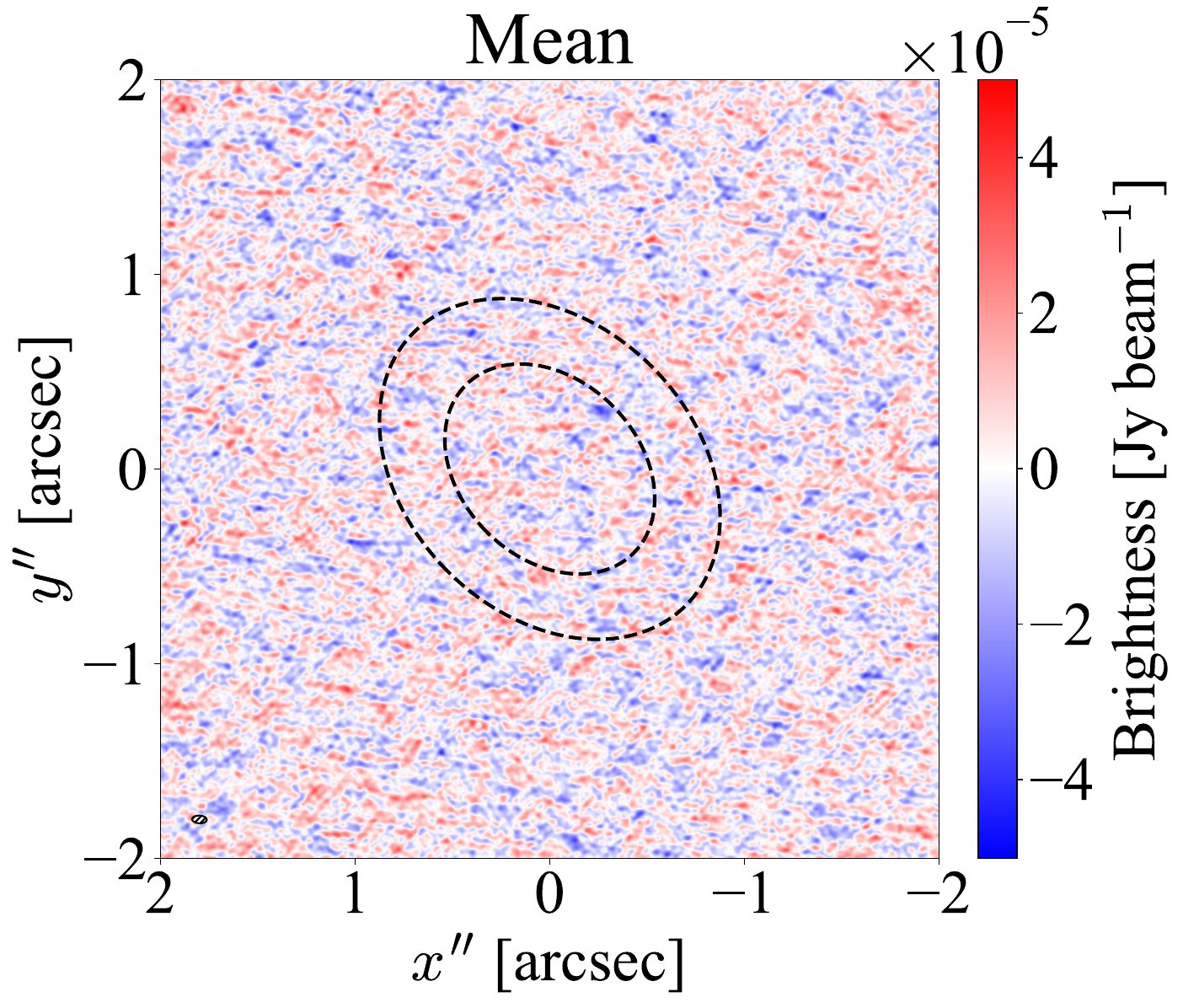}
\includegraphics[width=0.44\linewidth]{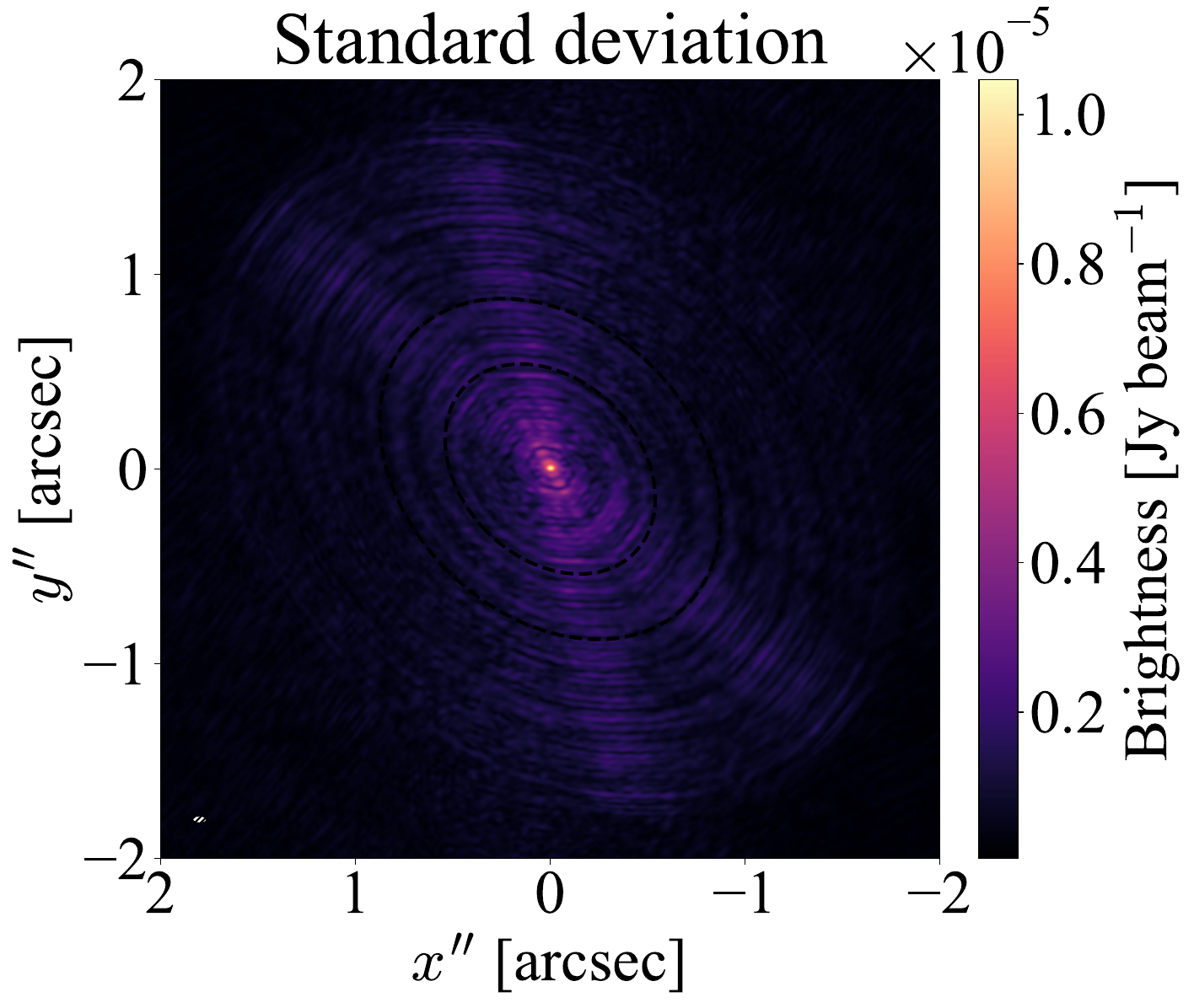}
\includegraphics[width=0.44\linewidth]{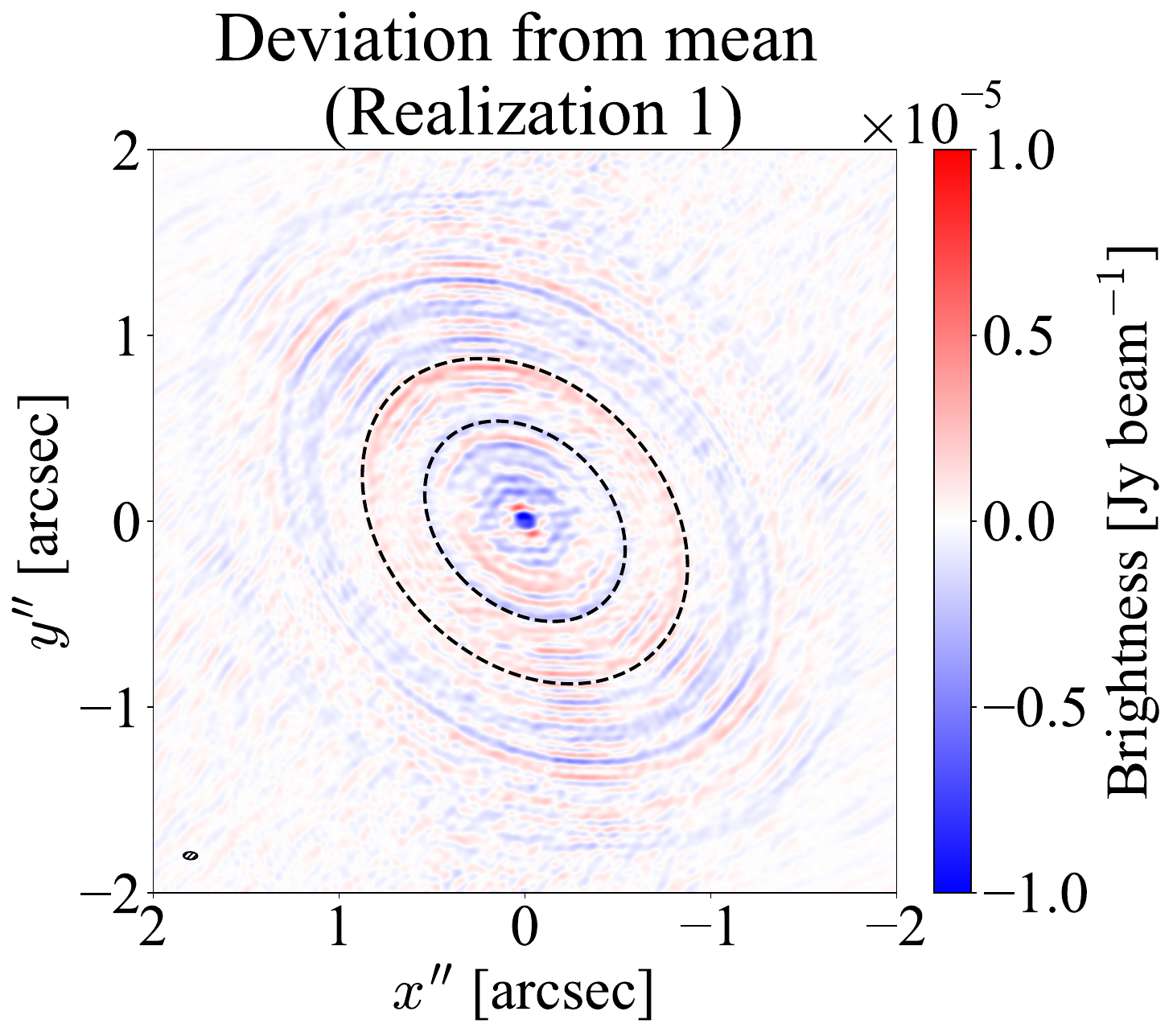}
\includegraphics[width=0.44\linewidth]{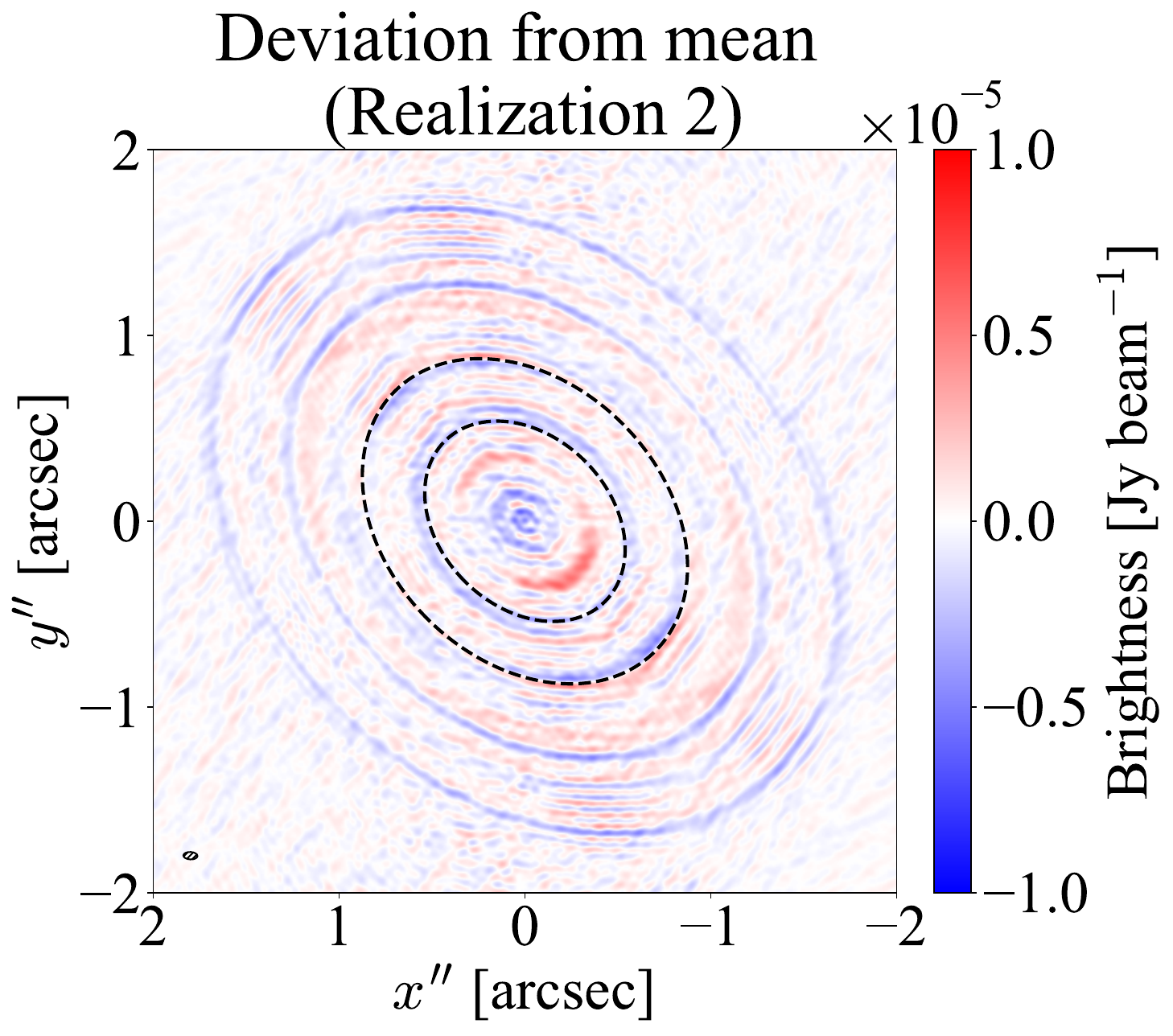}
\end{center}
\caption{Mean and variance of residual images produced with 10 samples of parameters derived from simulated data of AS 209 in Sec \ref{sec:sim}. Upper and left panels show the mean and standard deviation of the residual images. The lower panels show the difference maps of two random residual images from the mean residual image. }
\label{fig:sampled_residual}
\end{figure*}

Assuming the simulated data for AS 209 in Sec \ref{sec:sim}, we produced 10 different residual images by drawing samples from posterior distribution of $(\bm{g}, \bm{\theta}, \bm{a})$. We computed the mean and the standard deviation for 10 residual images, and calculated the differences of two random residual images from the mean image. Fig. \ref{fig:sampled_residual} shows the result. The mean residual image was consistent with that shown in Fig. \ref{fig:wrong_geo}. The image for the standard deviation was nearly axisymmetric, and the amplitude was $10^{-5}$ Jy beam$^{-1}$ at most in the innermost part; whereas, it was $2-4\times10^{-6}$ Jy beam$^{-1}$, which is a few times smaller than the observational error in the residual image, for most of the disc plane. In the lower panels, we show the differences from the mean image, and they are nearly axisymmetric as well. The axisymmetry would arise from the larger variances of the brightness profile than those of geometry and hyperparmaeters. Considering the small amplitudes and axisymmetry of the differential images, it is reasonable to conclude that their effects would be negligible in searching for non-axisymmetric structures, at least in the current observational setup. 

%-----------------------------Figure Start---------------------------
%-----------------------------Figure End------------------------------

\section{Comparison of residual images from two geometries }\label{sec:comp_residuals}
\begin{figure*}
\begin{center}
\includegraphics[width=0.42\linewidth]{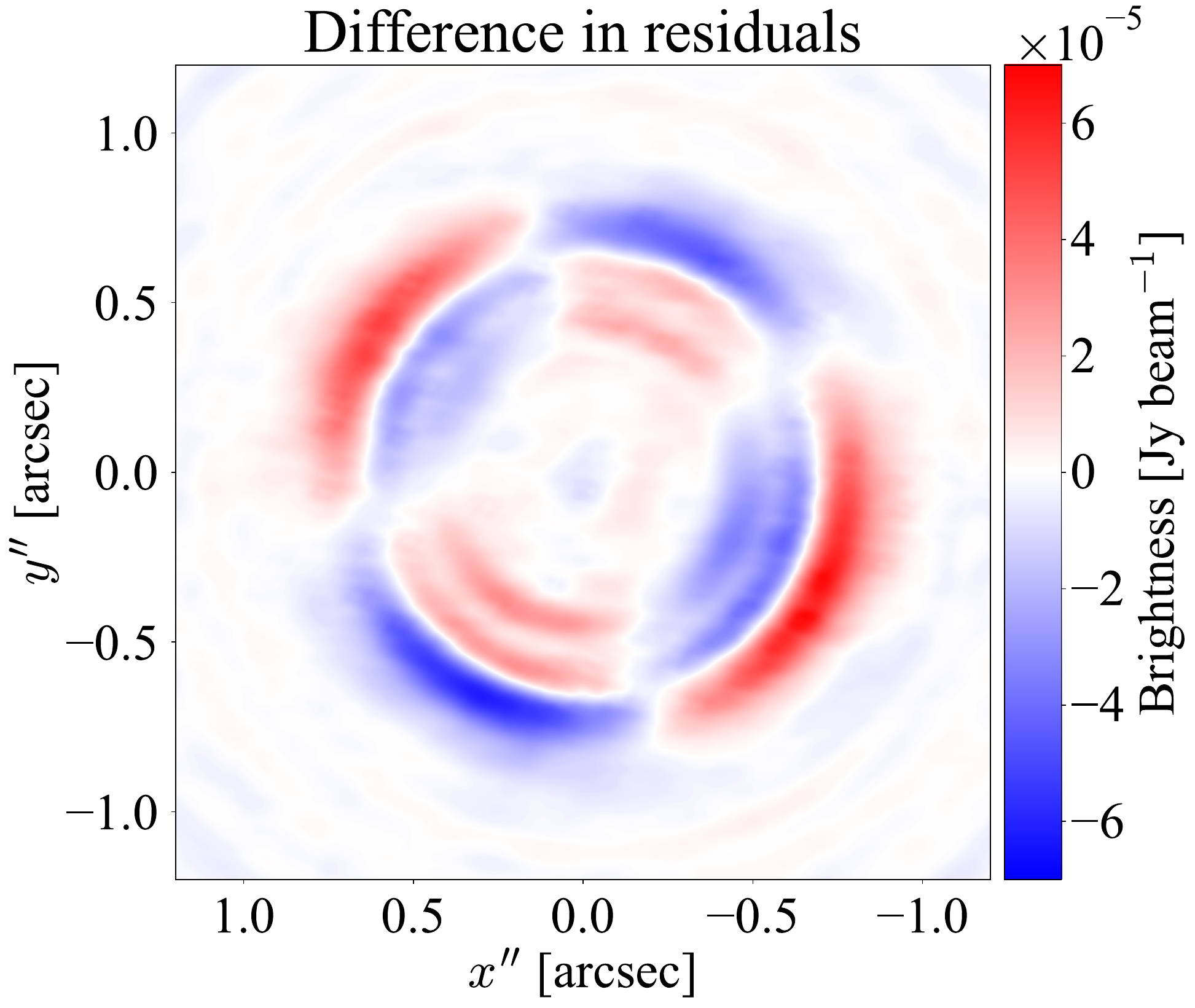}
\includegraphics[width=0.42\linewidth]{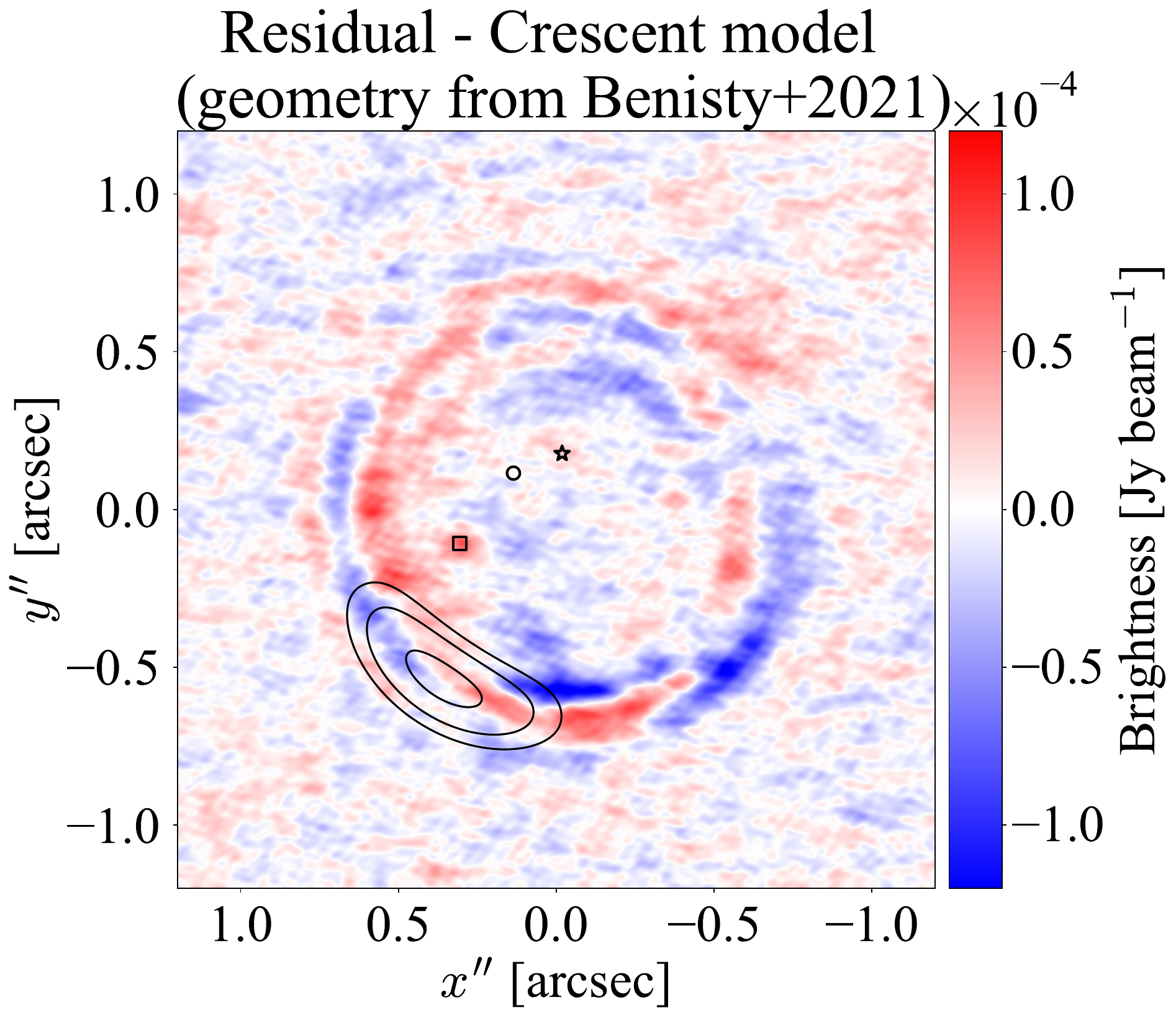}
\includegraphics[width=0.42\linewidth]{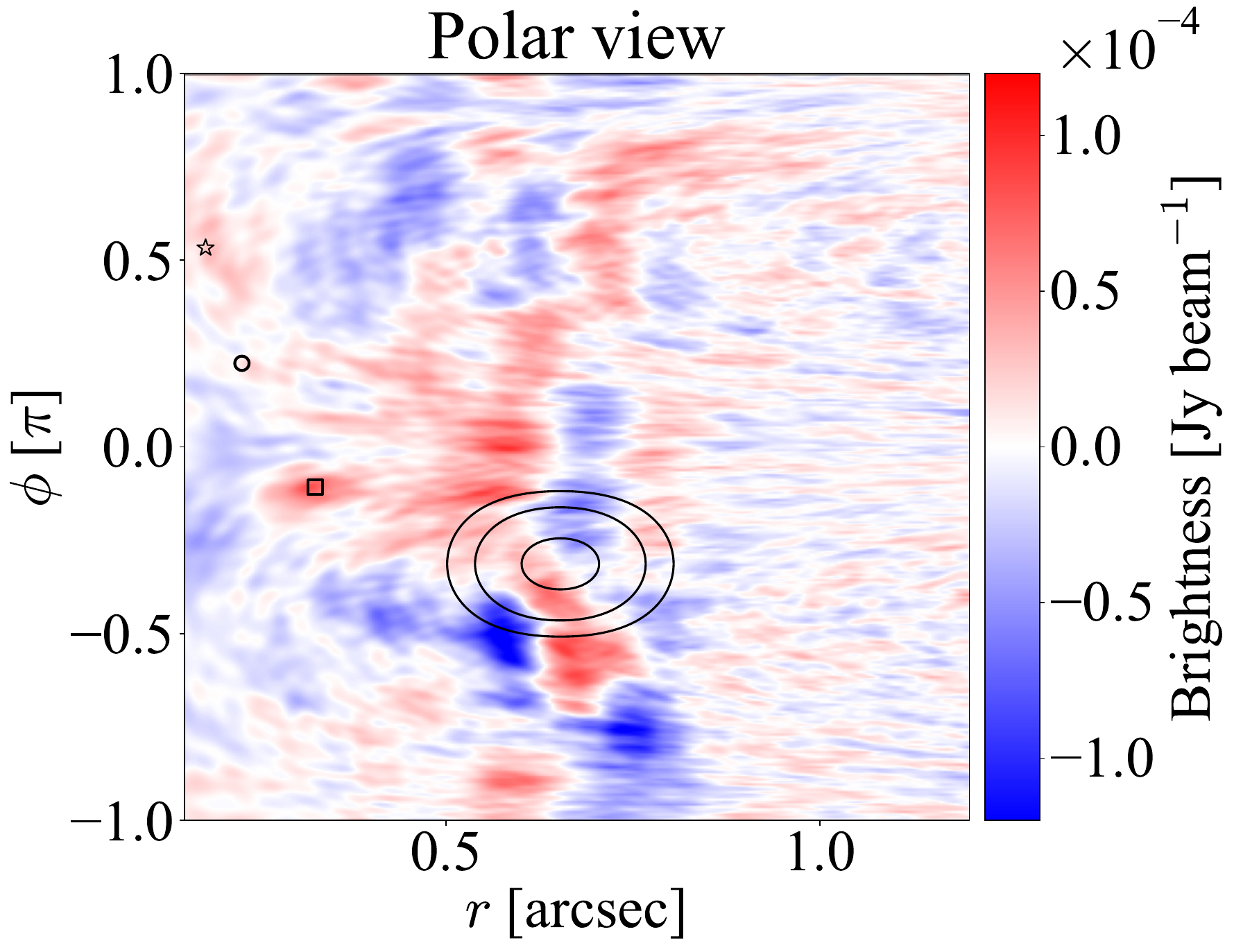}
\end{center}
\caption{Difference in residual images derived from geometry in our study and \protect\cite{benisty2021}, and residual images based on the geometry from \protect\cite{benisty2021} with the crescent model being subtracted. (upper left) Difference in the residual images based on geometries from two studies in Fig \ref{fig:pds70_model_obs}. (upper right) The residual image with the subtraction of the crescent model based on geometry from \protect\cite{benisty2021}. The format is same as that in Fig \ref{fig:pds70_cre_sub}. (lower) The image with the crescent mode being subtracted in polar coordinate.}   
\label{fig:pds_ben_comp}
\end{figure*}

The upper left panel in Fig \ref{fig:pds_ben_comp} illustrates the difference in the deprojected residual images derived from geometries in our study and \cite{benisty2021} (see Fig \ref{fig:pds70_model_obs}). We observed a noticeable $m=2$ pattern in the difference, primarily attributed to the discrepancy in PA by $1^{\circ}$. Additionally, adopting the geometry in \cite{benisty2021}, we also subtracted the crescent model from the residual image following the same procedure described in Sec \ref{sec:sub_cres}, while the model was re-optimized. In the upper right and lower panels in Fig \ref{fig:pds_ben_comp} present the subtracted residual images in a deprojected coordinate and a polar coordinate. 

The estimate of PA appears to be influenced by large-scale faint residuals rather than localized emission (also see Fig F1). Whether to adopt the removal of visibility of localized emission in a specific region as in \cite{benisty2021} might affect the estimation. However, both the removal and non-removal can equally introduce bias, because the removal can result in the loss of information in the data. Thus, although our estimate is obtained by optimization of the data, we have not conclusively determined which estimation is more accurate. 

A major difference is that the residual image derived from geometry in \cite{benisty2021} exhibits a wide leading arm extended from the crescent to the point around $(x'', y'') = (-0.75\arcsec, 0.4\arcsec)$. In addition, the significance of an arm (feature (d) in Sec \ref{sec:sub_cres}) appears to be diminished compared to the image in Fig \ref{fig:pds70_cre_sub}. On the other hand, all of the other features (a-c,e,f) described in Sec \ref{sec:sub_cres} are consistently identified in the two residual images with different geometries. 

% Don't change these lines
\bsp	% typesetting comment
\label{lastpage}
\end{document}